\documentclass[11pt]{article}
\usepackage[preprint]{acl}

\usepackage{times}
\usepackage{latexsym}
\usepackage{booktabs}
\usepackage{array}
\usepackage{arydshln}
\usepackage{multirow}
\usepackage{xcolor}
\usepackage{listings}
\usepackage{comment}
\usepackage[T1]{fontenc}
\usepackage[utf8]{inputenc}
\usepackage{microtype}
\usepackage{inconsolata}
\usepackage{graphicx}
\usepackage{float}
\usepackage{amsmath}
\usepackage{algorithm}
\usepackage{algpseudocode}
\algrenewcommand\algorithmicrequire{\textbf{Input:}}
\algrenewcommand\algorithmicensure{\textbf{Output:}}
\usepackage{amssymb}
\newcommand{\ddfail}{\mathsf{FAIL}}

\newcommand{\ddmean}{\operatorname{mean}}

\lstdefinestyle{appendixprompt}{
  basicstyle=\ttfamily\scriptsize,
  backgroundcolor=\color{gray!10},
  breaklines=true,
  columns=fullflexible,
  keepspaces=true,
  MoreSelectCharTable={\catcode"2019=\active},
  literate={’}{{\textquoteright}}1,
  frame=single,
  framerule=0pt,
  xleftmargin=0.5em,
  xrightmargin=0.5em,
  aboveskip=0.75em,
  belowskip=0.75em
}

\makeatletter
\newenvironment{promptdisplay}{\par\begingroup\def\@captype{figure}}{\par\endgroup}
\makeatother
\newcounter{optimizationstep}
\newlength{\optimizationsteplabelwidth}
\newcommand{\optimizationstepheading}[1]{%
  \refstepcounter{optimizationstep}%
  \paragraph{\mbox{%
    \makebox[\optimizationsteplabelwidth][l]{(\Roman{optimizationstep})}%
    \enspace #1}}}

\title{APO-DD: Dialogue Disentanglement via Reflective Optimization of Multi-Component Prompts}

\author{
  Naoki Takada, Tomohiro Okatsu, Yin Min Pa Pa, Katsunari Yoshioka, Tatsunori Mori \\
  Yokohama National University, Japan \\
  \texttt{\{takada-naoki-xs,mori-tomohiro-bm\}@ynu.jp} \\
  \texttt{\{yinminn-papa-jp,yoshioka,tmori\}@ynu.ac.jp}
}

\begin{document}
\maketitle

\begin{abstract}
In multi-party chat, interleaved dialogues make it difficult to identify the context of each utterance. Dialogue disentanglement separates these dialogues to support dialogue analysis and response generation. Prompt refinement can improve large language models' (LLMs') accuracy on this task, but manually finding effective prompts for each LLM requires extensive trial and error. We propose APO-DD, which adapts GEPA to optimize prompts using execution traces and reflections on failed decisions. It updates task instructions directly and chat representations through automatically generated formatting functions. We compare optimizing either component alone or both. With Qwen3.5-9B, the best optimized prompt achieves an average score of 68.7 across seven evaluation metrics on Ubuntu IRC, compared with 57.5 for the best manually refined prompt. It also outperforms manually refined prompts on other chat domains without further optimization. For this model, optimizing task instructions alone does not always improve accuracy over the initial prompts. Our results suggest that exploring chat representation as an optimization target may yield higher accuracy than optimizing task instructions alone. Our code is available at \url{https://anonymous.4open.science/r/DB6B}.
\end{abstract}

\section{Introduction}\label{sec:introduction}

\begin{figure}[h]
  \centering
  \includegraphics[width=\linewidth]{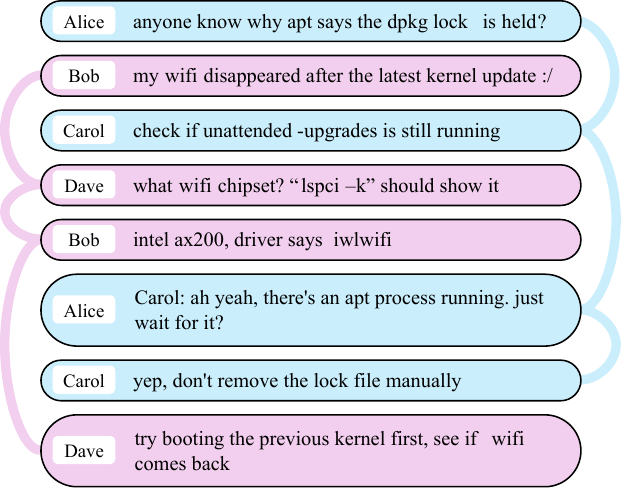}
  \caption{An example of dialogue disentanglement. Each dialogue cluster is shown in a different color.}
  \label{fig:intro}
\end{figure}

In multi-party chat, such as Slack and Internet Relay Chat (IRC), participants discuss different topics in parallel, so dialogues can become entangled. Such entanglement makes chats difficult to follow. It also complicates dialogue analysis, because unrelated utterances may be included in the same context. To address this problem, dialogue disentanglement was proposed~\citep{elsner2008you}. As shown in Figure~\ref{fig:intro}, dialogue disentanglement partitions an utterance sequence into coherent dialogue clusters connected by reply-to relations. This task is an important preprocessing step for downstream applications such as dialogue state tracking~\citep{ouyang2020dialogue, jacqmin2022you} and response generation~\citep{cai2022memory, gu2022hetermpc}.

Large language models (LLMs) show promise for dialogue disentanglement: manual prompt refinement enabled Gemini 2.5 Pro~\citep{comanici2025gemini} to substantially outperform conventional methods~\citep{takada2026rethinking}. However, protecting personal information may require local processing, and processing large amounts of chat data requires efficiency. These needs motivate improving accuracy with small, locally runnable LLMs, which still perform poorly~\citep{takada2026rethinking}. Prompt refinement offers a promising way to improve the accuracy: changing how chat data are presented can improve accuracy~\citep{takada2026rethinking}. However, effective designs vary across LLMs, requiring a search for prompts suited to the target model. Conducting this search manually needs extensive trial and error, motivating automatic prompt optimization. This leads to the following \textbf{Research Question}:

\noindent\emph{How can we automatically optimize prompts to improve disentanglement accuracy?}

To address this question, we propose APO-DD, an automatic prompt optimization framework for dialogue disentanglement. There are two LLM roles: the task LLM, which makes dialogue disentanglement decisions and whose accuracy we aim to improve, and the optimizer LLM, which updates the task LLM's prompt. Motivated by prior work (Section~\ref{sec:related_prompt_optimization}), we divide the task LLM's prompt into two components, the task instruction and the chat representation, and compare optimizing each component alone or both. The optimizer LLM proposes a new prompt using execution traces and reflections on failed decisions. Through repeated updates, we explore diverse prompts and select the one with the highest validation score as the optimized prompt. \noindent\textbf{Our contributions} are summarized as follows:
\begin{itemize}
  \item We propose APO-DD, a framework for automatically optimizing dialogue disentanglement prompts. With Qwen3.5-9B, the optimized prompt outperforms manually refined prompts across topics and chat platforms.
  \item We show that optimizing task instructions alone does not always improve accuracy, highlighting the importance of including chat representation in prompt optimization.
\end{itemize}

\section{Related Work}\label{sec:related_work}

\subsection{Dialogue Disentanglement}
Dialogue disentanglement has been studied under several names: conversation disentanglement~\citep{elsner2008you}, chat disentanglement~\citep{mehri2017chat}, thread detection~\citep{shen2006thread}, and thread extraction~\citep{adams2008topic}. This task has been approached in two stages. First, each utterance is classified as starting a new dialogue or replying to a specific utterance within the preceding context. Second, a clustering algorithm forms dialogue clusters from these predicted reply-to relations. Early methods inferred reply-to relations using handcrafted features such as shared words, time gaps, speaker identities, and user mentions~\citep{elsner2010disentangling}. Later work replaced these features with neural models using RNNs~\citep{mehri2017chat} and LSTMs~\citep{tan2019contextaware}. The release of a large annotated corpus~\citep{kummerfeld2019largescale} improved neural model accuracy by providing large amounts of training data. More recently, incorporating BERT~\citep{devlin2019bert} has further improved the accuracy~\citep{bhukar2023endtoend}.

LLMs have also been applied to dialogue disentanglement, but initial attempts underperformed conventional methods despite their strong performance across NLP tasks~\citep{brown2020language}. First experiments using GPT-4~\citep{openai2024gpt4} performed far below conventional non-LLM methods~\citep{li2024revisiting}, and fine-tuned Llama2-7B~\citep{touvron2023llama} did not close this gap~\citep{aselstyne2025revisiting}. \citet{pal2026disentangling} found that fine-tuned GPT-4o mini~\citep{openai2024gpt4omini} outperformed untuned GPT-4o~\citep{openai2024gpt4o}, but underperformed the non-LLM methods. Prompt refinement offers another way to improve accuracy without changing model weights. \citet{takada2026rethinking} demonstrated its effectiveness by manually refining chat representations. However, the study used simple task instructions and found that effective prompts differed across LLMs. This leaves room to explore both task instructions and chat representations for each LLM, but doing so manually requires extensive trial and error. We therefore use automatic prompt optimization to explore task instructions and chat representations and assess their effects on accuracy.

\subsection{Automatic Prompt Optimization}\label{sec:related_prompt_optimization}

Recent studies suggest that prompt optimization should consider shared task instructions and the input representation that varies with each input. ADO~\citep{lin2025ado} uses an additional LLM call to refine input representations and improve task performance. P3~\citep{zhang2025p3} jointly optimizes shared task instructions and additional instructions tailored to each input. This study finds that the two types of instructions interact and that optimizing either alone can be suboptimal. CFPO~\citep{liu2025prompt} jointly optimizes prompt content and format, including input representation and layout, outperforming content optimization alone. These findings motivate our optimization of two components: the task instruction and the chat representation.

We next consider how to update these two components. SAMMO~\citep{schnabel2024symbolic} uses separate operations to edit instructions and change data formats. It outperforms existing prompt optimization methods on classification tasks. UniPrompt~\citep{juneja2025task} similarly applies feedback-based edits to individual semantic sections. aPSF~\citep{liu2026adaptive} further shows that updating one semantic factor while keeping the others fixed is more effective than changing multiple factors in the same step. Based on these studies, we update only one component at each step: the task instruction or the chat representation.

Finally, we consider how to update the prompts and retain promising candidates. OPRO~\citep{yang2024large} uses previous prompts and their scores, whereas ProTeGi~\citep{pryzant2023automatic} generates feedback from errors. Trace~\citep{cheng2024trace} uses execution traces to guide updates. GEPA~\citep{agrawal2026gepa} combines reflection on execution traces with Pareto-based selection to optimize multiple prompts and retain diverse candidates. These studies suggest using LLM outputs and feedback on errors, not just evaluation scores, to guide prompt updates. We adopt GEPA because it can optimize multiple prompts and outperformed several existing prompt optimization methods. Using its framework, we treat the task instruction and chat representation as separate optimization targets and update one component at a time.

\section{APO-DD}

\subsection{Definition of Dialogue Disentanglement}

Let the chat data $C = (u_1, u_2, \dots, u_n)$ be a chronological sequence of $n$ utterances. Each utterance is represented as $u_i = (t_i, s_i, m_i)$, where $t_i$ is the timestamp, $s_i$ is the speaker ID, and $m_i$ is the message content. The output is a set of dialogue clusters $DC = \{d_1, d_2, \dots, d_p\}$. Each cluster $d_j$ contains utterances that belong to the same dialogue and are connected by reply-to relations. Dialogue disentanglement partitions $C$ into $DC$ so that every utterance belongs to one cluster, that is, $C = \bigcup_{d_j \in DC} d_j$ and $d_j \cap d_k = \emptyset$ for all $j \neq k$.

\subsection{LLM-Based Disentanglement}\label{sec:llm_disentanglement}

\begin{figure}[h]
  \centering
  \includegraphics[width=\linewidth]{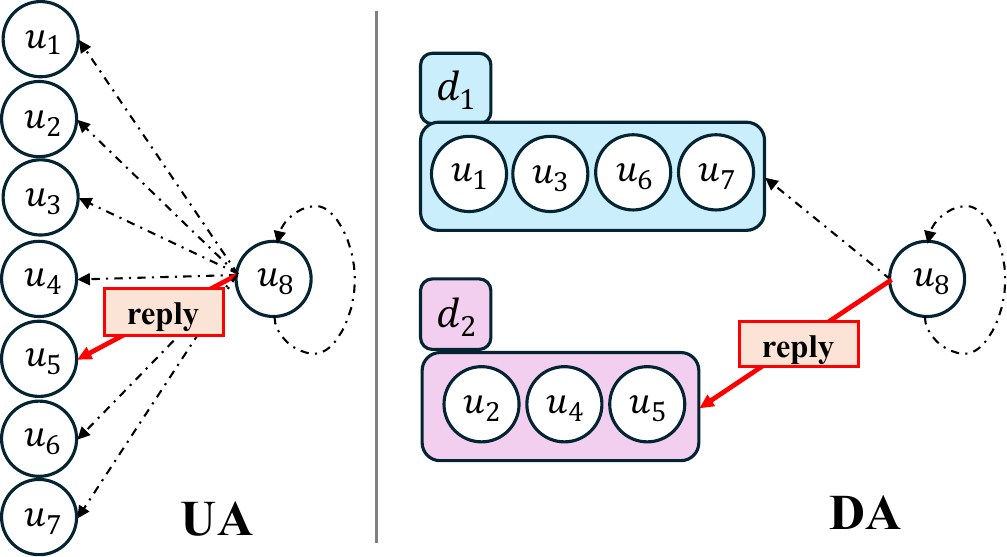}
  \caption{Overview of UA and DA.}
  \label{fig:ua_da}
\end{figure}

We adopt the LLM-based disentanglement framework of \citet{takada2026rethinking}. The LLM makes one decision for each utterance and we denote the target utterance of each decision by $u_{\mathrm{target}}$. Given $u_{\mathrm{target}}$ and its context, the LLM decides whether it starts a new dialogue or continues an existing one. If $u_{\mathrm{target}}$ replies to an existing dialogue, the LLM identifies the utterance or dialogue it replies to, depending on the assignment method. These decisions are used to construct dialogue clusters.

There are four assignment methods, as follows. \textbf{Utterance-level assignment (UA)} selects one preceding utterance as the parent of $u_{\mathrm{target}}$, whereas \textbf{dialogue-level assignment (DA)} selects one dialogue cluster formed by earlier decisions. Figure~\ref{fig:ua_da} illustrates this distinction. UA and DA use only the context preceding $u_{\mathrm{target}}$. Their extensions, \textbf{utterance-level assignment with subsequent context (UA+SC)} and \textbf{dialogue-level assignment with subsequent context (DA+SC)}, also provide the utterances following $u_{\mathrm{target}}$ as auxiliary evidence. We use all four methods because their effectiveness varies across LLMs~\citep{takada2026rethinking}.

\begin{figure*}[t]
  \centering
  \includegraphics[width=\textwidth]{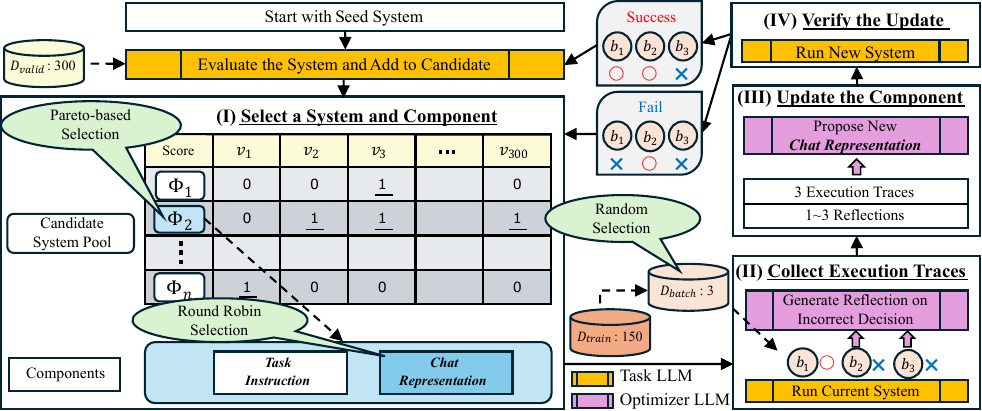}
  \caption{APO-DD optimization cycle. Pareto-based selection chooses a system $\Phi$ from the candidate system pool, and round robin selection chooses which component to update. In this example, the chat representation is selected. The optimizer LLM reflects on execution traces from $D_{\mathrm{batch}}$ to update the component. The updated system is added to the candidate system pool only if its task success rate on $D_{\mathrm{batch}}$ improves.}
  \label{fig:apo_dd}
\end{figure*}

\subsection{Scoring Decisions}\label{sec:scoring_assignment}

To assess whether prompt updates are effective during optimization, APO-DD scores each dialogue disentanglement decision based on the gold dialogue cluster $d_{\mathrm{gold}}$ containing $u_{\mathrm{target}}$. All four assignment methods determine whether $u_{\mathrm{target}}$ starts a new dialogue or continues an existing one. A decision to start a new dialogue is correct only if $u_{\mathrm{target}}$ is the first utterance in $d_{\mathrm{gold}}$. For a decision to continue an existing dialogue, the scoring criterion depends on the assignment method. In UA and UA+SC, the selected parent utterance must precede $u_{\mathrm{target}}$ and belong to $d_{\mathrm{gold}}$. In DA and DA+SC, the selected dialogue cluster must correspond to $d_{\mathrm{gold}}$. Each correct decision receives a score of 1; otherwise, it receives 0.

\subsection{Prompt Components}\label{sec:prompt_components}

APO-DD divides the task LLM's prompt into two components: $P=(p_{\mathrm{task}},p_{\mathrm{chat}})$. \textbf{Task instruction:} $p_{\mathrm{task}}$ is a text explaining the dialogue disentanglement task and specifies the evidence used for the decision. \textbf{Chat representation:} $p_{\mathrm{chat}}$ is the target utterance and its context formatted as text.

\subsection{Dialogue Disentanglement System}\label{sec:system_execution}

We introduce a chat formatting function $f_{\mathrm{chat}}$ to present $u_{\mathrm{target}}$ and its context from $C$, producing $p_{\mathrm{chat}}=f_{\mathrm{chat}}(C,u_{\mathrm{target}})$. We therefore define a dialogue disentanglement system $\Phi$ as follows:
\[
  \Phi=(p_{\mathrm{task}}, f_{\mathrm{chat}})
\]
Given chat data $C$ and a target utterance $u_{\mathrm{target}}$, $\Phi$ applies $f_{\mathrm{chat}}$ to obtain $p_{\mathrm{chat}}$ and calls the task LLM with $P=(p_{\mathrm{task}},p_{\mathrm{chat}})$.

\subsection{Optimization Cycle}

\paragraph{\mbox{Overview}}
APO-DD follows GEPA, combining reflection-based updates with Pareto-based candidate selection (Figure~\ref{fig:apo_dd}). Each run uses one assignment method (Section~\ref{sec:llm_disentanglement}) and starts with a seed system using a minimal prompt. The seed and accepted updates are evaluated on $D_{\mathrm{valid}}$ and added to the candidate system pool. Each iteration follows the four steps below to update the task instruction or chat representation. Algorithmic details and prompts are provided in Appendices~\ref{sec:algorithm_details} and~\ref{sec:prompt}.

\optimizationstepheading{Select a System and Component} 
To select a candidate system for update, we apply Pareto-based selection using each system's decision scores on $D_{\mathrm{valid}}$ (Section~\ref{sec:scoring_assignment}). First, the systems that attain the maximum score on at least one utterance in $D_{\mathrm{valid}}$ are selected from the candidate system pool. Next, redundant systems are removed if every utterance on which they make a correct decision also receives a correct decision from another remaining system. Then, one system is sampled with probability increasing with its number of correct decisions on $D_{\mathrm{valid}}$. This strategy preserves systems that perform well on different validation utterances. Finally, the task instruction or chat representation is selected as the update target in round robin order.

\optimizationstepheading{Collect Execution Traces}
The selected system $\Phi$ is run with reasoning enabled on each of the three utterances in a randomly sampled $D_{\mathrm{batch}}\subset D_{\mathrm{train}}$. Each execution trace records the task LLM input, its final answer, and whether the decision is correct. If all three decisions are correct, the batch is discarded and a new batch is sampled. For each incorrect decision, a natural language reflection is generated by the optimizer LLM from the task LLM's execution trace and the gold answer.

\optimizationstepheading{Propose an Update} 
For the system and component selected in step~(I), the optimizer LLM proposes an update based on the traces and reflections collected in step~(II). In the following, we use primes to denote updated elements. For a task instruction update, the proposed instruction text directly becomes $p'_{\mathrm{task}}$, producing $\Phi'=(p'_{\mathrm{task}},f_{\mathrm{chat}})$. For a chat representation update, the optimizer LLM proposes a representation to improve the task LLM's decision accuracy. We instruct the optimizer LLM to change only the presentation while preserving message text and candidate IDs, without generating task instructions or adding inferred information. In an additional call, the optimizer LLM generates $f'_{\mathrm{chat}}$ to format chat data according to the proposal. The generated formatting function is reused at inference time, so formatting new chat data requires no additional LLM call. The updated system is $\Phi'=(p_{\mathrm{task}},f'_{\mathrm{chat}})$.

\optimizationstepheading{Verify the Update}
$\Phi'$ is evaluated on the same $D_{\mathrm{batch}}$ with reasoning disabled. If its score exceeds $\Phi$'s score from step~(II), it is evaluated on $D_{\mathrm{valid}}$ and added to the candidate system pool with its validation score. Otherwise, it is discarded.

\section{Experimental Setup}\label{sec:experimental_setup}
To evaluate APO-DD, we compare disentanglement accuracy between optimized systems and those using seed or manually refined prompts. We also examine the effects of optimization targets and generalization across datasets and LLMs.

\subsection{LLM Configuration}\label{sec:models_hyperparameters}
We use Qwen3.5-9B~\citep{qwen3.5-9b} as the task LLM because it achieved the highest AVG with $\Phi_{\mathrm{manual}}$ on $D_{\mathrm{test}}$ among the evaluated models with $\leq$10B parameters (Appendix~\ref{sec:reference_llms}). We set \texttt{temperature} to 0. The optimizer LLM is Qwen3.8-Flash~\citep{qiu2026design}, with reasoning disabled and all other parameters left at the API defaults (\texttt{temperature}=0.7). To analyze generalization across LLMs (Section~\ref{sec:prompt_generalization}), we evaluate the systems optimized for Qwen3.5-9B with Gemma-4-12B-it~\citep{gemma4-12b}, without further optimization. Reasoning is disabled for both task LLMs in all benchmark evaluations.

Our best Qwen3.5-9B system runs on a 24\,GB RTX 3090 GPU (as detailed in Appendix~\ref{sec:app_local_inference}).

\subsection{Optimization Settings}

We apply APO-DD separately to each of the four assignment methods (Section~\ref{sec:llm_disentanglement}). The prompts used by the optimizer LLM to propose updates are based on GEPA's update prompt~\citep{agrawal2026gepa} and are provided in Appendix~\ref{sec:prompt}. We performed one optimization run per configuration, with a budget of 100k LLM calls fixed before examining test results. Section~\ref{sec:optimization_budget} analyzes how disentanglement accuracy changes as optimization progresses and discusses the potential for improvement beyond this budget.

All four assignment methods (Section~\ref{sec:llm_disentanglement}) use 50 preceding utterances. This number is based on our analysis of 47,500 utterances in Ubuntu IRC's \texttt{Train-500msg} split~\citep{kummerfeld2019largescale}: 90\%, 95\%, 99\%, and 99.9\% of reply relations span at most 11, 17, 49, and 241 utterances, respectively. The 50 preceding utterances include the utterance being replied to in approximately 99\% of reply relations and span roughly three times the 95th percentile distance. However, these reply distances may not generalize to other chat data, and including the parent does not guarantee sufficient context for a correct decision. UA+SC and DA+SC use 15 subsequent utterances. We chose this number because \citet{takada2026rethinking} reported that one model showed its largest improvement under UA+SC at this length, although the study found no consistent benefit across LLMs.

\subsection{Datasets}\label{sec:dataset}

We use the Ubuntu IRC dataset~\citep{kummerfeld2019largescale}. It contains real multi-party chat logs of technical discussions about Ubuntu. We use this dataset for optimization and evaluation as follows.

\textbf{\boldmath$D_{\mathrm{train}}$ (150 utterances)} is randomly sampled from utterances for which the seed system made incorrect decisions under UA on the \texttt{Train-100msg} and \texttt{Channel 2} splits. Each $D_{\mathrm{batch}}$ contains three utterances randomly sampled from $D_{\mathrm{train}}$. If all three decisions are correct, the batch is discarded without a prompt update. To reduce such cases, $D_{\mathrm{train}}$ is restricted to utterances for which the seed system made incorrect decisions.

\textbf{\boldmath$D_{\mathrm{valid}}$ (300 utterances)} is randomly sampled from the \texttt{Dev} split (2,500 utterances in 10 samples of 250). Before sampling $D_{\mathrm{train}}$ or $D_{\mathrm{valid}}$, we exclude system messages and utterances annotated as replies to utterances outside the preceding 50 utterances.

\textbf{\boldmath$D_{\mathrm{test}}$ (5,000 utterances)} contains the entire \texttt{Test} split. It is used for benchmark evaluation and analysis of the optimization budget, not for optimization.

To assess generalization across topics and chat platforms, we evaluate on two additional datasets. IRC New Domains~\citep{gouravajhala2023chat} contains 800 utterances on different topics. Slack Chats~\citep{chatterjee2020softwarerelated} contains 3,544 utterances on similar technical topics but from a different platform.

\subsection{Evaluated Systems}\label{sec:prompt_settings}
We evaluate the following dialogue disentanglement systems. $\Phi_{\mathrm{seed}}$ is the initial system for optimization and uses a minimal prompt. $\Phi_{\mathrm{opt}}$ is obtained by optimizing task instructions alone $\Phi_{\mathrm{opt}}$ (task), chat representation alone $\Phi_{\mathrm{opt}}$ (chat), or both components $\Phi_{\mathrm{opt}}$ (both). $\Phi_{\mathrm{manual}}$ adapts the manually refined prompt of \citet{takada2026rethinking}, which achieved high accuracy. The original manual prompt requests a reason field for each decision, increasing output length. To reduce generation time, we replace the original output instructions with a request to output only the reply target's ID or \texttt{new} if the target utterance starts a new dialogue. Prompt examples are provided in Appendix~\ref{sec:disentanglement_prompts}.

\subsection{Metrics}
Following prior studies, we compare predicted and gold dialogue clusters, focusing on clustering agreement and exact matches. We measure clustering agreement using Variation of Information (VI)~\citep{meila2007comparing}, Adjusted Rand Index (ARI)~\citep{hubert1985comparing}, One-to-One accuracy (1-1)~\citep{elsner2010disentangling}, and Shen-F1 (S-F)~\citep{shen2006thread}. An exact match occurs when a predicted cluster contains exactly the same utterances as a gold cluster. We measure exact matches using precision (P), recall (R), and F1~\citep{kummerfeld2019largescale}. Our main metric is AVG, the average of these seven scores, with higher values indicating better performance. All results are averaged over three evaluation runs with different seeds.

\subsection{Baselines}\label{sec:baselines}

We compare $\Phi_{\mathrm{opt}}$ with non-LLM and LLM-based methods. As non-LLM methods, we choose earlier machine learning models: Elsner~\citep{elsner2008you} and Feedforward+vote~\citep{kummerfeld2019largescale}; and more accurate BERT based models: DialBERT+conv~\citep{li2021dialbert}, StructuralBERT~\citep{ma2022structural}, StructuralBERT+DRL~\citep{bhukar2023endtoend}, and DiHRL~\citep{li2024revisiting}. For LLM-based methods, we compare fine-tuned Llama2-7B~\citep{aselstyne2025revisiting}, fine-tuned GPT-4o mini~\citep{pal2026disentangling}, and $\Phi_{\mathrm{manual}}$~\citep{takada2026rethinking} with Qwen3.5-9B.

\section{Results}\label{sec:results}

\subsection{Effects of Prompt Optimization}\label{sec:optimization_targets}

\begin{figure}[h]
  \centering
  \includegraphics[width=\linewidth]{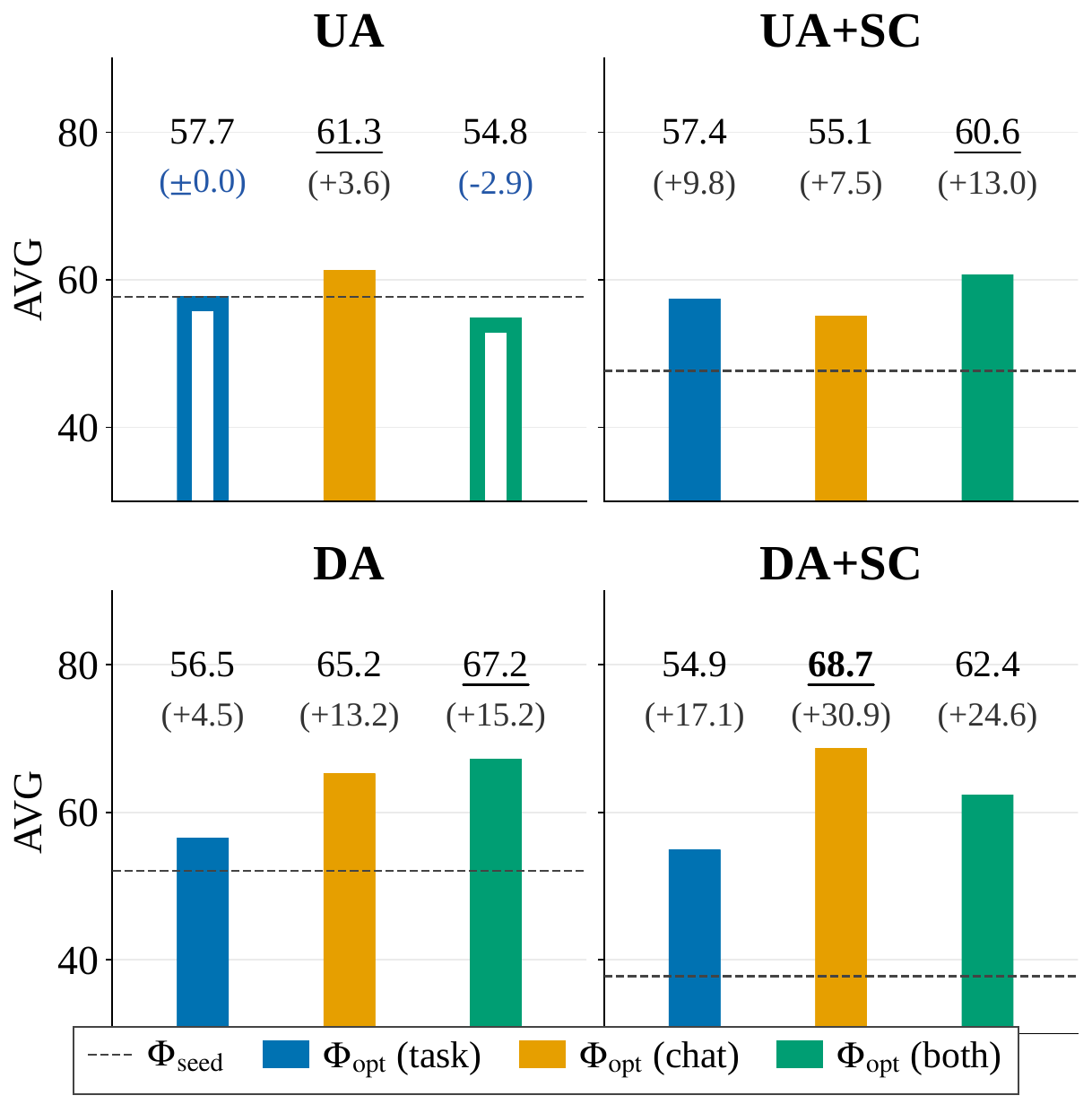}
  \caption{AVG on $D_{\mathrm{test}}$ when optimizing task instructions, chat representations, or both. $\Phi_{\mathrm{opt}}$ (chat) improves AVG across all assignment methods and achieves the highest score.}
  \label{fig:main_results}
\end{figure}

Figure~\ref{fig:main_results} compares AVG on $D_{\mathrm{test}}$ between $\Phi_{\mathrm{opt}}$ and $\Phi_{\mathrm{seed}}$ for each optimization target and assignment method. $\Phi_{\mathrm{opt}}$ (chat) improves AVG under every assignment method, whereas $\Phi_{\mathrm{opt}}$ (task) and $\Phi_{\mathrm{opt}}$ (both) do not consistently improve. Under UA, $\Phi_{\mathrm{opt}}$ (task) is unchanged from $\Phi_{\mathrm{seed}}$ because no updated system outperformed $\Phi_{\mathrm{seed}}$ on $D_{\mathrm{valid}}$, while $\Phi_{\mathrm{opt}}$ (both) performs worse. The best optimization target varies across assignment methods, but every method's highest AVG is achieved by $\Phi_{\mathrm{opt}}$ (chat) or $\Phi_{\mathrm{opt}}$ (both). These results highlight the importance of including chat representation in prompt optimization.

\begin{table*}[t]
  \centering
  \small
  \resizebox{\textwidth}{!}{%
  \begin{tabular}{@{}llcccccccc@{}}
    \toprule
    \multicolumn{2}{@{}l}{\textbf{Method}} & \textbf{VI} & \textbf{ARI} & \textbf{1-1} & \textbf{S-F} & \textbf{P} & \textbf{R} & \textbf{F1} & \textbf{AVG} \\
    \midrule
    \multicolumn{10}{@{}l}{\textbf{\emph{Non-LLM methods}}} \\
    \multicolumn{2}{@{}l}{Elsner~\cite{elsner2008you}}                        & 82.1 & -    & 51.4 & -    & 12.1 & 21.5 & 15.5 & - \\
    \multicolumn{2}{@{}l}{Feedforward+vote~\cite{kummerfeld2019largescale}}   & 91.5 & -    & 76.0 & -    & 36.3 & 39.7 & 38.0 & - \\
    \multicolumn{2}{@{}l}{DialBERT+conv~\cite{li2021dialbert}}                & 93.2 & 72.8 & 79.7 & -    & 42.1 & 47.9 & 44.8 & - \\
    \multicolumn{2}{@{}l}{StructuralBERT~\cite{ma2022structural}}             & 94.6 & 76.8 & \underline{84.2} & -    & \underline{\textbf{51.8}} & 51.8 & 51.7 & - \\
    \multicolumn{2}{@{}l}{StructuralBERT+DRL~\cite{bhukar2023endtoend}}       & \underline{\textbf{96.2}} & 78.5 & -    & -    & 51.5 & \underline{\textbf{52.3}} & \underline{\textbf{51.9}} & - \\
    \multicolumn{2}{@{}l}{DiHRL~\cite{li2024revisiting}}                      & 94.2 & \underline{81.1} & \underline{84.2} & \underline{87.5} & 48.0 & 49.9 & 48.9 & \underline{\textbf{70.5}} \\
    \midrule
    \multicolumn{10}{@{}l}{\textbf{\emph{LLM-based methods}}} \\
    Fine tuning~\citep{aselstyne2025revisiting}               & Llama2-7B     & -    & -    & 74.0 & 77.0 & -    & -    & -    & - \\
    Fine tuning~\citep{pal2026disentangling}                  & GPT-4o mini   & 92.5 & -    & 78.0 & -    & 42.6 & 48.5 & 45.3 & - \\
    $\Phi_{\mathrm{manual}}$, UA~\citep{takada2026rethinking} & Qwen3.5-9B    & 90.2 & 63.9 & 74.2 & 76.2 & 33.0 & 32.3 & 32.7 & 57.5 \\
    $\Phi_{\mathrm{opt}}$ (chat), DA+SC (Ours)                & Qwen3.5-9B    & \underline{94.5} & \underline{\textbf{83.4}} & \underline{\textbf{84.8}} & \underline{\textbf{88.2}} & 41.3 & 45.3 & 43.2 & \underline{68.7} \\
    \bottomrule
  \end{tabular}
  }
  \caption{Dialogue disentanglement performance on Ubuntu IRC. AVG is the average of seven metrics. Bold marks the best overall scores and underlining marks the best within each group. With Qwen3.5-9B, $\Phi_{\mathrm{opt}}$ (chat) outperforms $\Phi_{\mathrm{manual}}$ on all metrics and performs best on some metrics, but below DiHRL in AVG.}
  \label{tab:02_compare_with_sota}
\end{table*}



\subsection{Comparison with Baselines}\label{sec:baseline_comparison}
Table~\ref{tab:02_compare_with_sota} compares $\Phi_{\mathrm{opt}}$ with baselines on $D_{\mathrm{test}}$. Within each optimization run, we select $\Phi_{\mathrm{opt}}$ using $D_{\mathrm{valid}}$. For this comparison, we report the results for the settings of $\Phi_{\mathrm{opt}}$ and $\Phi_{\mathrm{manual}}$ that achieve the highest AVG on $D_{\mathrm{test}}$. $\Phi_{\mathrm{opt}}$ (chat) outperforms $\Phi_{\mathrm{manual}}$ on every metric and achieves the highest ARI, 1-1, and S-F among the compared methods. However, DiHRL has a higher AVG, and all BERT based methods outperform $\Phi_{\mathrm{opt}}$ (chat) on exact cluster match metrics (P, R, and F1).

\subsection{Generalization Across Chat Domains}\label{sec:generalization}

\begin{table}[H]
\centering
\begingroup
\fontsize{10}{11}\selectfont
\setlength{\tabcolsep}{1.5pt}
\begin{tabular*}{\linewidth}{@{\extracolsep{\fill}}lccccc@{}}
\toprule
\textbf{Method} & \textbf{\boldmath $\Phi_{\mathrm{seed}}$} & \textbf{\boldmath $\Phi_{\mathrm{manual}}$} & \textbf{\boldmath $\Phi_{\mathrm{opt}}$} & \textbf{\boldmath $\Phi_{\mathrm{opt}}$} & \textbf{\boldmath $\Phi_{\mathrm{opt}}$} \\
 & & & \textbf{(task)} & \textbf{(chat)} & \textbf{(both)} \\
\midrule
\multicolumn{6}{@{}l}{\textbf{\emph{IRC New Domains}}} \\
\addlinespace[1.5pt]
UA & $\underline{66.8}$ & \textcolor[HTML]{2457A7}{$62.1$} & \textcolor[HTML]{2457A7}{$\underline{66.8}$} & \textcolor[HTML]{2457A7}{$60.6$} & \textcolor[HTML]{2457A7}{$59.5$} \\
\addlinespace[1.5pt]
UA+SC & $45.3$ & $56.7$ & $59.7$ & $\underline{61.3}$ & $60.0$ \\
\addlinespace[1.5pt]
DA & $58.4$ & \textcolor[HTML]{2457A7}{$57.9$} & \textcolor[HTML]{2457A7}{$57.8$} & $\underline{65.3}$ & $60.3$ \\
\addlinespace[1.5pt]
DA+SC & $44.3$ & $62.8$ & $53.4$ & $\underline{\mathbf{67.9}}$ & $64.8$ \\
\addlinespace[1.5pt]
\midrule
\multicolumn{6}{@{}l}{\textbf{\emph{Slack Chats}}} \\
\addlinespace[1.5pt]
UA & $\underline{63.6}$ & \textcolor[HTML]{2457A7}{$62.4$} & \textcolor[HTML]{2457A7}{$\underline{63.6}$} & \textcolor[HTML]{2457A7}{$62.3$} & \textcolor[HTML]{2457A7}{$52.8$} \\
\addlinespace[1.5pt]
UA+SC & $43.0$ & $58.9$ & $\underline{63.3}$ & $57.2$ & $57.5$ \\
\addlinespace[1.5pt]
DA & $58.9$ & $\underline{62.9}$ & \textcolor[HTML]{2457A7}{$58.1$} & $62.9$ & \textcolor[HTML]{2457A7}{$56.3$} \\
\addlinespace[1.5pt]
DA+SC & $47.3$ & $65.7$ & $59.3$ & $\underline{\mathbf{66.9}}$ & $59.5$ \\
\addlinespace[1.5pt]
\bottomrule
\end{tabular*}
\endgroup
\caption{AVG across chat domains with Qwen3.5-9B, without further optimization. Bold marks the highest AVG per dataset and underlining the highest per row. Blue indicates no improvement over $\Phi_{\mathrm{seed}}$. $\Phi_{\mathrm{opt}}$ (chat) under DA+SC achieves the highest AVG on both datasets, as on $D_{\mathrm{test}}$.}
\label{tab:domain_transfer}
\end{table}

Table~\ref{tab:domain_transfer} reports AVG on IRC New Domains and Slack Chats with Qwen3.5-9B, without further optimization. Although improvements vary across assignment methods and datasets, $\Phi_{\mathrm{opt}}$ (chat) outperforms $\Phi_{\mathrm{seed}}$ more often than $\Phi_{\mathrm{opt}}$ (task) or $\Phi_{\mathrm{opt}}$ (both). In particular, $\Phi_{\mathrm{opt}}$ (chat) under DA+SC, which achieves the highest AVG on $D_{\mathrm{test}}$, also ranks first on both datasets, outperforming the best $\Phi_{\mathrm{manual}}$ on each. Thus, among the evaluated settings, $\Phi_{\mathrm{opt}}$ (chat) under DA+SC is the best-performing system for Qwen3.5-9B across all three datasets.

\subsection{Generalization Across LLMs}\label{sec:prompt_generalization}

\begin{figure}[H]
  \centering
  \includegraphics[width=\linewidth]{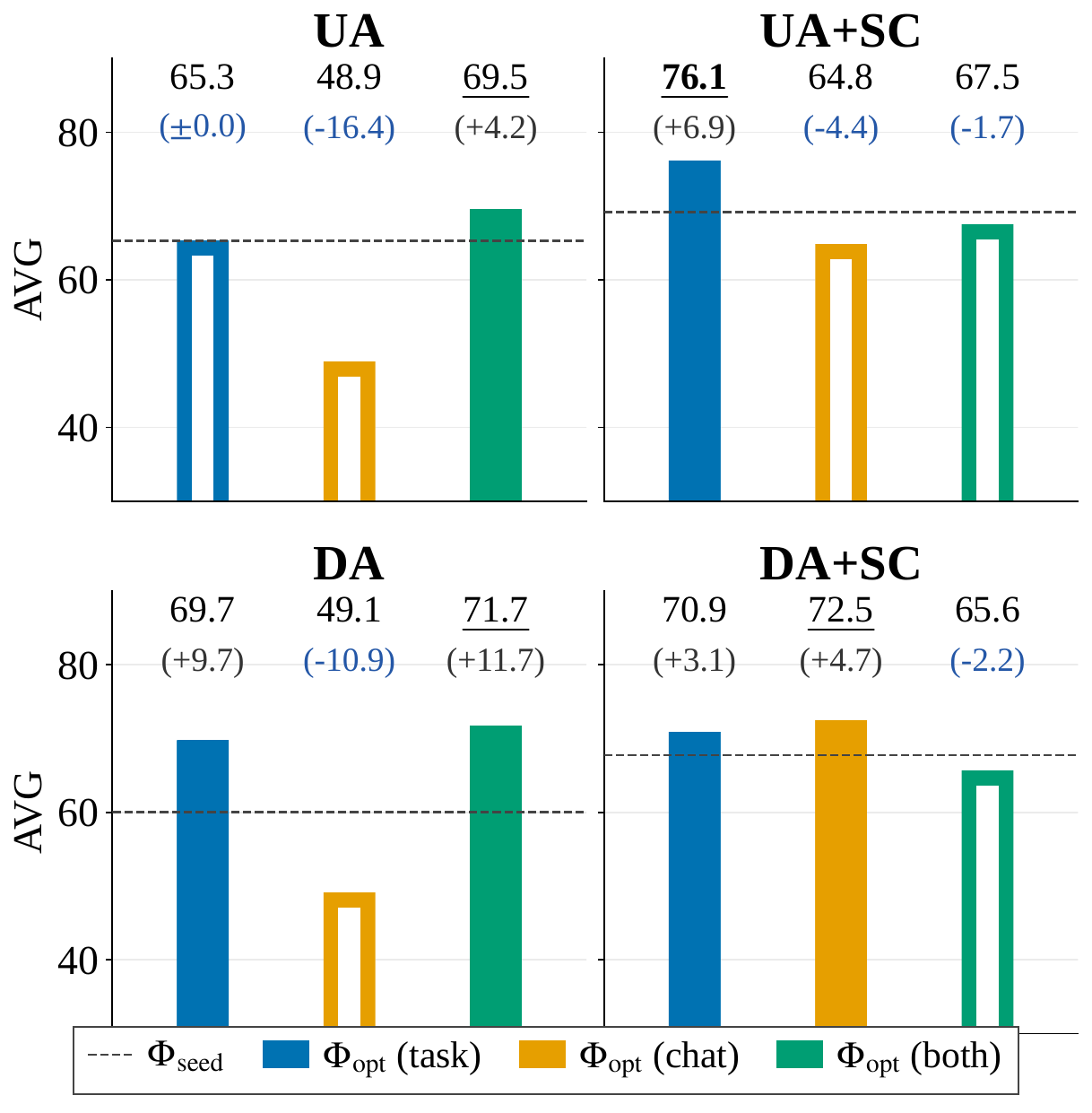}
  \caption{AVG on $D_{\mathrm{test}}$ with Gemma-4-12B-it using systems optimized for Qwen3.5-9B, without further optimization. Some systems fall below $\Phi_{\mathrm{seed}}$, indicating that improvements do not necessarily transfer across LLMs.}
  \label{fig:gemma_transfer}
\end{figure}

Figure~\ref{fig:gemma_transfer} reports AVG on $D_{\mathrm{test}}$ with Gemma-4-12B-it using systems optimized for Qwen3.5-9B, without further optimization. $\Phi_{\mathrm{opt}}$ (task) outperforms $\Phi_{\mathrm{seed}}$ except for the system under UA, which is the same as $\Phi_{\mathrm{seed}}$. In contrast, $\Phi_{\mathrm{opt}}$ (chat) falls below $\Phi_{\mathrm{seed}}$ under most assignment methods, despite consistent improvements with Qwen3.5-9B. Thus, improvements from prompt optimization do not necessarily transfer across LLMs.

\section{Analysis}
We examine the optimized prompts and analyze the optimization budget.

\subsection{Characteristics of Optimized Prompts}\label{sec:optimized_prompt_analysis}
We examine how APO-DD changed the prompt of $\Phi_{\mathrm{seed}}$, focusing on the DA+SC examples in Appendix~\ref{sec:disentanglement_prompts}. The seed task instruction briefly specifies the task and output format, whereas the task instruction in $\Phi_{\mathrm{opt}}$ (task) includes detailed rules, examples, and a decision procedure. For example, it directs the LLM to check prior interactions between speakers and avoid relying solely on keywords or temporal proximity. Despite these additions, $\Phi_{\mathrm{opt}}$ (task) achieves lower AVG than $\Phi_{\mathrm{opt}}$ (chat) on $D_{\mathrm{test}}$ with Qwen3.5-9B. This suggests that making task instructions more detailed is not always the best way to improve accuracy.

The seed chat representation groups utterances by dialogue ID and separates the target and subsequent utterances. $\Phi_{\mathrm{opt}}$ (chat) retains this structure but adds participant lists and marks subsequent utterances as non-selectable context. The optimizer LLM is instructed to propose updates only to the chat representation and its explanation, excluding task instructions. Despite these restrictions, the chat representation in this DA+SC example contained brief task instructions telling the task LLM to consider semantic links, not just matching speakers. With this chat representation, $\Phi_{\mathrm{opt}}$ (chat) under DA+SC achieves the highest AVG on all three datasets in the Qwen3.5-9B experiments. Each of the other assignment methods also achieves its highest AVG on $D_{\mathrm{test}}$ with $\Phi_{\mathrm{opt}}$ (chat) or $\Phi_{\mathrm{opt}}$ (both), rather than $\Phi_{\mathrm{opt}}$ (task). These results support exploring chat representation as an optimization target for dialogue disentanglement.

\subsection{Optimization Budget}\label{sec:optimization_budget}
\begin{figure}[h]
  \centering
  \includegraphics[width=\linewidth]{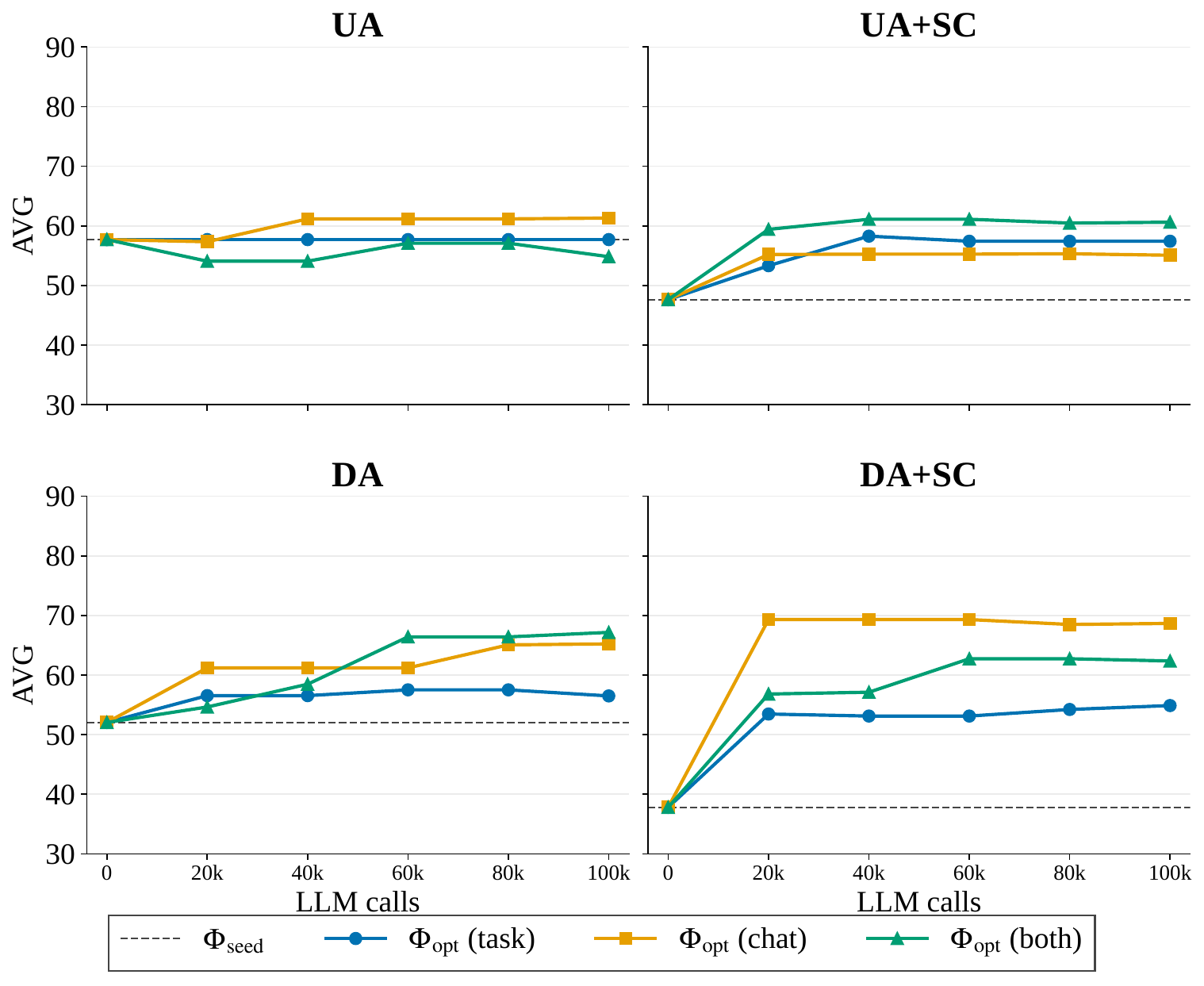}
  \caption{AVG on $D_{\mathrm{test}}$ as the optimization budget increases. Substantial improvements occur up to 80k calls, with little change from 80k to 100k.}
  \label{fig:optimization_progress}
\end{figure}

Figure~\ref{fig:optimization_progress} provides a retrospective analysis of AVG on $D_{\mathrm{test}}$ under the preset 100k call budget. At each checkpoint, we evaluate the system with the highest $D_{\mathrm{valid}}$ score among candidates validated within the call limit. Most gains occur before 80k calls, with little change from 80k to 100k. This does not imply convergence. In extended runs, a single improvement on $D_{\mathrm{valid}}$ sometimes required roughly 100k additional LLM calls (Appendix~\ref{sec:extended_optimization}, Figure~\ref{fig:extended_optimization}). Further improvements on $D_{\mathrm{valid}}$ are possible, but optimization until convergence may be costly. These observations do not identify an optimal budget. Future work includes exploring larger budgets or achieving high accuracy with fewer LLM calls.

\section{Conclusion}
We proposed APO-DD, a framework that adapts GEPA for automatic prompt optimization in dialogue disentanglement. With Qwen3.5-9B and reasoning disabled, $\Phi_{\mathrm{opt}}$ (chat) under DA+SC achieved the highest AVG among the evaluated settings on all three datasets, outperforming the best $\Phi_{\mathrm{manual}}$ on each without further optimization. On the main benchmark, $\Phi_{\mathrm{opt}}$ (chat) achieved an AVG of 68.7 without fine-tuning, compared with 57.5 for the best $\Phi_{\mathrm{manual}}$. The optimized system's AVG was also close to the highest reported for the compared non-LLM methods. These results support exploring chat representation as an optimization target rather than restricting optimization to task instructions. However, improvements did not consistently transfer to Gemma-4-12B-it, and our results do not identify a universally best assignment method or optimization target. Future work could explore more efficient optimization with predictable budget requirements. Another direction to further improve accuracy is supervised fine-tuning with optimized prompts.

\section{Limitations}

\paragraph{GEPA Configuration}
We did not systematically tune the optimization budget, the sizes of $D_{\mathrm{train}}$, $D_{\mathrm{batch}}$, and $D_{\mathrm{valid}}$, or the numbers of preceding and subsequent utterances in each prompt. GEPA also allows users to define trace collection and use. Our configuration enables reasoning during trace collection but uses final answers, not reasoning text, for reflection generation and prompt updates. We did not evaluate alternative trace configurations, and suitable settings may differ across LLMs and chat data.

\paragraph{Variability Across Optimization Runs}
We performed only one optimization run per configuration. Each selected system is evaluated three times. The confidence intervals reflect variation across these evaluations, not across optimization runs.

\paragraph{Scope of Chat Representation Optimization}
Although the optimizer LLM is instructed not to include task instructions in the chat representation, generated chat representations can contain brief task instruction. We did not evaluate their contribution separately from formatting changes. 

\paragraph{Limited LLM Coverage}
We optimized prompts only for Qwen3.5-9B, selected because it achieved the highest AVG with $\Phi_{\mathrm{manual}}$ among the evaluated models with $\leq$10B parameters (Appendix~\ref{sec:reference_llms}). We evaluated transfer of the optimized systems to Gemma-4-12B-it without further optimization. Whether optimizing prompts for other LLMs, including larger models, yields similar improvements remains untested.

\paragraph{Comparison with Non-LLM Methods}
We compare non-LLM methods using their reported results on $D_{\mathrm{test}}$, as we could not find publicly available trained models for StructuralBERT, StructuralBERT+DRL, or DiHRL, and found accessible code only for StructuralBERT (Appendix~\ref{sec:baseline_availability}). We also could not find reported results for these baselines on IRC New Domains or Slack Chats and did not evaluate them on these datasets. Comparisons with non-LLM methods are therefore limited to $D_{\mathrm{test}}$.

\paragraph{Potential Pretraining Overlap}
All datasets used in this study are publicly available and may have been included in model pretraining. Our results may therefore not fully reflect generalization to unseen data.

\section{Ethical Considerations}
We use three publicly available chat datasets: Ubuntu IRC~\citep{kummerfeld2019largescale}, IRC New Domains~\citep{gouravajhala2023chat}, and Slack Chats~\citep{chatterjee2020softwarerelated}. Ubuntu IRC and IRC New Domains are licensed under CC BY 4.0. Slack Chats is used only for evaluation and is not redistributed. No private or confidential chat data were used. Processing independently collected chat data requires careful privacy protection, as they may contain personal or sensitive information. Such data should be processed locally, or anonymized before transmission to an external API.

\subsection{AI Tool Usage}
We used ChatGPT to assist with coding and translation. The authors take responsibility for the paper and code.

\section*{Acknowledgments}

This paper is based on results obtained from a project, JPNP24003, commissioned by the New Energy and Industrial Technology Development Organization (NEDO).

\bibliography{custom,custom_zotero}

\clearpage
\appendix
\setcounter{table}{0}
\renewcommand{\thetable}{A\arabic{table}}
\renewcommand{\theHtable}{appendix.\arabic{table}}
\setcounter{figure}{0}
\renewcommand{\thefigure}{A\arabic{figure}}
\renewcommand{\theHfigure}{appendix.\arabic{figure}}
\setcounter{algorithm}{0}
\renewcommand{\thealgorithm}{A\arabic{algorithm}}
\providecommand{\theHalgorithm}{}
\renewcommand{\theHalgorithm}{appendix.\arabic{algorithm}}
\setcounter{equation}{0}
\renewcommand{\theequation}{A\arabic{equation}}
\renewcommand{\theHequation}{appendix.\arabic{equation}}

\section{Detailed Algorithms}\label{sec:algorithm_details}
The assignment method and ID convention are fixed for each run. If communication fails, the LLM call is retried without limit. Failed communication attempts do not increase the call count.

\subsection{Dialogue Disentanglement}\label{sec:disentanglement_algorithm}
Algorithm~\ref{alg:dd-disentangle} runs a selected system without prompt updates. The function $f_{\mathrm{chat}}(C,u_{\mathrm{target}})$ constructs the context and converts it to chat text using the selected seed, manual, or optimized system's formatter. The displayed context construction applies to seed and optimized systems: up to 50 preceding utterances and, for UA+SC and DA+SC, up to 15 subsequent utterances. UA+SC with $\Phi_{\mathrm{manual}}$ may include candidate utterances earlier than this 50-utterance window. $C_{\mathrm{pre}}$ and $C_{\mathrm{sub}}$ contain the selected preceding and subsequent utterances; $DC_{\mathrm{pre}}$ groups the selected preceding utterances by dialogue. \textsc{TaskLLM} calls the LLM and parses its output text. Outputs other than \texttt{new} or a selectable candidate ID are treated as $\ddfail$. Invalid outputs are retried at most twice with different random seeds. Communication failures do not advance the loop counter $r$.

In Algorithm~\ref{alg:dd-disentangle}, indices start at 1: $i$ is the target utterance's position in $C$, $k$ indexes preceding or subsequent utterances, $j$ is the next dialogue index, and $r$ counts LLM output attempts. Prompt IDs follow the selected system's code or template, rather than necessarily matching these indices. $\hat{u}$ is the utterance identified by the returned ID; $\hat{d}$ refers to the corresponding dialogue in $DC$, whose update is reflected in $DC$. For DA and DA+SC, the displayed dialogue ID identifies the original dialogue in $DC$, not just its portion in $DC_{\mathrm{pre}}$.

\begin{algorithm}[htbp]
\small
\caption{Dialogue disentanglement with a selected system}
\label{alg:dd-disentangle}
\begin{algorithmic}[1]
\Require $C=(u_1,\ldots,u_n)$; $\Phi=(p_{\mathrm{task}},f_{\mathrm{chat}})$
\Statex \hspace{\algorithmicindent} $DC=\emptyset$ (shared with $f_{\mathrm{chat}}$)
\Statex \hspace{\algorithmicindent} $\mu\in\{\texttt{UA},\texttt{UA+SC},\texttt{DA},\texttt{DA+SC}\}$
\Ensure $DC=\{d_1,\ldots,d_p\}$ partitioning $C$
\Statex
\State $d_1\gets\{u_1\}$; $DC\gets\{d_1\}$; $j\gets2$
\If{$\mu\in\{\texttt{UA},\texttt{UA+SC}\}$}
    \For{$i=2,\ldots,|C|$}
        \State $u_{\mathrm{target}}\gets u_i$
        \State $p_{\mathrm{chat}}\gets f_{\mathrm{chat}}(C,u_{\mathrm{target}})$
        \State $P\gets(p_{\mathrm{task}},p_{\mathrm{chat}})$
        \For{$r=1,\ldots,3$}
            \Statex \Comment{If the LLM output is invalid, change the seed and retry.}
            \State $\alpha\gets\Call{TaskLLM}{P}$
            \Statex \Comment{utterance ID in $p_{\mathrm{chat}}$, \texttt{new}, or $\ddfail$}
            \If{$\alpha\ne\ddfail$}
                \State \textbf{break}
            \EndIf
        \EndFor
        \If{$\alpha\in\{\texttt{new},\ddfail\}$}
            \State $d_j\gets\{u_{\mathrm{target}}\}$
            \State $DC\gets DC\cup\{d_j\}$; $j\gets j+1$
        \Else
            \Statex \Comment{$\hat{u}$: utterance with ID $\alpha$ in $p_{\mathrm{chat}}$}
            \State $\hat{d}\gets\hat{d}\cup\{u_{\mathrm{target}}\}$ where $\hat{d}\in DC,\ \hat{u}\in\hat{d}$
        \EndIf
    \EndFor
\Else
    \For{$i=2,\ldots,|C|$}
        \State $u_{\mathrm{target}}\gets u_i$
        \State $p_{\mathrm{chat}}\gets f_{\mathrm{chat}}(C,u_{\mathrm{target}})$
        \State $P\gets(p_{\mathrm{task}},p_{\mathrm{chat}})$
        \For{$r=1,\ldots,3$}
            \Statex \Comment{If the LLM output is invalid, change the seed and retry.}
            \State $\alpha\gets\Call{TaskLLM}{P}$
            \Statex \Comment{dialogue ID in $p_{\mathrm{chat}}$, \texttt{new}, or $\ddfail$}
            \If{$\alpha\ne\ddfail$}
                \State \textbf{break}
            \EndIf
        \EndFor
        \If{$\alpha\in\{\texttt{new},\ddfail\}$}
            \State $d_j\gets\{u_{\mathrm{target}}\}$
            \State $DC\gets DC\cup\{d_j\}$; $j\gets j+1$
        \Else
            \Statex \Comment{$\hat{d}$: dialogue in $DC$ with ID $\alpha$ in $p_{\mathrm{chat}}$}
            \State $\hat{d}\gets\hat{d}\cup\{u_{\mathrm{target}}\}$ where $\hat{d}\in DC$
        \EndIf
    \EndFor
\EndIf
\State \Return $DC$
\Statex
\Function{$f_{\mathrm{chat}}$}{$C,u_{\mathrm{target}}$}
    \State $i\gets\operatorname{index}_C(u_{\mathrm{target}})$\label{alg:dd-prepare-start}
    \State $C_{\mathrm{pre}}\gets\{u_k\in C:\max(1,i-50)\le k<i\}$
    \Statex \hspace{\algorithmicindent}\Comment{50 preceding utterances, or all if fewer}
    \State $C_{\mathrm{sub}}\gets\emptyset$
    \If{$\mu\in\{\texttt{UA+SC},\texttt{DA+SC}\}$}
        \State $C_{\mathrm{sub}}\gets\{u_k\in C:i<k\le\min(|C|,i+15)\}$
    \EndIf
    \If{$\mu\in\{\texttt{UA},\texttt{UA+SC}\}$}
        \State \Return chat text for $(C_{\mathrm{pre}},u_{\mathrm{target}},C_{\mathrm{sub}})$
    \Else
        \State $DC_{\mathrm{pre}}\gets\{d\cap C_{\mathrm{pre}}\mid d\in DC,\ d\cap C_{\mathrm{pre}}\ne\emptyset\}$\label{alg:dd-prepare-end}
        \State \Return chat text for $(DC_{\mathrm{pre}},u_{\mathrm{target}},C_{\mathrm{sub}})$
    \EndIf
\EndFunction
\end{algorithmic}
\end{algorithm}

\subsection{APO-DD}\label{sec:apo_dd_algorithm}
In Algorithms~\ref{alg:dd-disentangle}--\ref{alg:dd-render}, $\mu$ is the assignment method and $\alpha$ the parsed decision. Algorithm~\ref{alg:dd-optimize} uses an ordered list $\mathcal K$ of permitted components (\textit{task instruction} or \textit{chat representation}, denoted $\mathrm{task}$ and $\mathrm{chat}$), the candidate system pool $\mathcal P$ in insertion order, a call limit $B$, a random seed $\sigma$, and an iteration index $\ell$ starting at 0. The shared counter $N$ records cumulative LLM calls (Appendix~\ref{sec:app_call_budget}).

For each system $\Phi$, $\mathbf v_\Phi$ stores its binary validation scores, $b_\Phi$ the call count at validation completion, $q_\Phi$ the next component index in $\mathcal K$, and $\rho_\Phi\in\{0,1\}$ whether a code repair attempt is available. Subscripts identify system objects, not prompt texts. An all-correct batch during execution trace collection is skipped without advancing $q_\Phi$. Otherwise, component selection advances $q_\Phi$ and copies it to $q_{\Phi'}$, even if the subsequent update fails or is discarded. \textsc{DeepCopy} copies the prompt and code; $q_\Phi$ is copied explicitly. \textsc{UniformSample} samples without replacement, and \textsc{SaveState} saves the complete optimization state.

Square brackets denote ordered lists, $\Vert$ concatenation, and a list exponent its repetition count. Dots in a returned tuple mark unused elements. Score entries are indexed by $z$. The operator $\operatorname*{arg\,max}^{\mathrm{first}}$ selects the earliest inserted maximizer. Batch verification compares scores $\mathbf v_1$ with reasoning disabled against trace collection scores $\mathbf v_0$ with reasoning enabled; the latter are not recomputed with reasoning disabled.

\begin{algorithm}[htbp]
\small
\caption{APO-DD optimization cycle}
\label{alg:dd-optimize}
\begin{algorithmic}[1]
\Require $D_{\mathrm{train}}$, $D_{\mathrm{valid}}$; $\mu$, $\mathcal K$, $B$, $\sigma$
\Ensure $\Phi_{\mathrm{opt}}$
\State $N\gets0$; $\ell\gets-1$
\State $\Phi_{\mathrm{seed}}\gets\Call{LoadSeed}{\mu}$
\State $q_{\Phi_{\mathrm{seed}}}\gets1$; $\rho_{\Phi_{\mathrm{seed}}}\gets0$
\State $(\mathbf v_{\Phi_{\mathrm{seed}}},\cdot,\cdot)\gets$
\Statex \hspace{\algorithmicindent}$\Call{Evaluate}{\Phi_{\mathrm{seed}},D_{\mathrm{valid}},0}$
\State $b_{\Phi_{\mathrm{seed}}}\gets N$; $\mathcal P\gets[\Phi_{\mathrm{seed}}]$
\While{$N<B$}
    \State $\ell\gets\ell+1$
    \State $\Phi\gets\Call{SelectSystem}{\mathcal P,D_{\mathrm{valid}},\sigma,\ell}$
    \State $D_{\mathrm{batch}}\gets\Call{UniformSample}{D_{\mathrm{train}},3}$
    \State $\Phi'\gets\Call{DeepCopy}{\Phi}$
    \State $q_{\Phi'}\gets q_\Phi$; $\rho_{\Phi'}\gets0$
    \State $(\mathbf v_0,\mathcal T,\mathcal F)\gets$
    \Statex \hspace{\algorithmicindent}$\Call{Evaluate}{\Phi',D_{\mathrm{batch}},1}$
    \If{$\sum_z v_{0,z}=|D_{\mathrm{batch}}|$}
        \State $\Call{SaveState}{}$; \textbf{continue}
    \EndIf
    \State $\kappa\gets\mathcal K[q_\Phi]$
    \State $q_\Phi\gets1+(q_\Phi\bmod|\mathcal K|)$
    \State $q_{\Phi'}\gets q_\Phi$
    \State $\Phi'\gets\Call{Update}{\Phi',\kappa,\mathcal T,\mathcal F}$
    \If{$\Phi'=\ddfail$}
        \State $\Call{SaveState}{}$; \textbf{continue}
    \EndIf
    \State $E\gets\Call{Evaluate}{\Phi',D_{\mathrm{batch}},0}$
    \If{$E=\ddfail$}
        \State $\Call{SaveState}{}$; \textbf{continue}
    \EndIf
    \State $(\mathbf v_1,\cdot,\cdot)\gets E$
    \If{$\sum_z v_{1,z}\le\sum_z v_{0,z}$}
        \State $\Call{SaveState}{}$; \textbf{continue}
    \EndIf
    \State $E\gets\Call{Evaluate}{\Phi',D_{\mathrm{valid}},0}$
    \If{$E\ne\ddfail$}
        \State $(\mathbf v_{\Phi'},\cdot,\cdot)\gets E$; $b_{\Phi'}\gets N$
        \State $\mathcal P\gets\mathcal P\Vert[\Phi']$
    \EndIf
    \State $\Call{SaveState}{}$
\EndWhile
\State $\displaystyle\Phi_{\mathrm{opt}}\gets
    \operatorname*{arg\,max}_{\substack{\Phi\in\mathcal P\\b_\Phi\le B}}^{\mathrm{first}}
    \sum_z v_{\Phi,z}$
\State \Return $\Phi_{\mathrm{opt}}$
\end{algorithmic}
\end{algorithm}

\paragraph{Pareto-based System Selection}
We use GEPA's candidate selection code unchanged. In Algorithm~\ref{alg:dd-select}, $\mathcal L_z$ contains all systems attaining the maximum score on validation instance $z$, including ties at zero. $Q$ is the temporary list of retained systems, $\Psi$ a system to remove, and $W$ the sampling list, with each system repeated according to its number of appearances in the retained $\mathcal L_z$. Filtering does not change $\mathcal P$ or its scores.

\textsc{FirstSeen} traverses the sets in instance order and retains each system's first occurrence, using Python's set traversal order. \textsc{StableSort} sorts by ascending mean score without reordering ties. \textsc{First} returns the first list element, or $\ddfail$ for an empty list. \textsc{Remove} deletes the specified system from $Q$ while preserving the order of the remaining systems. \textsc{Draw} uses Python's \texttt{random.Random} with the seed string $\operatorname{str}(\sigma)\Vert\texttt{:candidate:}\Vert\operatorname{str}(\ell)$ and applies \texttt{choice} to the stored system IDs in $W$.

\begin{algorithm}[htbp]
\small
\caption{Pareto-based system selection}
\label{alg:dd-select}
\begin{algorithmic}[1]
\Require $\mathcal P$, $(\mathbf v_\Phi)_{\Phi\in\mathcal P}$; $D_{\mathrm{valid}}$, $\sigma$, $\ell$
\Ensure $\Phi\in\mathcal P$
\Function{SelectSystem}{$\mathcal P,D_{\mathrm{valid}},\sigma,\ell$}
    \For{$z=1,\ldots,|D_{\mathrm{valid}}|$}
        \State $\displaystyle\mathcal L_z\gets
            \operatorname*{arg\,max}_{\Phi\in\mathcal P}v_{\Phi,z}$
    \EndFor
    \State $Q\gets\Call{FirstSeen}{(\mathcal L_z)_z}$
    \State $Q\gets\Call{StableSort}{Q,\Phi\mapsto\ddmean(\mathbf v_\Phi)}$
    \Loop
        \State $\Psi\gets\text{\textsc{First}}([\Phi\in Q:$
        \Statex \hspace{\algorithmicindent}$\forall z,\ \Phi\in\mathcal L_z\Rightarrow$
        \Statex \hspace{\algorithmicindent}$(\mathcal L_z\cap\operatorname{set}(Q))\setminus\{\Phi\}\ne\emptyset])$
        \If{$\Psi=\ddfail$}
            \State \textbf{break}
        \EndIf
        \State $Q\gets\Call{Remove}{Q,\Psi}$
    \EndLoop
    \For{$z=1,\ldots,|D_{\mathrm{valid}}|$}
        \State $\mathcal L_z\gets\mathcal L_z\cap\operatorname{set}(Q)$
    \EndFor
    \State $W\gets[]$
    \For{$\Phi\in\Call{FirstSeen}{(\mathcal L_z)_z}$}
        \State $W\gets W\Vert[\Phi]^{|\{z:\Phi\in\mathcal L_z\}|}$
    \EndFor
    \State \Return $\Call{Draw}{W,\sigma,\ell}$
\EndFunction
\end{algorithmic}
\end{algorithm}

\paragraph{Evaluation and Code Repair}
Algorithm~\ref{alg:dd-evaluate} processes instances $x=(C,u_{\mathrm{target}},DC_{\mathrm{gold}})$ in stored order, where $DC_{\mathrm{gold}}$ contains the gold dialogue clusters. With $\eta=1$, the task LLM uses reasoning, execution traces are collected, and the optimizer LLM generates reflections on incorrect decisions. With $\eta=0$, reasoning is disabled for batch verification and validation. The result $E=(\mathbf v,\mathcal T,\mathcal F)$ contains binary scores, traces, and separate reflections; $\mathcal T$ and $\mathcal F$ are empty when $\eta=0$. Here, $\nu$ denotes a binary score and $\mathbf{1}[\cdot]$ the indicator function.

\textsc{TaskLLM}$(P,X,\eta)$ returns the parsed decision $\alpha$ and final answer text $y$. Reflections use $(P,y,g)$, and prompt updates use traces $(P,y,\nu)$ together with the reflections. $\mathcal A_\mu(X)$ contains only IDs of preceding utterances for UA and UA+SC, or IDs of their dialogue groups for DA and DA+SC. Invalid outputs are recorded as $\alpha=\ddfail$ and scored 0 without retries. \textsc{Resolve}$(X,\alpha)$ returns the utterance $\hat u$ identified by $\alpha$ for UA and UA+SC. For DA and DA+SC, it returns the original gold dialogue cluster $\hat d\in DC_{\mathrm{gold}}$ containing the displayed subset identified by $\alpha$, not the subset itself. \textsc{GoldID}$(X,u_{\mathrm{target}},d_{\mathrm{gold}},\mu)$ selects the most recent preceding utterance in $d_{\mathrm{gold}}$ among the displayed candidates. It returns that utterance's displayed ID for UA and UA+SC, or the corresponding displayed dialogue ID for DA and DA+SC. If $u_{\mathrm{target}}$ is the first utterance in $d_{\mathrm{gold}}$, it returns \texttt{new}. This value, $g$, is passed to reflection. For UA and UA+SC, scoring also accepts other preceding utterances in the same gold dialogue cluster.

If chat formatting fails during execution trace collection ($\eta=1$), \textsc{FailedBatch} replaces the entire batch result with zero scores and one failure trace $(x,\epsilon,0)$ per instance, where $\epsilon$ describes the formatting failure. \textsc{ReflectFailure} calls the optimizer LLM once per instance using the current system, the instance, and the error information. Both lists follow the stored instance order. Thus, all three scores are 0, and the three failure traces and reflections are passed to the selected component update. These traces contain error information, not task LLM outputs. Formatting failures during batch verification or validation ($\eta=0$) instead return $\ddfail$.

Generated chat formatting functions sometimes raised errors, so we introduced code repair. Code repair changes only the working copy, not a stored candidate. The flag $\delta=1$ indicates repair during the current \textsc{Render} call. Algorithm~\ref{alg:dd-evaluate} first prepares all inputs, storing $(x,p_{\mathrm{chat}},X)$ in the ordered list $\mathcal H$. Successful repair clears this list and restarts input preparation on the entire $D$. Only after preparation succeeds are the task LLM calls, scoring, and decision reflections performed. Unrecoverable formatting failures follow the failure handling above. Repair during validation does not repeat batch verification.

\begin{algorithm}[htbp]
\small
\caption{Decision scoring, execution trace collection, and reflection generation}
\label{alg:dd-evaluate}
\begin{algorithmic}[1]
\Require $\Phi=(p_{\mathrm{task}},f_{\mathrm{chat}})$; $D\in\{D_{\mathrm{batch}},D_{\mathrm{valid}}\}$;
\Statex \hspace{\algorithmicindent} $\eta\in\{0,1\}$
\Ensure $E=(\mathbf v,\mathcal T,\mathcal F)$ or $\ddfail$
\Function{Evaluate}{$\Phi,D,\eta$}
    \Repeat
        \State $\mathcal H\gets[]$; $\delta\gets0$
        \For{$x\in D$}
            \State $(C,u_{\mathrm{target}},DC_{\mathrm{gold}})\gets x$
            \State $(p_{\mathrm{chat}},\delta,X,\epsilon)\gets$
            \Statex \hspace{\algorithmicindent}$\Call{Render}{\Phi,C,u_{\mathrm{target}},DC_{\mathrm{gold}}}$
            \If{$p_{\mathrm{chat}}=\ddfail$}
                \If{$\eta=1$}
                    \State \Return $\Call{FailedBatch}{\Phi,D,\epsilon}$
                \EndIf
                \State \Return $\ddfail$
            \EndIf
            \If{$\delta=1$}
                \State \textbf{break}
            \EndIf
            \State $\mathcal H\gets\mathcal H\Vert[(x,p_{\mathrm{chat}},X)]$
        \EndFor
    \Until{$\delta=0$}
    \State $\mathbf v\gets[]$; $\mathcal T\gets[]$; $\mathcal F\gets[]$
    \For{$(x,p_{\mathrm{chat}},X)\in\mathcal H$}
        \State $(C,u_{\mathrm{target}},DC_{\mathrm{gold}})\gets x$
        \State $i\gets\operatorname{index}_C(u_{\mathrm{target}})$
        \State $P\gets(p_{\mathrm{task}},p_{\mathrm{chat}})$
        \State $(\alpha,y)\gets\Call{TaskLLM}{P,X,\eta}$
        \Statex \Comment{$y$: final answer text only}
        \If{$\alpha\notin\mathcal A_\mu(X)\cup\{\texttt{new}\}$}
            \State $\alpha\gets\ddfail$
        \EndIf
        \State $\{d_{\mathrm{gold}}\}\gets\{d\in DC_{\mathrm{gold}}:u_{\mathrm{target}}\in d\}$
        \State $\nu\gets0$
        \If{$\alpha=\texttt{new}$}
            \State $\nu\gets\mathbf{1}[i=\min\{k:u_k\in d_{\mathrm{gold}}\}]$
        \ElsIf{$\alpha\ne\ddfail$}
            \If{$\mu\in\{\texttt{UA},\texttt{UA+SC}\}$}
                \State $\hat u\gets\Call{Resolve}{X,\alpha}$
                \State $\nu\gets\mathbf{1}[\hat u\in d_{\mathrm{gold}}]$
            \Else
                \State $\hat d\gets\Call{Resolve}{X,\alpha}$
                \State $\nu\gets\mathbf{1}[\hat d=d_{\mathrm{gold}}]$
            \EndIf
        \EndIf
        \State $\mathbf v\gets\mathbf v\Vert[\nu]$
        \If{$\eta=1$}
            \State $\mathcal T\gets\mathcal T\Vert[(P,y,\nu)]$
            \If{$\nu=0$}
                \State $g\gets\Call{GoldID}{X,u_{\mathrm{target}},d_{\mathrm{gold}},\mu}$
                \State $\mathcal F\gets\mathcal F\Vert[(x,\Call{Reflect}{P,y,g})]$
            \EndIf
        \EndIf
    \EndFor
    \State \Return $(\mathbf v,\mathcal T,\mathcal F)$
\EndFunction
\Statex
\Function{FailedBatch}{$\Phi,D,\epsilon$}
    \State $\mathcal T\gets[(x,\epsilon,0):x\in D]$
    \State $\mathcal F\gets[(x,\Call{ReflectFailure}{\Phi,x,\epsilon}):x\in D]$
    \State \Return $([0]^{|D|},\mathcal T,\mathcal F)$
\EndFunction
\end{algorithmic}
\end{algorithm}

Algorithm~\ref{alg:dd-revise} updates the selected component $\kappa\in\mathcal K$ using the traces and reflections, including those from \textsc{FailedBatch}. \textsc{Reflect}, \textsc{ReflectFailure}, \textsc{UpdateTask}, \textsc{ProposeChat}, \textsc{GenerateCode}, and \textsc{RepairCode} denote the corresponding optimizer LLM calls. $\pi$ is the proposed chat representation used for code generation. Assignments to $\Phi[p_{\mathrm{task}}]$ or $\Phi[f_{\mathrm{chat}}]$ update the corresponding field of the working copy; $q_\Phi$ remains unchanged. \textsc{LoadCode} compiles and loads the generated function, returning $\emptyset$ on success or error information otherwise. Initial code generation sets $\rho_\Phi=1$. Compilation and execution share one repair attempt, which sets $\rho_\Phi=0$ when used. Thus, at most two code versions are generated per iteration: the initial code and one repair.

In Algorithm~\ref{alg:dd-render}, the input schema $\mathcal S$, execution time limit $\Delta$, and assignment method $\mu$ are fixed shared settings. \textsc{Encode} constructs $X$ from the selected chat data using $\mathcal S$, preserving candidate IDs and their associations with original utterances or dialogues. $DC$ is shared with $f_{\mathrm{chat}}$ and set to $DC_{\mathrm{gold}}$. \textsc{FixedTemplate} identifies a seed or manual formatting template. \textsc{CompileAndRun} runs the current $f_{\mathrm{chat}}(C,u_{\mathrm{target}})$ using $X$ within $\Delta$. \textsc{Render} returns $(p_{\mathrm{chat}},\delta,X,\epsilon)$, where $\epsilon$ is empty on success and describes the error or content loss on failure; $\mathrm{TIMEOUT}$ denotes a timeout. \textsc{NoContentLoss} checks whether the message text of each utterance in $X$ appears in $p_{\mathrm{chat}}$. Here, $|f_{\mathrm{chat}}|$ denotes source code length. Repair requires $\rho_\Phi=1$ and sets it to 0; timeouts, empty source code, and loss of message content are rejected without repair.

\begin{algorithm}[htbp]
\small
\caption{Updating the task instruction or chat representation}
\label{alg:dd-revise}
\begin{algorithmic}[1]
\Require $\Phi=(p_{\mathrm{task}},f_{\mathrm{chat}})$; $\kappa\in\{\mathrm{task},\mathrm{chat}\}$;
\Statex \hspace{\algorithmicindent} $\mathcal T$, $\mathcal F$; $\mathcal S$
\Ensure $\Phi$ or $\ddfail$
\Function{Update}{$\Phi,\kappa,\mathcal T,\mathcal F$}
    \If{$\kappa=\mathrm{task}$}
        \State $p'_{\mathrm{task}}\gets\Call{UpdateTask}{p_{\mathrm{task}},\mathcal T,\mathcal F}$
        \If{$|p'_{\mathrm{task}}|=0$}
            \State \Return $\ddfail$
        \EndIf
        \State $\Phi[p_{\mathrm{task}}]\gets p'_{\mathrm{task}}$
    \Else
        \State $\pi\gets\Call{ProposeChat}{\Phi,\mathcal T,\mathcal F}$
        \State $f'_{\mathrm{chat}}\gets\Call{GenerateCode}{\pi,\mathcal S}$
        \State $\rho_\Phi\gets1$
        \If{$|f'_{\mathrm{chat}}|=0$}
            \State \Return $\ddfail$
        \EndIf
        \State $\epsilon\gets\Call{LoadCode}{f'_{\mathrm{chat}}}$
        \If{$\epsilon\ne\emptyset$}
            \State $f'_{\mathrm{chat}}\gets\Call{RepairCode}{f'_{\mathrm{chat}},\epsilon}$
            \State $\rho_\Phi\gets0$
            \If{$|f'_{\mathrm{chat}}|=0\lor\Call{LoadCode}{f'_{\mathrm{chat}}}\ne\emptyset$}
                \State \Return $\ddfail$
            \EndIf
        \EndIf
        \State $\Phi[f_{\mathrm{chat}}]\gets f'_{\mathrm{chat}}$
    \EndIf
    \State \Return $\Phi$
\EndFunction
\end{algorithmic}
\end{algorithm}

\begin{algorithm}[htbp]
\small
\caption{Chat formatting, code repair, and content checks}
\label{alg:dd-render}
\begin{algorithmic}[1]
\Require $\Phi=(p_{\mathrm{task}},f_{\mathrm{chat}})$; $C$, $u_{\mathrm{target}}$;
\Statex \hspace{\algorithmicindent} $DC_{\mathrm{gold}}$, $\mathcal S$, $\Delta$
\Ensure $(p_{\mathrm{chat}},\delta,X,\epsilon)$, with $p_{\mathrm{chat}}=\ddfail$ on failure
\Function{Render}{$\Phi,C,u_{\mathrm{target}},DC_{\mathrm{gold}}$}
    \State $\delta\gets0$; $DC\gets DC_{\mathrm{gold}}$
    \State $i\gets\operatorname{index}_C(u_{\mathrm{target}})$
    \State $C_{\mathrm{pre}}\gets\{u_k\in C:\max(1,i-50)\le k<i\}$
    \State $C_{\mathrm{sub}}\gets\emptyset$
    \If{$\mu\in\{\texttt{UA+SC},\texttt{DA+SC}\}$}
        \State $C_{\mathrm{sub}}\gets\{u_k\in C:i<k\le\min(|C|,i+15)\}$
    \EndIf
    \If{$\mu\in\{\texttt{UA},\texttt{UA+SC}\}$}
        \State $X\gets\Call{Encode}{\mathcal S,C_{\mathrm{pre}},u_{\mathrm{target}},C_{\mathrm{sub}}}$
    \Else
        \State $DC_{\mathrm{pre}}\gets\{d\cap C_{\mathrm{pre}}\mid d\in DC,\ d\cap C_{\mathrm{pre}}\ne\emptyset\}$
        \State $X\gets\Call{Encode}{\mathcal S,DC_{\mathrm{pre}},u_{\mathrm{target}},C_{\mathrm{sub}}}$
    \EndIf
    \If{$\Call{FixedTemplate}{f_{\mathrm{chat}}}$}
        \State \Return $(f_{\mathrm{chat}}(C,u_{\mathrm{target}}),\delta,X,\emptyset)$
    \EndIf
    \Loop
        \State $(p_{\mathrm{chat}},\epsilon)\gets$
        \Statex \hspace{\algorithmicindent}$\text{\textsc{CompileAndRun}}(\Phi[f_{\mathrm{chat}}],C,u_{\mathrm{target}},$
        \Statex \hspace{\algorithmicindent}$X,\Delta)$
        \If{$\epsilon=\emptyset$}
            \If{$\Call{NoContentLoss}{p_{\mathrm{chat}},X}$}
                \State \Return $(p_{\mathrm{chat}},\delta,X,\emptyset)$
            \Else
                \State $\epsilon\gets\text{message content loss}$
                \State \Return $(\ddfail,\delta,X,\epsilon)$
            \EndIf
        \EndIf
        \If{$\epsilon=\mathrm{TIMEOUT}\lor|\Phi[f_{\mathrm{chat}}]|=0$}
            \State \Return $(\ddfail,\delta,X,\epsilon)$
        \EndIf
        \If{$\rho_\Phi=0$}
            \State \Return $(\ddfail,\delta,X,\epsilon)$
        \EndIf
        \State $f'_{\mathrm{chat}}\gets\Call{RepairCode}{\Phi[f_{\mathrm{chat}}],\epsilon}$
        \State $\Phi[f_{\mathrm{chat}}]\gets f'_{\mathrm{chat}}$
        \State $\rho_\Phi\gets0$; $\delta\gets1$
    \EndLoop
\EndFunction
\end{algorithmic}
\end{algorithm}

\paragraph{Scope of Chat Representation Updates}
When updating $f_{\mathrm{chat}}$, $p_{\mathrm{task}}$ remains unchanged. The update prompt restricts changes to presentation, excluding task instructions and inferred information. The code generation prompt also excludes output instructions, analysis, and decision logic. Algorithm~\ref{alg:dd-render} checks preservation of message content.

\clearpage
\section{Implementation Details}\label{sec:implementation_details}

We use GEPA version 0.1.1 without its merge operation~\citep{agrawal2026gepa}.

For DA and DA+SC, the task LLM input groups preceding utterances by gold dialogue cluster during execution trace collection, batch verification, and evaluation on $D_{\mathrm{valid}}$. During benchmark evaluation on $D_{\mathrm{test}}$, IRC New Domains, and Slack Chats, these methods instead use dialogue clusters formed by earlier decisions.

\subsection{Local Inference}\label{sec:app_local_inference}
Qwen3.5-9B~\citep{qwen3.5-9b} is the primary task LLM. Gemma-4-12B-it~\citep{gemma4-12b} is used to evaluate the transfer of systems optimized for Qwen3.5-9B, without further optimization (Section~\ref{sec:prompt_generalization}). Open-weight LLMs are run locally using vLLM, including those in Appendix~\ref{sec:reference_llms}. For optimization and the reported benchmark results, we used NVIDIA RTX PRO 5000 and RTX 6000 Ada GPUs, each with 48\,GB of memory. We set \texttt{max\_model\_len} to 40,000 tokens during prompt optimization and 15,000 tokens during benchmark evaluation.

We also verified that $\Phi_{\mathrm{opt}}$ (chat) under DA+SC, optimized with Qwen3.8-Flash and selected at the 100k checkpoint, runs with Qwen3.5-9B on a single NVIDIA GeForce RTX 3090 GPU with 24\,GB of memory. With 50 preceding and 15 subsequent utterances and reasoning disabled, the combined input and output length was at most 5,000 tokens per decision in our evaluation. For this check, we used the original Hugging Face model in BF16 without quantization, with vLLM's \texttt{max\_model\_len}=10,000 and \texttt{temperature}=0. Our released code and optimized prompts allow this system to be run without repeating optimization (Appendix~\ref{sec:all_prompts}).

\subsection{Task LLM Settings}\label{sec:app_task_settings}
For Qwen3.5-9B as the task LLM, we set \texttt{temperature} to 0. All benchmark evaluations of $\Phi_{\mathrm{seed}}$, $\Phi_{\mathrm{manual}}$, and $\Phi_{\mathrm{opt}}$ on $D_{\mathrm{test}}$, IRC New Domains, and Slack Chats disable reasoning, including evaluations with Gemma-4-12B-it.

During optimization, the task LLM uses reasoning for batch evaluation before prompt updates, with a limit of 2k tokens. Reasoning is disabled for batch verification and evaluation on $D_{\mathrm{valid}}$. When the reasoning limit is reached, we append the following fixed suffix before continuing generation, following \citet{yang2025qwen3}: \texttt{Considering the limited time by the user, I have to give the solution based on the thinking directly now.\textbackslash n\textless/think\textgreater\textbackslash n\textbackslash n}

\subsection{Optimizer LLM Settings}
\paragraph{Qwen3.8-Flash}
For the experiments in Section~\ref{sec:results}, we used Qwen3.8-Flash~\citep{qiu2026design} through OpenRouter from September to October 2026. We explicitly disabled reasoning with \texttt{reasoning=\{"enabled":false\}} and left all other generation parameters unspecified, using the defaults provided by the API.

\subsection{LLM Call Counting}\label{sec:app_call_budget}
The counter $N$ counts LLM calls during optimization: task LLM calls for execution trace collection, batch verification, and evaluation on $D_{\mathrm{valid}}$; optimizer LLM calls for reflections, prompt updates, code generation, and code repair; and continuation requests after reasoning is stopped. Code repair calls remain counted when input preparation restarts.

\section{Availability of Prior Models and Code}\label{sec:baseline_availability}
Based on our search, we could not find publicly available trained dialogue disentanglement models for StructuralBERT~\citep{ma2022structural}, StructuralBERT+DRL~\citep{bhukar2023endtoend}, or DiHRL~\citep{li2024revisiting}. We found accessible code for StructuralBERT,\footnote{\url{https://github.com/xbmxb/StructureCharacterization4DD}} but the code URL provided for StructuralBERT+DRL returned HTTP 404 when checked on October 8, 2026.\footnote{\url{https://github.com/karan121bhukar/RL-ConvDisentanglement}} For DiHRL, the \texttt{WHULACC/DiHRL} repository was empty when checked on October 10, 2026.\footnote{\url{https://github.com/WHULACC/DiHRL}} We also found a separate DiHRL implementation repository,\footnote{\url{https://github.com/Aayushjshah/dihrl_slack}} but it contained no trained model.

We also searched for the fine-tuned Llama2-7B model of \citet{aselstyne2025revisiting} and the fine-tuned GPT-4o mini model of \citet{pal2026disentangling}, but could not find publicly available versions of these fine-tuned models.

\section{Detailed Evaluation Results}\label{sec:detailed_results}
Tables~\ref{tab:app_qwen35_by_qwen38flash_100k}--\ref{tab:app_gemma4_by_qwen38flash_100k} report all three optimization targets and four assignment methods on $D_{\mathrm{test}}$, IRC New Domains, and Slack Chats at the 100k checkpoint, using the selection rule in Appendix~\ref{sec:app_optimization_checkpoints}. Table~\ref{tab:app_manual_ubuntu_irc_reference} provides additional results with $\Phi_{\mathrm{manual}}$ on $D_{\mathrm{test}}$.

In these tables, all values are averaged over three evaluation runs. Subscripts ($\pm$) give 95\% confidence interval half-widths calculated using Student's t distribution with two degrees of freedom. AVG is the average of the seven metric scores; its interval is calculated from the three averages computed separately for each run, not from the metric intervals. The confidence intervals reflect variation across these evaluations on each fixed dataset, not across optimization runs.

For each dataset, bold marks the highest in each column within a table. Underlining marks the highest across assignment methods for each optimization target, or for each LLM in Table~\ref{tab:app_manual_ubuntu_irc_reference}.

\subsection{Optimization with Qwen3.8-Flash}
Table~\ref{tab:app_qwen35_by_qwen38flash_100k} reports system evaluation results with Qwen3.8-Flash as the optimizer LLM and Qwen3.5-9B as the task LLM.

\subsection{Transfer to Gemma-4-12B-it}
Table~\ref{tab:app_gemma4_by_qwen38flash_100k} reports the performance of systems optimized for Qwen3.5-9B with Qwen3.8-Flash and evaluated with Gemma-4-12B-it without further optimization.

\subsection{Manually Refined Prompts Across LLMs}\label{sec:reference_llms}
Table~\ref{tab:app_manual_ubuntu_irc_reference} compares $\Phi_{\mathrm{manual}}$ under UA, UA+SC, DA, and DA+SC on $D_{\mathrm{test}}$ with additional LLMs.
These experiments were conducted in October 2026 with reasoning disabled for all models. Open-weight LLMs were run using vLLM with \texttt{temperature}=0. GPT-5.6-Sol was accessed through OpenRouter; its API did not allow temperature to be specified, so we used the API default rather than setting it to zero.

\begin{table*}[p]
  \centering
  \begingroup
  \small
  \setlength{\tabcolsep}{2.5pt}
  \begin{tabular*}{\textwidth}{@{\extracolsep{\fill}}llrrrrrrrr@{}}
    \toprule
    \textbf{System} & \textbf{\shortstack[l]{Assignment\\method}} & \textbf{VI} & \textbf{ARI} & \textbf{1-1} & \textbf{S-F} & \textbf{P} & \textbf{R} & \textbf{F1} & \textbf{AVG} \\
    \midrule
    \multicolumn{10}{@{}l}{\textbf{\boldmath$D_{\mathrm{test}}$}} \\
    $\Phi_{\mathrm{opt}}$ (task) & UA & $90.2_{\pm 0.1}$ & $66.0_{\pm 1.0}$ & $74.7_{\pm 0.4}$ & $77.5_{\pm 0.3}$ & $\underline{30.8}_{\pm 0.5}$ & $\underline{33.0}_{\pm 0.0}$ & $\underline{31.8}_{\pm 0.3}$ & $\underline{57.7}_{\pm 0.4}$ \\
     & UA+SC & $\underline{91.2}_{\pm 0.2}$ & $63.0_{\pm 4.8}$ & $75.3_{\pm 1.5}$ & $\underline{81.0}_{\pm 1.3}$ & $28.6_{\pm 1.3}$ & $32.5_{\pm 2.1}$ & $30.4_{\pm 1.7}$ & $57.4_{\pm 0.8}$ \\
     & DA & $90.5_{\pm 0.4}$ & $67.7_{\pm 3.4}$ & $76.6_{\pm 1.5}$ & $79.0_{\pm 1.3}$ & $25.7_{\pm 0.7}$ & $28.8_{\pm 0.8}$ & $27.2_{\pm 0.6}$ & $56.5_{\pm 0.9}$ \\
     & DA+SC & $90.8_{\pm 0.4}$ & $\underline{70.4}_{\pm 2.4}$ & $\underline{77.8}_{\pm 1.0}$ & $80.7_{\pm 1.3}$ & $20.6_{\pm 0.7}$ & $22.3_{\pm 0.8}$ & $21.5_{\pm 0.7}$ & $54.9_{\pm 0.8}$ \\
    \noalign{\vskip 3pt}
    \cdashline{1-10}
    \noalign{\vskip 3pt}
    $\Phi_{\mathrm{opt}}$ (chat) & UA & $91.7_{\pm 0.1}$ & $68.2_{\pm 0.1}$ & $76.9_{\pm 0.3}$ & $81.8_{\pm 0.2}$ & $32.3_{\pm 0.8}$ & $41.9_{\pm 1.1}$ & $36.5_{\pm 0.9}$ & $61.3_{\pm 0.5}$ \\
     & UA+SC & $90.2_{\pm 0.2}$ & $62.2_{\pm 0.8}$ & $74.5_{\pm 0.5}$ & $79.1_{\pm 0.6}$ & $23.8_{\pm 1.4}$ & $29.5_{\pm 1.8}$ & $26.3_{\pm 1.5}$ & $55.1_{\pm 0.9}$ \\
     & DA & $93.4_{\pm 0.6}$ & $71.5_{\pm 4.1}$ & $81.1_{\pm 1.0}$ & $84.6_{\pm 1.5}$ & $39.4_{\pm 0.6}$ & $44.6_{\pm 0.4}$ & $41.9_{\pm 0.5}$ & $65.2_{\pm 1.0}$ \\
     & DA+SC & $\underline{\mathbf{94.5}}_{\pm 0.3}$ & $\underline{\mathbf{83.4}}_{\pm 3.5}$ & $\underline{\mathbf{84.8}}_{\pm 1.3}$ & $\underline{\mathbf{88.2}}_{\pm 1.0}$ & $\underline{41.3}_{\pm 0.3}$ & $\underline{\mathbf{45.3}}_{\pm 0.8}$ & $\underline{\mathbf{43.2}}_{\pm 0.2}$ & $\underline{\mathbf{68.7}}_{\pm 0.9}$ \\
    \noalign{\vskip 3pt}
    \cdashline{1-10}
    \noalign{\vskip 3pt}
    $\Phi_{\mathrm{opt}}$ (both) & UA & $89.5_{\pm 0.2}$ & $63.4_{\pm 1.1}$ & $72.2_{\pm 0.5}$ & $74.7_{\pm 0.3}$ & $28.5_{\pm 0.4}$ & $27.6_{\pm 0.7}$ & $28.0_{\pm 0.6}$ & $54.8_{\pm 0.5}$ \\
     & UA+SC & $92.3_{\pm 0.0}$ & $69.7_{\pm 0.9}$ & $78.5_{\pm 0.1}$ & $81.8_{\pm 0.2}$ & $33.5_{\pm 0.4}$ & $34.5_{\pm 0.4}$ & $34.0_{\pm 0.3}$ & $60.6_{\pm 0.2}$ \\
     & DA & $\underline{93.7}_{\pm 0.1}$ & $\underline{81.3}_{\pm 0.2}$ & $\underline{82.3}_{\pm 0.1}$ & $\underline{83.9}_{\pm 0.1}$ & $\underline{\mathbf{45.0}}_{\pm 1.2}$ & $\underline{41.0}_{\pm 1.1}$ & $\underline{42.9}_{\pm 1.1}$ & $\underline{67.2}_{\pm 0.5}$ \\
     & DA+SC & $92.1_{\pm 0.2}$ & $74.9_{\pm 2.1}$ & $77.4_{\pm 0.5}$ & $80.7_{\pm 0.2}$ & $34.7_{\pm 1.7}$ & $39.7_{\pm 1.9}$ & $37.0_{\pm 1.7}$ & $62.4_{\pm 0.3}$ \\
    \midrule
    \multicolumn{10}{@{}l}{\textbf{IRC New Domains}} \\
    $\Phi_{\mathrm{opt}}$ (task) & UA & $\underline{91.7}_{\pm 0.1}$ & $67.2_{\pm 1.1}$ & $\underline{77.9}_{\pm 0.7}$ & $\underline{79.1}_{\pm 0.3}$ & $\underline{\mathbf{54.2}}_{\pm 1.2}$ & $\underline{\mathbf{47.1}}_{\pm 0.8}$ & $\underline{\mathbf{50.4}}_{\pm 0.9}$ & $\underline{66.8}_{\pm 0.4}$ \\
     & UA+SC & $90.9_{\pm 0.3}$ & $\underline{70.9}_{\pm 5.9}$ & $75.2_{\pm 1.7}$ & $78.2_{\pm 1.2}$ & $37.4_{\pm 3.4}$ & $31.2_{\pm 2.1}$ & $34.0_{\pm 2.6}$ & $59.7_{\pm 0.4}$ \\
     & DA & $89.8_{\pm 0.2}$ & $64.9_{\pm 5.7}$ & $74.1_{\pm 2.4}$ & $75.6_{\pm 1.5}$ & $36.9_{\pm 1.9}$ & $29.9_{\pm 2.3}$ & $33.0_{\pm 2.2}$ & $57.8_{\pm 1.9}$ \\
     & DA+SC & $88.7_{\pm 0.9}$ & $63.8_{\pm 8.3}$ & $72.0_{\pm 3.3}$ & $76.1_{\pm 2.5}$ & $24.7_{\pm 2.8}$ & $24.3_{\pm 3.4}$ & $24.5_{\pm 3.1}$ & $53.4_{\pm 0.8}$ \\
    \noalign{\vskip 3pt}
    \cdashline{1-10}
    \noalign{\vskip 3pt}
    $\Phi_{\mathrm{opt}}$ (chat) & UA & $91.4_{\pm 0.3}$ & $69.4_{\pm 2.1}$ & $78.7_{\pm 0.6}$ & $82.1_{\pm 0.6}$ & $32.8_{\pm 0.0}$ & $35.9_{\pm 0.0}$ & $34.3_{\pm 0.0}$ & $60.6_{\pm 0.5}$ \\
     & UA+SC & $91.4_{\pm 0.1}$ & $73.7_{\pm 0.3}$ & $78.6_{\pm 0.3}$ & $80.8_{\pm 0.2}$ & $35.2_{\pm 1.6}$ & $34.6_{\pm 2.1}$ & $34.9_{\pm 1.8}$ & $61.3_{\pm 0.8}$ \\
     & DA & $92.6_{\pm 0.5}$ & $74.5_{\pm 7.6}$ & $79.9_{\pm 3.0}$ & $83.2_{\pm 2.0}$ & $42.9_{\pm 2.2}$ & $41.8_{\pm 2.7}$ & $42.4_{\pm 2.5}$ & $65.3_{\pm 1.0}$ \\
     & DA+SC & $\underline{\mathbf{93.3}}_{\pm 0.5}$ & $\underline{\mathbf{82.5}}_{\pm 1.3}$ & $\underline{\mathbf{83.3}}_{\pm 0.7}$ & $\underline{\mathbf{85.6}}_{\pm 0.7}$ & $\underline{44.3}_{\pm 2.1}$ & $\underline{42.9}_{\pm 2.3}$ & $\underline{43.6}_{\pm 2.2}$ & $\underline{\mathbf{67.9}}_{\pm 1.3}$ \\
    \noalign{\vskip 3pt}
    \cdashline{1-10}
    \noalign{\vskip 3pt}
    $\Phi_{\mathrm{opt}}$ (both) & UA & $89.5_{\pm 0.2}$ & $64.0_{\pm 0.4}$ & $73.8_{\pm 0.7}$ & $73.7_{\pm 0.6}$ & $45.0_{\pm 3.8}$ & $32.6_{\pm 2.7}$ & $37.8_{\pm 3.1}$ & $59.5_{\pm 1.5}$ \\
     & UA+SC & $90.4_{\pm 0.5}$ & $70.5_{\pm 6.4}$ & $76.3_{\pm 3.1}$ & $78.2_{\pm 2.6}$ & $37.1_{\pm 1.6}$ & $33.0_{\pm 1.6}$ & $34.9_{\pm 1.6}$ & $60.0_{\pm 1.1}$ \\
     & DA & $89.0_{\pm 0.2}$ & $61.8_{\pm 1.0}$ & $70.2_{\pm 0.6}$ & $70.6_{\pm 0.7}$ & $\underline{53.5}_{\pm 2.2}$ & $34.8_{\pm 1.4}$ & $\underline{42.2}_{\pm 1.6}$ & $60.3_{\pm 1.1}$ \\
     & DA+SC & $\underline{91.1}_{\pm 0.2}$ & $\underline{78.4}_{\pm 3.3}$ & $\underline{79.7}_{\pm 0.6}$ & $\underline{81.7}_{\pm 0.5}$ & $41.9_{\pm 2.7}$ & $\underline{40.0}_{\pm 1.6}$ & $40.9_{\pm 2.1}$ & $\underline{64.8}_{\pm 1.5}$ \\
    \midrule
    \multicolumn{10}{@{}l}{\textbf{Slack Chats}} \\
    $\Phi_{\mathrm{opt}}$ (task) & UA & $\underline{93.9}_{\pm 0.1}$ & $77.9_{\pm 0.5}$ & $81.6_{\pm 0.4}$ & $83.8_{\pm 0.4}$ & $\underline{\mathbf{38.1}}_{\pm 1.3}$ & $\underline{33.9}_{\pm 1.1}$ & $\underline{35.9}_{\pm 1.2}$ & $\underline{63.6}_{\pm 0.7}$ \\
     & UA+SC & $93.8_{\pm 0.3}$ & $\underline{\mathbf{80.7}}_{\pm 0.8}$ & $\underline{83.5}_{\pm 0.4}$ & $\underline{86.1}_{\pm 0.7}$ & $33.3_{\pm 3.2}$ & $32.6_{\pm 3.3}$ & $32.9_{\pm 3.2}$ & $63.3_{\pm 1.6}$ \\
     & DA & $91.9_{\pm 0.7}$ & $70.2_{\pm 4.3}$ & $78.2_{\pm 2.0}$ & $80.0_{\pm 1.9}$ & $30.7_{\pm 2.4}$ & $26.9_{\pm 2.3}$ & $28.7_{\pm 2.4}$ & $58.1_{\pm 2.2}$ \\
     & DA+SC & $92.2_{\pm 0.2}$ & $73.8_{\pm 0.3}$ & $78.9_{\pm 0.7}$ & $83.4_{\pm 0.6}$ & $26.8_{\pm 2.8}$ & $31.3_{\pm 4.8}$ & $28.8_{\pm 3.6}$ & $59.3_{\pm 1.8}$ \\
    \noalign{\vskip 3pt}
    \cdashline{1-10}
    \noalign{\vskip 3pt}
    $\Phi_{\mathrm{opt}}$ (chat) & UA & $93.4_{\pm 0.2}$ & $77.8_{\pm 2.0}$ & $82.5_{\pm 0.7}$ & $86.5_{\pm 0.5}$ & $29.5_{\pm 0.6}$ & $34.7_{\pm 0.6}$ & $31.9_{\pm 0.6}$ & $62.3_{\pm 0.6}$ \\
     & UA+SC & $91.4_{\pm 0.1}$ & $71.4_{\pm 0.7}$ & $76.8_{\pm 0.3}$ & $81.1_{\pm 0.3}$ & $24.8_{\pm 0.6}$ & $28.4_{\pm 0.6}$ & $26.5_{\pm 0.6}$ & $57.2_{\pm 0.3}$ \\
     & DA & $93.6_{\pm 0.3}$ & $78.6_{\pm 4.0}$ & $82.3_{\pm 1.4}$ & $85.8_{\pm 1.5}$ & $31.6_{\pm 1.7}$ & $35.3_{\pm 1.9}$ & $33.3_{\pm 1.8}$ & $62.9_{\pm 0.2}$ \\
     & DA+SC & $\underline{\mathbf{94.1}}_{\pm 0.1}$ & $\underline{80.5}_{\pm 0.3}$ & $\underline{\mathbf{85.2}}_{\pm 0.0}$ & $\underline{\mathbf{88.5}}_{\pm 0.3}$ & $\underline{36.7}_{\pm 0.7}$ & $\underline{\mathbf{43.6}}_{\pm 0.6}$ & $\underline{\mathbf{39.8}}_{\pm 0.7}$ & $\underline{\mathbf{66.9}}_{\pm 0.4}$ \\
    \noalign{\vskip 3pt}
    \cdashline{1-10}
    \noalign{\vskip 3pt}
    $\Phi_{\mathrm{opt}}$ (both) & UA & $90.4_{\pm 0.7}$ & $70.7_{\pm 2.1}$ & $74.8_{\pm 2.1}$ & $75.8_{\pm 1.9}$ & $22.7_{\pm 1.0}$ & $16.4_{\pm 0.6}$ & $19.0_{\pm 0.7}$ & $52.8_{\pm 0.8}$ \\
     & UA+SC & $\underline{92.3}_{\pm 0.3}$ & $75.4_{\pm 1.3}$ & $\underline{80.9}_{\pm 1.1}$ & $83.1_{\pm 0.8}$ & $23.6_{\pm 1.4}$ & $23.7_{\pm 1.1}$ & $23.6_{\pm 1.3}$ & $57.5_{\pm 1.0}$ \\
     & DA & $91.5_{\pm 0.2}$ & $69.1_{\pm 0.6}$ & $74.6_{\pm 0.4}$ & $76.3_{\pm 0.3}$ & $\underline{31.7}_{\pm 2.6}$ & $23.8_{\pm 2.3}$ & $27.2_{\pm 2.4}$ & $56.3_{\pm 1.0}$ \\
     & DA+SC & $92.0_{\pm 0.2}$ & $\underline{77.1}_{\pm 0.4}$ & $80.0_{\pm 0.2}$ & $\underline{83.6}_{\pm 0.3}$ & $26.9_{\pm 1.5}$ & $\underline{29.0}_{\pm 1.1}$ & $\underline{27.9}_{\pm 1.3}$ & $\underline{59.5}_{\pm 0.5}$ \\
    \bottomrule
  \end{tabular*}
  \endgroup
  \caption{System evaluation with Qwen3.8-Flash as the optimizer LLM and Qwen3.5-9B as the task LLM.}
  \label{tab:app_qwen35_by_qwen38flash_100k}
\end{table*}

\begin{table*}[p]
  \centering
  \begingroup
  \small
  \setlength{\tabcolsep}{2.5pt}
  \begin{tabular*}{\textwidth}{@{\extracolsep{\fill}}llrrrrrrrr@{}}
    \toprule
    \textbf{System} & \textbf{\shortstack[l]{Assignment\\method}} & \textbf{VI} & \textbf{ARI} & \textbf{1-1} & \textbf{S-F} & \textbf{P} & \textbf{R} & \textbf{F1} & \textbf{AVG} \\
    \midrule
    \multicolumn{10}{@{}l}{\textbf{\boldmath$D_{\mathrm{test}}$}} \\
    $\Phi_{\mathrm{opt}}$ (task) & UA & $93.7_{\pm 0.2}$ & $70.1_{\pm 2.2}$ & $80.4_{\pm 0.9}$ & $86.1_{\pm 0.6}$ & $37.4_{\pm 0.9}$ & $47.5_{\pm 0.8}$ & $41.9_{\pm 0.7}$ & $65.3_{\pm 0.4}$ \\
     & UA+SC & $\underline{\mathbf{96.4}}_{\pm 0.2}$ & $\underline{\mathbf{87.6}}_{\pm 0.7}$ & $\underline{\mathbf{88.7}}_{\pm 0.6}$ & $\underline{\mathbf{91.6}}_{\pm 0.6}$ & $\underline{\mathbf{54.1}}_{\pm 1.0}$ & $\underline{\mathbf{58.4}}_{\pm 0.4}$ & $\underline{\mathbf{56.2}}_{\pm 0.7}$ & $\underline{\mathbf{76.1}}_{\pm 0.1}$ \\
     & DA & $94.5_{\pm 0.3}$ & $77.1_{\pm 3.4}$ & $83.8_{\pm 1.0}$ & $88.5_{\pm 0.7}$ & $43.9_{\pm 1.5}$ & $52.3_{\pm 1.5}$ & $47.7_{\pm 1.4}$ & $69.7_{\pm 1.1}$ \\
     & DA+SC & $95.1_{\pm 0.1}$ & $77.7_{\pm 0.4}$ & $84.7_{\pm 0.4}$ & $89.3_{\pm 0.2}$ & $46.9_{\pm 3.0}$ & $53.0_{\pm 3.2}$ & $49.7_{\pm 3.1}$ & $70.9_{\pm 1.5}$ \\
    \noalign{\vskip 3pt}
    \cdashline{1-10}
    \noalign{\vskip 3pt}
    $\Phi_{\mathrm{opt}}$ (chat) & UA & $86.6_{\pm 0.2}$ & $37.9_{\pm 0.4}$ & $62.1_{\pm 0.1}$ & $72.0_{\pm 0.1}$ & $20.5_{\pm 0.4}$ & $37.0_{\pm 0.4}$ & $26.4_{\pm 0.4}$ & $48.9_{\pm 0.2}$ \\
     & UA+SC & $93.1_{\pm 0.4}$ & $71.1_{\pm 3.7}$ & $78.2_{\pm 2.0}$ & $84.6_{\pm 1.5}$ & $36.8_{\pm 0.5}$ & $48.2_{\pm 0.0}$ & $41.7_{\pm 0.3}$ & $64.8_{\pm 1.2}$ \\
     & DA & $86.7_{\pm 0.4}$ & $36.6_{\pm 2.1}$ & $62.2_{\pm 1.5}$ & $71.2_{\pm 1.7}$ & $21.6_{\pm 2.0}$ & $37.9_{\pm 2.0}$ & $27.5_{\pm 2.1}$ & $49.1_{\pm 1.6}$ \\
     & DA+SC & $\underline{95.4}_{\pm 0.3}$ & $\underline{84.6}_{\pm 0.6}$ & $\underline{86.2}_{\pm 0.6}$ & $\underline{89.6}_{\pm 0.4}$ & $\underline{48.0}_{\pm 4.7}$ & $\underline{53.1}_{\pm 4.0}$ & $\underline{50.4}_{\pm 4.4}$ & $\underline{72.5}_{\pm 2.1}$ \\
    \noalign{\vskip 3pt}
    \cdashline{1-10}
    \noalign{\vskip 3pt}
    $\Phi_{\mathrm{opt}}$ (both) & UA & $94.6_{\pm 0.2}$ & $79.4_{\pm 0.9}$ & $84.3_{\pm 0.7}$ & $87.9_{\pm 0.6}$ & $43.4_{\pm 1.4}$ & $50.1_{\pm 1.4}$ & $46.5_{\pm 1.4}$ & $69.5_{\pm 0.8}$ \\
     & UA+SC & $94.5_{\pm 0.5}$ & $80.6_{\pm 2.5}$ & $83.8_{\pm 1.2}$ & $87.5_{\pm 1.1}$ & $40.6_{\pm 1.4}$ & $43.4_{\pm 0.7}$ & $41.9_{\pm 1.0}$ & $67.5_{\pm 0.8}$ \\
     & DA & $\underline{95.2}_{\pm 0.1}$ & $\underline{82.2}_{\pm 0.3}$ & $\underline{85.1}_{\pm 0.3}$ & $\underline{89.4}_{\pm 0.4}$ & $\underline{47.1}_{\pm 0.9}$ & $\underline{53.1}_{\pm 0.4}$ & $\underline{49.9}_{\pm 0.7}$ & $\underline{71.7}_{\pm 0.3}$ \\
     & DA+SC & $94.0_{\pm 0.2}$ & $74.2_{\pm 1.7}$ & $80.7_{\pm 0.3}$ & $85.8_{\pm 0.3}$ & $36.9_{\pm 0.8}$ & $46.7_{\pm 2.1}$ & $41.2_{\pm 1.3}$ & $65.6_{\pm 0.5}$ \\
    \midrule
    \multicolumn{10}{@{}l}{\textbf{IRC New Domains}} \\
    $\Phi_{\mathrm{opt}}$ (task) & UA & $93.5_{\pm 0.2}$ & $75.7_{\pm 5.6}$ & $80.2_{\pm 1.3}$ & $84.6_{\pm 0.8}$ & $48.5_{\pm 3.4}$ & $56.5_{\pm 2.7}$ & $52.2_{\pm 3.1}$ & $70.1_{\pm 0.6}$ \\
     & UA+SC & $93.7_{\pm 0.5}$ & $70.8_{\pm 2.9}$ & $81.5_{\pm 1.4}$ & $83.9_{\pm 1.3}$ & $\underline{\mathbf{59.0}}_{\pm 3.2}$ & $55.8_{\pm 1.6}$ & $57.4_{\pm 2.3}$ & $71.7_{\pm 1.8}$ \\
     & DA & $93.7_{\pm 0.2}$ & $75.1_{\pm 4.3}$ & $82.6_{\pm 0.7}$ & $86.2_{\pm 1.5}$ & $52.9_{\pm 1.6}$ & $\underline{\mathbf{58.5}}_{\pm 0.8}$ & $55.5_{\pm 1.2}$ & $72.1_{\pm 0.5}$ \\
     & DA+SC & $\underline{94.5}_{\pm 0.2}$ & $\underline{\mathbf{81.7}}_{\pm 1.2}$ & $\underline{85.4}_{\pm 0.7}$ & $\underline{87.5}_{\pm 0.9}$ & $58.1_{\pm 3.5}$ & $58.0_{\pm 2.1}$ & $\underline{\mathbf{58.0}}_{\pm 2.5}$ & $\underline{74.7}_{\pm 0.7}$ \\
    \noalign{\vskip 3pt}
    \cdashline{1-10}
    \noalign{\vskip 3pt}
    $\Phi_{\mathrm{opt}}$ (chat) & UA & $84.2_{\pm 0.2}$ & $39.5_{\pm 1.7}$ & $59.3_{\pm 1.1}$ & $69.2_{\pm 1.5}$ & $20.0_{\pm 0.4}$ & $38.2_{\pm 0.8}$ & $26.3_{\pm 0.5}$ & $48.1_{\pm 0.4}$ \\
     & UA+SC & $92.8_{\pm 0.4}$ & $69.7_{\pm 1.3}$ & $79.8_{\pm 1.2}$ & $84.7_{\pm 1.0}$ & $45.8_{\pm 1.2}$ & $53.8_{\pm 0.0}$ & $49.5_{\pm 0.7}$ & $68.0_{\pm 0.3}$ \\
     & DA & $84.5_{\pm 0.6}$ & $39.4_{\pm 2.8}$ & $59.6_{\pm 2.7}$ & $68.9_{\pm 2.3}$ & $19.0_{\pm 0.7}$ & $35.3_{\pm 0.0}$ & $24.7_{\pm 0.6}$ & $47.4_{\pm 1.3}$ \\
     & DA+SC & $\underline{94.2}_{\pm 0.7}$ & $\underline{80.4}_{\pm 4.8}$ & $\underline{85.2}_{\pm 2.6}$ & $\underline{86.9}_{\pm 1.3}$ & $\underline{56.8}_{\pm 0.6}$ & $\underline{54.7}_{\pm 2.8}$ & $\underline{55.7}_{\pm 1.6}$ & $\underline{73.4}_{\pm 1.8}$ \\
    \noalign{\vskip 3pt}
    \cdashline{1-10}
    \noalign{\vskip 3pt}
    $\Phi_{\mathrm{opt}}$ (both) & UA & $93.2_{\pm 0.3}$ & $70.5_{\pm 0.8}$ & $81.8_{\pm 1.0}$ & $84.2_{\pm 1.0}$ & $53.0_{\pm 1.3}$ & $53.3_{\pm 2.3}$ & $53.1_{\pm 1.8}$ & $69.9_{\pm 1.2}$ \\
     & UA+SC & $92.5_{\pm 0.4}$ & $76.0_{\pm 4.4}$ & $81.0_{\pm 2.6}$ & $82.8_{\pm 2.1}$ & $43.9_{\pm 2.9}$ & $41.8_{\pm 1.4}$ & $42.9_{\pm 1.9}$ & $65.8_{\pm 0.6}$ \\
     & DA & $\underline{\mathbf{94.7}}_{\pm 0.4}$ & $\underline{80.4}_{\pm 4.9}$ & $\underline{\mathbf{87.1}}_{\pm 1.4}$ & $\underline{\mathbf{89.0}}_{\pm 1.2}$ & $\underline{57.7}_{\pm 2.3}$ & $\underline{57.6}_{\pm 1.4}$ & $\underline{57.7}_{\pm 1.8}$ & $\underline{\mathbf{74.9}}_{\pm 0.4}$ \\
     & DA+SC & $92.0_{\pm 0.5}$ & $74.5_{\pm 2.4}$ & $79.8_{\pm 3.4}$ & $83.4_{\pm 1.6}$ & $39.0_{\pm 3.6}$ & $42.8_{\pm 2.1}$ & $40.8_{\pm 2.6}$ & $64.6_{\pm 2.2}$ \\
    \midrule
    \multicolumn{10}{@{}l}{\textbf{Slack Chats}} \\
    $\Phi_{\mathrm{opt}}$ (task) & UA & $93.9_{\pm 0.1}$ & $77.2_{\pm 0.3}$ & $81.7_{\pm 0.6}$ & $86.9_{\pm 0.4}$ & $36.7_{\pm 0.0}$ & $48.2_{\pm 0.0}$ & $41.7_{\pm 0.0}$ & $66.6_{\pm 0.2}$ \\
     & UA+SC & $\underline{\mathbf{95.5}}_{\pm 0.3}$ & $\underline{\mathbf{85.5}}_{\pm 1.6}$ & $\underline{87.4}_{\pm 1.2}$ & $\underline{\mathbf{90.8}}_{\pm 1.0}$ & $\underline{\mathbf{42.1}}_{\pm 0.4}$ & $\underline{\mathbf{49.4}}_{\pm 0.6}$ & $\underline{\mathbf{45.4}}_{\pm 0.3}$ & $\underline{\mathbf{70.9}}_{\pm 0.5}$ \\
     & DA & $94.1_{\pm 0.2}$ & $79.6_{\pm 2.9}$ & $82.9_{\pm 1.2}$ & $88.1_{\pm 0.9}$ & $36.6_{\pm 1.8}$ & $48.8_{\pm 1.7}$ & $41.8_{\pm 1.8}$ & $67.4_{\pm 1.4}$ \\
     & DA+SC & $94.9_{\pm 0.2}$ & $81.0_{\pm 1.1}$ & $84.5_{\pm 0.9}$ & $88.7_{\pm 0.6}$ & $39.8_{\pm 1.3}$ & $47.9_{\pm 0.6}$ & $43.5_{\pm 1.0}$ & $68.6_{\pm 0.4}$ \\
    \noalign{\vskip 3pt}
    \cdashline{1-10}
    \noalign{\vskip 3pt}
    $\Phi_{\mathrm{opt}}$ (chat) & UA & $84.0_{\pm 0.1}$ & $42.2_{\pm 0.7}$ & $54.6_{\pm 0.5}$ & $66.6_{\pm 0.5}$ & $11.1_{\pm 0.2}$ & $26.0_{\pm 0.6}$ & $15.5_{\pm 0.3}$ & $42.9_{\pm 0.2}$ \\
     & UA+SC & $93.1_{\pm 0.2}$ & $77.0_{\pm 0.4}$ & $79.6_{\pm 0.4}$ & $85.4_{\pm 0.2}$ & $30.6_{\pm 0.8}$ & $41.8_{\pm 0.6}$ & $35.3_{\pm 0.8}$ & $63.3_{\pm 0.5}$ \\
     & DA & $85.3_{\pm 0.4}$ & $47.6_{\pm 1.8}$ & $60.5_{\pm 2.2}$ & $70.9_{\pm 2.2}$ & $14.6_{\pm 1.1}$ & $29.8_{\pm 2.3}$ & $19.6_{\pm 1.4}$ & $46.9_{\pm 1.3}$ \\
     & DA+SC & $\underline{95.4}_{\pm 0.2}$ & $\underline{\mathbf{85.5}}_{\pm 0.4}$ & $\underline{\mathbf{87.8}}_{\pm 0.3}$ & $\underline{\mathbf{90.8}}_{\pm 0.3}$ & $\underline{40.9}_{\pm 1.6}$ & $\underline{47.5}_{\pm 2.3}$ & $\underline{43.9}_{\pm 1.9}$ & $\underline{70.2}_{\pm 1.0}$ \\
    \noalign{\vskip 3pt}
    \cdashline{1-10}
    \noalign{\vskip 3pt}
    $\Phi_{\mathrm{opt}}$ (both) & UA & $93.3_{\pm 0.2}$ & $76.8_{\pm 1.4}$ & $81.2_{\pm 1.2}$ & $86.2_{\pm 0.9}$ & $30.1_{\pm 0.1}$ & $38.8_{\pm 0.0}$ & $33.9_{\pm 0.1}$ & $62.9_{\pm 0.5}$ \\
     & UA+SC & $93.6_{\pm 0.3}$ & $81.0_{\pm 1.0}$ & $83.0_{\pm 0.8}$ & $86.2_{\pm 0.9}$ & $32.6_{\pm 1.3}$ & $34.7_{\pm 1.7}$ & $33.6_{\pm 1.5}$ & $63.5_{\pm 0.9}$ \\
     & DA & $\underline{94.8}_{\pm 0.2}$ & $\underline{82.1}_{\pm 2.1}$ & $\underline{85.0}_{\pm 0.7}$ & $\underline{89.4}_{\pm 0.5}$ & $\underline{37.8}_{\pm 1.3}$ & $\underline{46.6}_{\pm 1.3}$ & $\underline{41.7}_{\pm 1.3}$ & $\underline{68.2}_{\pm 1.0}$ \\
     & DA+SC & $93.2_{\pm 0.6}$ & $77.7_{\pm 2.8}$ & $81.3_{\pm 2.0}$ & $86.7_{\pm 1.6}$ & $27.6_{\pm 1.8}$ & $36.6_{\pm 1.9}$ & $31.5_{\pm 1.9}$ & $62.1_{\pm 1.6}$ \\
    \bottomrule
  \end{tabular*}
  \endgroup
  \caption{Transfer to Gemma-4-12B-it of systems optimized for Qwen3.5-9B with Qwen3.8-Flash, without further optimization.}
  \label{tab:app_gemma4_by_qwen38flash_100k}
\end{table*}

\begin{table*}[p]
  \centering
  \begingroup
  \small
  \setlength{\tabcolsep}{2.5pt}
  \begin{tabular*}{\textwidth}{@{\extracolsep{\fill}}llrrrrrrrr@{}}
    \toprule
    \textbf{LLM} & \textbf{\shortstack[l]{Assignment\\method}} & \textbf{VI} & \textbf{ARI} & \textbf{1-1} & \textbf{S-F} & \textbf{P} & \textbf{R} & \textbf{F1} & \textbf{AVG} \\
    \midrule
    K2-Horizon-7B & UA & $72.5_{\pm 0.0}$ & $6.6_{\pm 0.3}$ & $31.6_{\pm 0.1}$ & $40.7_{\pm 0.1}$ & $2.5_{\pm 0.3}$ & $5.9_{\pm 0.7}$ & $3.6_{\pm 0.4}$ & $23.3_{\pm 0.2}$ \\
     & UA+SC & $70.6_{\pm 0.2}$ & $3.3_{\pm 0.3}$ & $28.0_{\pm 0.3}$ & $36.3_{\pm 0.4}$ & $2.6_{\pm 0.1}$ & $4.4_{\pm 0.4}$ & $3.3_{\pm 0.2}$ & $21.2_{\pm 0.3}$ \\
     & DA & $79.3_{\pm 1.9}$ & $29.6_{\pm 3.2}$ & $47.4_{\pm 2.2}$ & $54.1_{\pm 3.2}$ & $8.3_{\pm 2.3}$ & $11.4_{\pm 3.6}$ & $9.6_{\pm 2.8}$ & $34.2_{\pm 2.6}$ \\
     & DA+SC & $\underline{81.9}_{\pm 1.1}$ & $\underline{37.3}_{\pm 1.3}$ & $\underline{54.1}_{\pm 1.6}$ & $\underline{60.8}_{\pm 1.6}$ & $\underline{14.4}_{\pm 1.6}$ & $\underline{16.2}_{\pm 1.8}$ & $\underline{15.2}_{\pm 1.7}$ & $\underline{40.0}_{\pm 1.0}$ \\
    \noalign{\vskip 3pt}
    \cdashline{1-10}
    \noalign{\vskip 3pt}
    Ling-3.0-tiny (8B-A1B) & UA & $71.6_{\pm 0.0}$ & $23.8_{\pm 1.8}$ & $38.2_{\pm 0.4}$ & $42.3_{\pm 0.4}$ & $1.1_{\pm 0.1}$ & $1.4_{\pm 0.0}$ & $1.2_{\pm 0.0}$ & $25.6_{\pm 0.4}$ \\
     & UA+SC & $65.7_{\pm 1.2}$ & $17.5_{\pm 2.4}$ & $32.4_{\pm 1.5}$ & $31.0_{\pm 2.2}$ & $1.1_{\pm 0.5}$ & $0.7_{\pm 0.4}$ & $0.8_{\pm 0.4}$ & $21.3_{\pm 1.2}$ \\
     & DA & $\underline{73.2}_{\pm 0.8}$ & $\underline{24.5}_{\pm 2.2}$ & $\underline{41.5}_{\pm 1.2}$ & $\underline{45.8}_{\pm 2.5}$ & $\underline{2.2}_{\pm 1.5}$ & $\underline{2.4}_{\pm 1.8}$ & $\underline{2.3}_{\pm 1.6}$ & $\underline{27.4}_{\pm 1.2}$ \\
     & DA+SC & $67.1_{\pm 1.7}$ & $21.0_{\pm 4.1}$ & $34.2_{\pm 3.8}$ & $35.9_{\pm 3.6}$ & $1.8_{\pm 2.4}$ & $1.2_{\pm 1.8}$ & $1.5_{\pm 2.0}$ & $23.2_{\pm 1.1}$ \\
    \noalign{\vskip 3pt}
    \cdashline{1-10}
    \noalign{\vskip 3pt}
    Ornith-1.5-9B & UA & $87.1_{\pm 0.1}$ & $47.8_{\pm 0.5}$ & $66.0_{\pm 0.4}$ & $72.6_{\pm 0.7}$ & $20.0_{\pm 2.0}$ & $26.9_{\pm 2.5}$ & $22.9_{\pm 2.2}$ & $49.0_{\pm 1.0}$ \\
     & UA+SC & $85.5_{\pm 0.1}$ & $39.9_{\pm 1.0}$ & $61.1_{\pm 1.0}$ & $69.4_{\pm 0.4}$ & $14.0_{\pm 0.7}$ & $20.7_{\pm 1.1}$ & $16.7_{\pm 0.8}$ & $43.9_{\pm 0.4}$ \\
     & DA & $\underline{88.7}_{\pm 0.3}$ & $\underline{56.0}_{\pm 2.5}$ & $\underline{71.4}_{\pm 3.1}$ & $\underline{74.7}_{\pm 2.0}$ & $\underline{24.1}_{\pm 2.9}$ & $\underline{27.3}_{\pm 3.1}$ & $\underline{25.6}_{\pm 3.0}$ & $\underline{52.6}_{\pm 2.1}$ \\
     & DA+SC & $87.2_{\pm 1.0}$ & $51.4_{\pm 7.2}$ & $67.1_{\pm 2.3}$ & $71.3_{\pm 2.8}$ & $20.0_{\pm 2.0}$ & $21.3_{\pm 2.8}$ & $20.6_{\pm 2.4}$ & $48.4_{\pm 1.3}$ \\
    \noalign{\vskip 3pt}
    \cdashline{1-10}
    \noalign{\vskip 3pt}
    gpt-oss-120B & UA & $89.1_{\pm 1.1}$ & $64.9_{\pm 12.9}$ & $71.4_{\pm 4.0}$ & $76.4_{\pm 3.9}$ & $18.4_{\pm 1.9}$ & $\underline{20.0}_{\pm 3.2}$ & $\underline{19.2}_{\pm 2.5}$ & $\underline{51.3}_{\pm 4.2}$ \\
     & UA+SC & $87.2_{\pm 0.4}$ & $57.7_{\pm 6.8}$ & $68.2_{\pm 1.3}$ & $73.2_{\pm 1.9}$ & $17.0_{\pm 1.6}$ & $17.7_{\pm 2.0}$ & $17.3_{\pm 1.8}$ & $48.3_{\pm 0.8}$ \\
     & DA & $88.5_{\pm 0.3}$ & $58.6_{\pm 4.0}$ & $69.7_{\pm 0.9}$ & $75.5_{\pm 0.2}$ & $15.6_{\pm 2.8}$ & $16.8_{\pm 2.2}$ & $16.2_{\pm 2.6}$ & $48.7_{\pm 1.7}$ \\
     & DA+SC & $\underline{89.4}_{\pm 1.0}$ & $\underline{66.9}_{\pm 11.4}$ & $\underline{72.7}_{\pm 4.9}$ & $\underline{77.1}_{\pm 4.3}$ & $\underline{18.9}_{\pm 4.3}$ & $16.3_{\pm 4.3}$ & $17.5_{\pm 4.3}$ & $\underline{51.3}_{\pm 4.5}$ \\
    \noalign{\vskip 3pt}
    \cdashline{1-10}
    \noalign{\vskip 3pt}
    Qwen3.5-9B & UA & $\underline{90.2}_{\pm 0.0}$ & $\underline{63.9}_{\pm 1.4}$ & $\underline{74.2}_{\pm 0.7}$ & $76.2_{\pm 0.7}$ & $\underline{33.0}_{\pm 0.5}$ & $32.3_{\pm 0.4}$ & $\underline{32.7}_{\pm 0.1}$ & $\underline{57.5}_{\pm 0.4}$ \\
     & UA+SC & $86.7_{\pm 0.2}$ & $47.8_{\pm 1.4}$ & $65.8_{\pm 1.3}$ & $68.8_{\pm 1.0}$ & $24.4_{\pm 0.1}$ & $24.8_{\pm 0.7}$ & $24.6_{\pm 0.4}$ & $49.0_{\pm 0.7}$ \\
     & DA & $89.8_{\pm 0.6}$ & $57.0_{\pm 3.1}$ & $72.4_{\pm 1.6}$ & $75.7_{\pm 2.2}$ & $31.9_{\pm 2.6}$ & $\underline{33.4}_{\pm 2.8}$ & $\underline{32.7}_{\pm 2.7}$ & $56.1_{\pm 2.1}$ \\
     & DA+SC & $89.9_{\pm 0.5}$ & $60.3_{\pm 7.0}$ & $73.3_{\pm 2.9}$ & $\underline{76.5}_{\pm 2.1}$ & $30.0_{\pm 1.1}$ & $31.7_{\pm 1.5}$ & $30.8_{\pm 1.3}$ & $56.1_{\pm 1.4}$ \\
    \noalign{\vskip 3pt}
    \cdashline{1-10}
    \noalign{\vskip 3pt}
    Qwen3.5-35B-A3B & UA & $\underline{95.8}_{\pm 0.2}$ & $\underline{85.8}_{\pm 0.2}$ & $87.2_{\pm 0.5}$ & $\underline{90.0}_{\pm 0.2}$ & $\underline{49.9}_{\pm 1.9}$ & $\underline{50.4}_{\pm 1.9}$ & $\underline{50.1}_{\pm 1.8}$ & $\underline{72.7}_{\pm 0.9}$ \\
     & UA+SC & $\underline{95.8}_{\pm 0.3}$ & $85.1_{\pm 0.6}$ & $\underline{87.6}_{\pm 0.2}$ & $\underline{90.0}_{\pm 0.2}$ & $\underline{49.9}_{\pm 3.4}$ & $50.1_{\pm 3.5}$ & $50.0_{\pm 3.4}$ & $72.6_{\pm 1.5}$ \\
     & DA & $93.8_{\pm 0.7}$ & $74.4_{\pm 9.9}$ & $82.3_{\pm 1.5}$ & $84.8_{\pm 1.7}$ & $43.8_{\pm 1.2}$ & $44.4_{\pm 1.5}$ & $44.1_{\pm 0.9}$ & $66.8_{\pm 2.1}$ \\
     & DA+SC & $94.9_{\pm 0.3}$ & $78.6_{\pm 2.0}$ & $84.8_{\pm 1.2}$ & $87.6_{\pm 1.2}$ & $47.5_{\pm 4.9}$ & $49.0_{\pm 4.9}$ & $48.3_{\pm 4.9}$ & $70.1_{\pm 2.5}$ \\
    \noalign{\vskip 3pt}
    \cdashline{1-10}
    \noalign{\vskip 3pt}
    Qwen3.8-27B & UA & $94.6_{\pm 0.1}$ & $73.4_{\pm 1.0}$ & $81.9_{\pm 0.2}$ & $87.8_{\pm 0.2}$ & $42.2_{\pm 0.6}$ & $49.7_{\pm 0.8}$ & $45.6_{\pm 0.7}$ & $67.9_{\pm 0.2}$ \\
     & UA+SC & $96.2_{\pm 0.0}$ & $85.4_{\pm 0.0}$ & $87.6_{\pm 0.1}$ & $91.3_{\pm 0.1}$ & $\underline{51.8}_{\pm 0.2}$ & $55.0_{\pm 0.4}$ & $\underline{53.3}_{\pm 0.2}$ & $74.4_{\pm 0.1}$ \\
     & DA & $95.8_{\pm 0.3}$ & $84.9_{\pm 7.2}$ & $87.0_{\pm 1.9}$ & $90.8_{\pm 1.2}$ & $48.5_{\pm 1.3}$ & $54.1_{\pm 1.2}$ & $51.2_{\pm 1.2}$ & $73.2_{\pm 1.3}$ \\
     & DA+SC & $\underline{96.5}_{\pm 0.1}$ & $\underline{89.4}_{\pm 0.1}$ & $\underline{88.6}_{\pm 0.2}$ & $\underline{92.0}_{\pm 0.1}$ & $51.2_{\pm 1.2}$ & $\underline{55.6}_{\pm 1.1}$ & $\underline{53.3}_{\pm 1.1}$ & $\underline{75.2}_{\pm 0.5}$ \\
    \noalign{\vskip 3pt}
    \cdashline{1-10}
    \noalign{\vskip 3pt}
    Gemma-4-12B-it & UA & $94.6_{\pm 0.1}$ & $77.5_{\pm 0.6}$ & $84.5_{\pm 0.3}$ & $\underline{88.3}_{\pm 0.3}$ & $43.8_{\pm 0.7}$ & $48.7_{\pm 0.7}$ & $46.1_{\pm 0.7}$ & $69.1_{\pm 0.4}$ \\
     & UA+SC & $\underline{95.0}_{\pm 0.2}$ & $\underline{79.6}_{\pm 0.7}$ & $\underline{85.2}_{\pm 0.2}$ & $88.2_{\pm 0.3}$ & $\underline{46.4}_{\pm 1.7}$ & $\underline{49.6}_{\pm 1.2}$ & $\underline{48.0}_{\pm 1.5}$ & $\underline{70.3}_{\pm 0.8}$ \\
     & DA & $93.5_{\pm 0.5}$ & $71.1_{\pm 4.5}$ & $80.6_{\pm 1.1}$ & $84.1_{\pm 1.0}$ & $41.4_{\pm 1.5}$ & $46.9_{\pm 2.5}$ & $44.0_{\pm 1.9}$ & $65.9_{\pm 1.6}$ \\
     & DA+SC & $93.8_{\pm 0.6}$ & $77.3_{\pm 3.1}$ & $81.6_{\pm 2.1}$ & $84.5_{\pm 1.5}$ & $38.7_{\pm 1.8}$ & $43.7_{\pm 1.4}$ & $41.0_{\pm 1.6}$ & $65.8_{\pm 1.7}$ \\
    \noalign{\vskip 3pt}
    \cdashline{1-10}
    \noalign{\vskip 3pt}
    Gemma-4-26B-A4B-it & UA & $95.1_{\pm 0.5}$ & $81.2_{\pm 3.3}$ & $85.3_{\pm 1.2}$ & $88.2_{\pm 1.1}$ & $49.1_{\pm 1.5}$ & $53.6_{\pm 2.1}$ & $51.3_{\pm 1.8}$ & $72.0_{\pm 0.7}$ \\
     & UA+SC & $95.5_{\pm 0.1}$ & $82.7_{\pm 1.7}$ & $\underline{86.8}_{\pm 0.5}$ & $\underline{89.5}_{\pm 0.5}$ & $49.5_{\pm 5.3}$ & $\underline{54.8}_{\pm 6.3}$ & $52.0_{\pm 5.8}$ & $73.0_{\pm 2.5}$ \\
     & DA & $94.9_{\pm 1.0}$ & $81.5_{\pm 8.0}$ & $84.6_{\pm 3.7}$ & $87.1_{\pm 3.4}$ & $49.4_{\pm 3.9}$ & $52.9_{\pm 3.3}$ & $51.1_{\pm 3.6}$ & $71.6_{\pm 3.6}$ \\
     & DA+SC & $\underline{95.6}_{\pm 0.4}$ & $\underline{86.2}_{\pm 4.5}$ & $86.7_{\pm 1.5}$ & $88.6_{\pm 1.0}$ & $\underline{50.4}_{\pm 1.8}$ & $54.2_{\pm 2.5}$ & $\underline{52.2}_{\pm 2.1}$ & $\underline{73.4}_{\pm 1.3}$ \\
    \noalign{\vskip 3pt}
    \cdashline{1-10}
    \noalign{\vskip 3pt}
    Gemma-4-31B-it & UA & $96.3_{\pm 0.0}$ & $85.8_{\pm 0.0}$ & $88.3_{\pm 0.0}$ & $91.3_{\pm 0.0}$ & $52.2_{\pm 0.0}$ & $56.6_{\pm 0.0}$ & $54.3_{\pm 0.0}$ & $75.0_{\pm 0.0}$ \\
     & UA+SC & $96.7_{\pm 0.0}$ & $87.8_{\pm 0.0}$ & $89.3_{\pm 0.0}$ & $92.1_{\pm 0.0}$ & $55.1_{\pm 0.0}$ & $60.6_{\pm 0.0}$ & $57.7_{\pm 0.0}$ & $77.1_{\pm 0.0}$ \\
     & DA & $96.9_{\pm 0.0}$ & $90.6_{\pm 0.0}$ & $\underline{\mathbf{90.5}}_{\pm 0.0}$ & $\underline{\mathbf{92.5}}_{\pm 0.0}$ & $\underline{57.7}_{\pm 0.0}$ & $\underline{63.4}_{\pm 0.0}$ & $\underline{60.4}_{\pm 0.0}$ & $\underline{78.9}_{\pm 0.0}$ \\
     & DA+SC & $\underline{\mathbf{97.1}}_{\pm 0.0}$ & $\underline{\mathbf{91.9}}_{\pm 0.0}$ & $90.4_{\pm 0.0}$ & $\underline{\mathbf{92.5}}_{\pm 0.0}$ & $56.1_{\pm 0.0}$ & $\underline{63.4}_{\pm 0.0}$ & $59.5_{\pm 0.0}$ & $78.7_{\pm 0.0}$ \\
    \noalign{\vskip 3pt}
    \cdashline{1-10}
    \noalign{\vskip 3pt}
    GPT-5.6-Sol & UA & $97.0_{\pm 0.3}$ & $\underline{91.6}_{\pm 1.5}$ & $90.2_{\pm 1.2}$ & $92.0_{\pm 1.1}$ & $57.6_{\pm 0.4}$ & $63.3_{\pm 1.1}$ & $60.3_{\pm 0.7}$ & $78.9_{\pm 0.9}$ \\
     & UA+SC & $\underline{\mathbf{97.1}}_{\pm 0.2}$ & $\underline{91.6}_{\pm 1.0}$ & $\underline{\mathbf{90.5}}_{\pm 0.9}$ & $\underline{92.1}_{\pm 0.7}$ & $\underline{\mathbf{58.8}}_{\pm 0.5}$ & $\underline{\mathbf{65.2}}_{\pm 1.1}$ & $\underline{\mathbf{61.8}}_{\pm 0.7}$ & $\underline{\mathbf{79.6}}_{\pm 0.7}$ \\
     & DA & $96.3_{\pm 0.1}$ & $89.1_{\pm 0.4}$ & $87.5_{\pm 0.2}$ & $89.3_{\pm 0.3}$ & $52.9_{\pm 0.4}$ & $59.2_{\pm 0.0}$ & $55.9_{\pm 0.2}$ & $75.7_{\pm 0.1}$ \\
     & DA+SC & $96.6_{\pm 0.2}$ & $91.2_{\pm 0.8}$ & $88.6_{\pm 0.8}$ & $90.5_{\pm 0.7}$ & $54.9_{\pm 1.7}$ & $62.2_{\pm 1.8}$ & $58.3_{\pm 1.7}$ & $77.5_{\pm 1.1}$ \\
    \bottomrule
  \end{tabular*}
  \endgroup
  \caption{Results with $\Phi_{\mathrm{manual}}$ on $D_{\mathrm{test}}$.}
  \label{tab:app_manual_ubuntu_irc_reference}
\end{table*}

\section{Optimization Progress}\label{sec:extended_optimization}
\subsection{Checkpoint Selection}\label{sec:app_optimization_checkpoints}
At checkpoints every 20k calls through 100k, we select the candidate with the highest total score on $D_{\mathrm{valid}}$ among those whose full validation has finished within the call limit. Ties select the earliest candidate added to the candidate system pool. The 100k selection is $\Phi_{\mathrm{opt}}$ used in the reported results.

\subsection{Validation Beyond 100k Calls}
\begin{figure*}[p]
  \centering
  \includegraphics[width=\textwidth]{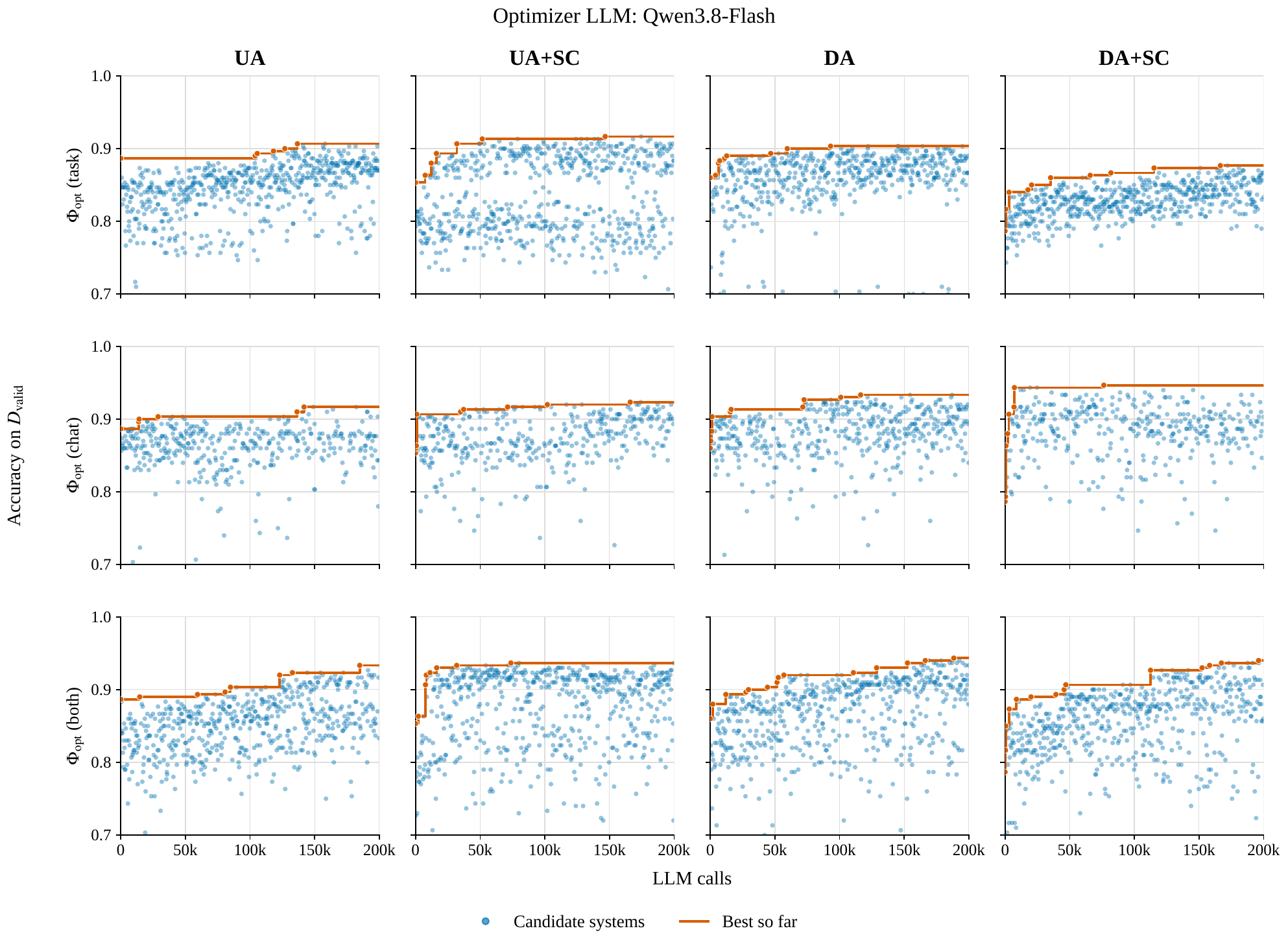}
  \caption{Decision accuracy on $D_{\mathrm{valid}}$ during prompt optimization with Qwen3.8-Flash as the optimizer LLM and Qwen3.5-9B as the task LLM. Blue points show candidate systems; orange lines show the best scores so far. Some runs improve beyond 100k calls after long intervals without improvement.}
  \label{fig:extended_optimization}
\end{figure*}

Figure~\ref{fig:extended_optimization} shows decision accuracy on $D_{\mathrm{valid}}$ against cumulative LLM calls with Qwen3.8-Flash as the optimizer LLM and Qwen3.5-9B as the task LLM. Rows correspond to optimization targets (task, chat, and both), and columns to assignment methods (UA, UA+SC, DA, and DA+SC). Blue points represent candidate systems, and orange lines track the highest score found so far. Some runs improve beyond 100k calls after long intervals without improvement.

\section{Input Token Usage}\label{sec:input_tokens}
\begin{figure}[H]
  \centering
  \includegraphics[width=\linewidth]{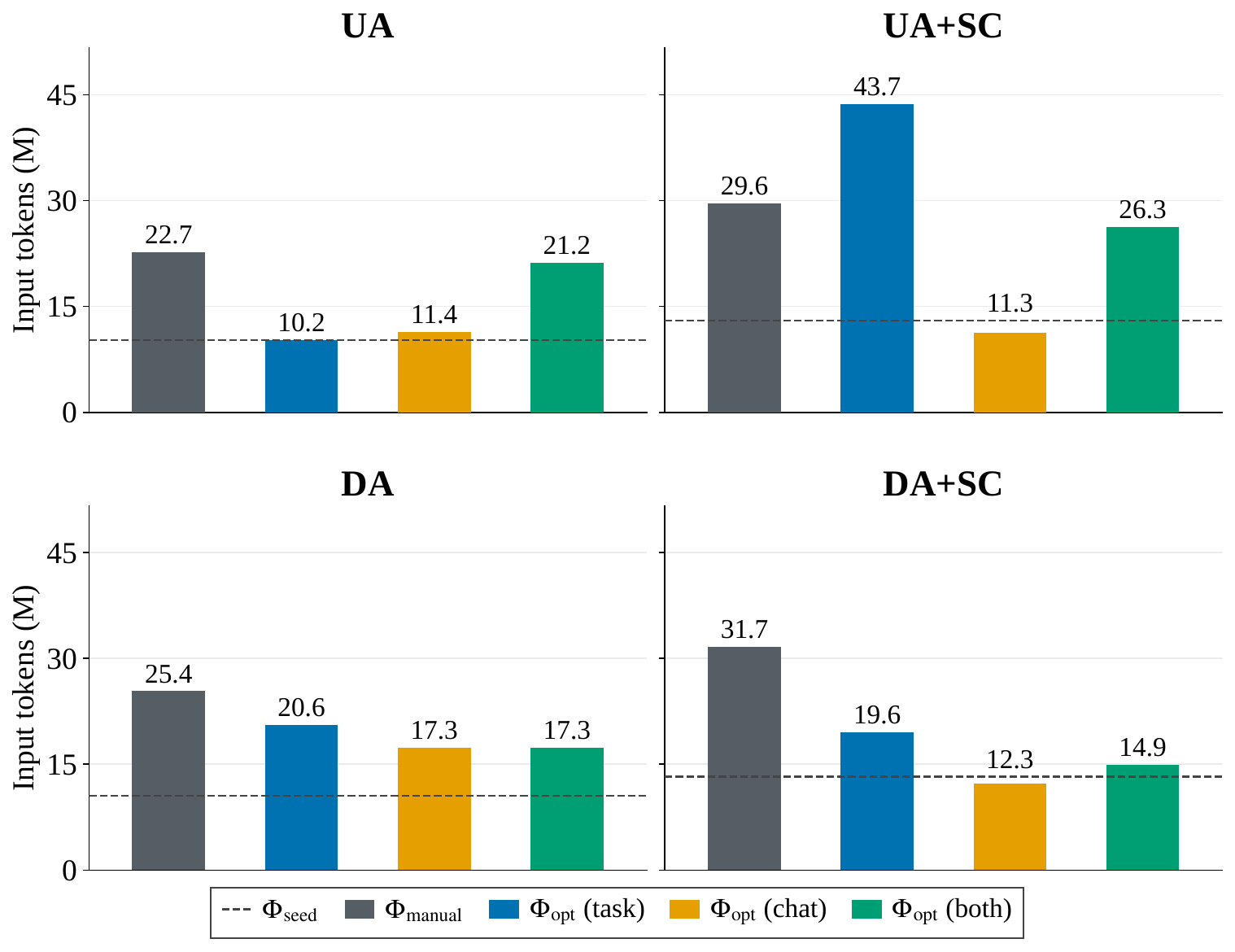}
  \caption{Total input tokens sent to Qwen3.5-9B for all of $D_{\mathrm{test}}$. Dashed lines indicate $\Phi_{\mathrm{seed}}$; M denotes millions. Optimizing chat representation alone generally uses fewer input tokens than optimizing task instructions or both components.}
  \label{fig:input_tokens}
\end{figure}

Figure~\ref{fig:input_tokens} reports the total input tokens sent to Qwen3.5-9B to evaluate all of $D_{\mathrm{test}}$ with each system configuration. These totals measure evaluation input volume, not output tokens or calls made during prompt optimization. Compared with $\Phi_{\mathrm{opt}}$ (task), $\Phi_{\mathrm{opt}}$ (chat) uses fewer input tokens under every assignment method except UA.

\clearpage
\section{Comparison of Optimizer LLMs}\label{sec:stronger_optimizer}
As an additional experiment, we compare Qwen3.5-9B with the main optimizer LLM, Qwen3.8-Flash, keeping Qwen3.5-9B as the task LLM. Both use the 100k checkpoint selection rule in Appendix~\ref{sec:app_optimization_checkpoints}.

\subsection{Qwen3.5-9B Optimizer Settings}
With Qwen3.5-9B as the optimizer LLM, reasoning is enabled for up to 10k tokens. At this limit, generation continues after inserting the fixed suffix described in Appendix~\ref{sec:app_task_settings}. We use the officially recommended sampling parameters for general tasks with reasoning enabled~\citep{qwen3.5-9b}: \texttt{temperature}=1.0, \texttt{top\_p}=0.95, \texttt{top\_k}=20, \texttt{min\_p}=0.0, \texttt{presence\_penalty}=1.5, and \texttt{repetition\_penalty}=1.0.

\subsection{Results}
Table~\ref{tab:stronger_optimizer} compares AVG on $D_{\mathrm{test}}$. Qwen3.8-Flash yields higher mean AVG across assignment methods for all three optimization targets, but not for every configuration. 

\begin{table}[h]
\centering
\small
\setlength{\tabcolsep}{5pt}
\begin{tabular}{@{}lccccc@{}}
\toprule
\textbf{\shortstack[l]{Optimizer\\LLM}} & \textbf{UA} & \textbf{UA+SC} & \textbf{DA} & \textbf{DA+SC} & \textbf{Mean} \\
\midrule
\multicolumn{6}{@{}l}{$\Phi_{\mathrm{opt}}$ (task)} \\
Qwen3.5-9B & 53.4 & 49.0 & 55.6 & \textbf{59.5} & 54.4 \\
Qwen3.8-Flash & \textbf{57.7} & \textbf{57.4} & \textbf{56.5} & 54.9 & \textbf{56.6} \\
\noalign{\vskip 2pt}
\cdashline{2-6}
\noalign{\vskip 2pt}
$\Delta$ & +4.3 & +8.4 & +0.9 & -4.6 & +2.2 \\
\midrule
\multicolumn{6}{@{}l}{$\Phi_{\mathrm{opt}}$ (chat)} \\
Qwen3.5-9B    & 60.1 & \textbf{60.8} & 58.1 & 53.0 & 58.0 \\
Qwen3.8-Flash & \textbf{61.3} & 55.1 & \textbf{65.2} & \textbf{68.7} & \textbf{62.6} \\
\noalign{\vskip 2pt}
\cdashline{2-6}
\noalign{\vskip 2pt}
$\Delta$ & +1.2 & -5.7 & +7.1 & +15.7 & +4.6 \\
\midrule
\multicolumn{6}{@{}l}{$\Phi_{\mathrm{opt}}$ (both)} \\
Qwen3.5-9B    & \textbf{55.3} & 57.2 & 61.3 & 58.0 & 58.0 \\
Qwen3.8-Flash & 54.8 & \textbf{60.6} & \textbf{67.2} & \textbf{62.4} & \textbf{61.2} \\
\noalign{\vskip 2pt}
\cdashline{2-6}
\noalign{\vskip 2pt}
$\Delta$ & -0.5 & +3.4 & +5.9 & +4.4 & +3.2 \\
\bottomrule
\end{tabular}
\caption{AVG on $D_{\mathrm{test}}$ with Qwen3.5-9B as the task LLM when changing the optimizer LLM. The stronger optimizer LLM, Qwen3.8-Flash, performs better on average, although some configurations yield lower scores.}
\label{tab:stronger_optimizer}
\end{table}

Table~\ref{tab:app_qwen35_by_qwen35_100k} reports the 100k results on $D_{\mathrm{test}}$, IRC New Domains, and Slack Chats, using the reporting conventions in Appendix~\ref{sec:detailed_results}. Figure~\ref{fig:app_qwen35_main_results} summarizes AVG on $D_{\mathrm{test}}$. Under UA, $\Phi_{\mathrm{opt}}$ (task) and $\Phi_{\mathrm{opt}}$ (both) perform worse than $\Phi_{\mathrm{seed}}$, whereas $\Phi_{\mathrm{opt}}$ (chat) improves AVG. All three optimization targets improve AVG under the other assignment methods.

\begin{table*}[p]
  \centering
  \begingroup
  \small
  \setlength{\tabcolsep}{2.5pt}
  \begin{tabular*}{\textwidth}{@{\extracolsep{\fill}}llrrrrrrrr@{}}
    \toprule
    \textbf{System} & \textbf{\shortstack[l]{Assignment\\method}} & \textbf{VI} & \textbf{ARI} & \textbf{1-1} & \textbf{S-F} & \textbf{P} & \textbf{R} & \textbf{F1} & \textbf{AVG} \\
    \midrule
    \multicolumn{10}{@{}l}{\textbf{\boldmath$D_{\mathrm{test}}$}} \\
    $\Phi_{\mathrm{opt}}$ (task) & UA & $89.6_{\pm 0.4}$ & $59.1_{\pm 2.4}$ & $72.6_{\pm 1.3}$ & $75.6_{\pm 0.9}$ & $25.3_{\pm 1.6}$ & $25.9_{\pm 1.4}$ & $25.6_{\pm 1.5}$ & $53.4_{\pm 0.2}$ \\
     & UA+SC & $87.2_{\pm 0.2}$ & $59.5_{\pm 1.6}$ & $69.5_{\pm 0.8}$ & $72.5_{\pm 0.5}$ & $18.0_{\pm 2.8}$ & $18.4_{\pm 2.6}$ & $18.2_{\pm 2.7}$ & $49.0_{\pm 1.5}$ \\
     & DA & $90.5_{\pm 0.8}$ & $70.0_{\pm 6.1}$ & $76.1_{\pm 2.2}$ & $78.2_{\pm 1.6}$ & $23.8_{\pm 2.0}$ & $25.8_{\pm 2.9}$ & $24.8_{\pm 2.4}$ & $55.6_{\pm 2.6}$ \\
     & DA+SC & $\underline{91.5}_{\pm 1.1}$ & $\underline{\mathbf{73.9}}_{\pm 10.8}$ & $\underline{\mathbf{79.9}}_{\pm 3.5}$ & $\underline{81.4}_{\pm 3.2}$ & $\underline{29.2}_{\pm 1.6}$ & $\underline{30.8}_{\pm 2.5}$ & $\underline{30.0}_{\pm 2.0}$ & $\underline{59.5}_{\pm 3.4}$ \\
    \noalign{\vskip 3pt}
    \cdashline{1-10}
    \noalign{\vskip 3pt}
    $\Phi_{\mathrm{opt}}$ (chat) & UA & $91.6_{\pm 0.3}$ & $66.2_{\pm 3.5}$ & $77.9_{\pm 0.7}$ & $80.9_{\pm 0.9}$ & $\underline{33.6}_{\pm 1.0}$ & $\underline{35.9}_{\pm 0.8}$ & $\underline{34.7}_{\pm 0.9}$ & $60.1_{\pm 0.6}$ \\
     & UA+SC & $\underline{\mathbf{92.1}}_{\pm 0.2}$ & $\underline{73.3}_{\pm 0.4}$ & $\underline{78.9}_{\pm 1.1}$ & $\underline{\mathbf{83.1}}_{\pm 0.9}$ & $30.8_{\pm 0.4}$ & $34.7_{\pm 1.1}$ & $32.6_{\pm 0.7}$ & $\underline{60.8}_{\pm 0.5}$ \\
     & DA & $90.6_{\pm 0.6}$ & $64.9_{\pm 3.7}$ & $76.2_{\pm 1.3}$ & $79.3_{\pm 0.8}$ & $29.4_{\pm 0.5}$ & $34.6_{\pm 1.1}$ & $31.8_{\pm 0.8}$ & $58.1_{\pm 1.1}$ \\
     & DA+SC & $89.7_{\pm 0.8}$ & $64.8_{\pm 13.8}$ & $71.2_{\pm 4.3}$ & $73.8_{\pm 3.1}$ & $24.3_{\pm 2.0}$ & $23.3_{\pm 1.6}$ & $23.8_{\pm 1.8}$ & $53.0_{\pm 3.9}$ \\
    \noalign{\vskip 3pt}
    \cdashline{1-10}
    \noalign{\vskip 3pt}
    $\Phi_{\mathrm{opt}}$ (both) & UA & $90.2_{\pm 0.4}$ & $59.1_{\pm 4.2}$ & $73.5_{\pm 1.5}$ & $77.0_{\pm 1.3}$ & $28.3_{\pm 1.1}$ & $29.9_{\pm 1.4}$ & $29.1_{\pm 1.2}$ & $55.3_{\pm 1.0}$ \\
     & UA+SC & $90.4_{\pm 0.3}$ & $60.3_{\pm 4.2}$ & $74.2_{\pm 2.0}$ & $76.8_{\pm 1.3}$ & $32.0_{\pm 2.3}$ & $34.0_{\pm 2.2}$ & $33.0_{\pm 2.3}$ & $57.2_{\pm 2.1}$ \\
     & DA & $\underline{91.8}_{\pm 0.5}$ & $\underline{73.0}_{\pm 1.2}$ & $\underline{76.8}_{\pm 1.4}$ & $\underline{78.3}_{\pm 1.1}$ & $36.2_{\pm 0.8}$ & $\underline{\mathbf{36.4}}_{\pm 1.6}$ & $\underline{\mathbf{36.3}}_{\pm 1.2}$ & $\underline{\mathbf{61.3}}_{\pm 1.1}$ \\
     & DA+SC & $90.3_{\pm 0.5}$ & $65.0_{\pm 1.4}$ & $72.7_{\pm 1.3}$ & $73.9_{\pm 1.2}$ & $\underline{\mathbf{38.4}}_{\pm 4.1}$ & $31.5_{\pm 2.1}$ & $34.6_{\pm 2.9}$ & $58.0_{\pm 1.6}$ \\
    \midrule
    \multicolumn{10}{@{}l}{\textbf{IRC New Domains}} \\
    $\Phi_{\mathrm{opt}}$ (task) & UA & $88.8_{\pm 0.2}$ & $65.2_{\pm 0.8}$ & $69.4_{\pm 0.5}$ & $71.4_{\pm 0.5}$ & $32.4_{\pm 1.8}$ & $\underline{24.5}_{\pm 1.4}$ & $\underline{27.9}_{\pm 1.5}$ & $54.2_{\pm 0.9}$ \\
     & UA+SC & $\underline{89.8}_{\pm 1.2}$ & $\underline{71.3}_{\pm 6.2}$ & $\underline{73.7}_{\pm 4.2}$ & $\underline{75.1}_{\pm 3.4}$ & $32.7_{\pm 1.9}$ & $23.0_{\pm 0.8}$ & $27.0_{\pm 0.9}$ & $\underline{56.1}_{\pm 2.6}$ \\
     & DA & $86.3_{\pm 0.2}$ & $55.6_{\pm 1.8}$ & $66.0_{\pm 0.5}$ & $67.6_{\pm 0.8}$ & $\underline{33.0}_{\pm 3.6}$ & $22.3_{\pm 3.6}$ & $26.6_{\pm 3.7}$ & $51.0_{\pm 1.3}$ \\
     & DA+SC & $85.9_{\pm 0.8}$ & $56.1_{\pm 1.9}$ & $65.6_{\pm 1.2}$ & $65.5_{\pm 1.4}$ & $30.7_{\pm 4.5}$ & $18.7_{\pm 2.8}$ & $23.2_{\pm 3.4}$ & $49.4_{\pm 1.7}$ \\
    \noalign{\vskip 3pt}
    \cdashline{1-10}
    \noalign{\vskip 3pt}
    $\Phi_{\mathrm{opt}}$ (chat) & UA & $92.6_{\pm 0.2}$ & $67.0_{\pm 0.3}$ & $78.7_{\pm 0.6}$ & $81.0_{\pm 0.4}$ & $\underline{\mathbf{53.5}}_{\pm 2.3}$ & $\underline{\mathbf{50.9}}_{\pm 1.6}$ & $\underline{\mathbf{52.2}}_{\pm 1.9}$ & $\underline{\mathbf{68.0}}_{\pm 1.0}$ \\
     & UA+SC & $\underline{\mathbf{92.9}}_{\pm 0.2}$ & $\underline{75.3}_{\pm 1.8}$ & $\underline{\mathbf{82.1}}_{\pm 0.6}$ & $\underline{\mathbf{83.7}}_{\pm 0.7}$ & $48.2_{\pm 0.8}$ & $44.6_{\pm 2.7}$ & $46.3_{\pm 1.8}$ & $67.6_{\pm 1.1}$ \\
     & DA & $89.7_{\pm 0.8}$ & $66.1_{\pm 3.0}$ & $75.2_{\pm 3.2}$ & $78.2_{\pm 3.0}$ & $30.9_{\pm 2.1}$ & $30.8_{\pm 0.8}$ & $30.9_{\pm 1.4}$ & $57.4_{\pm 2.0}$ \\
     & DA+SC & $88.5_{\pm 1.0}$ & $59.6_{\pm 2.0}$ & $67.6_{\pm 3.1}$ & $71.1_{\pm 2.3}$ & $32.8_{\pm 2.2}$ & $27.5_{\pm 2.1}$ & $29.9_{\pm 2.1}$ & $53.9_{\pm 2.1}$ \\
    \noalign{\vskip 3pt}
    \cdashline{1-10}
    \noalign{\vskip 3pt}
    $\Phi_{\mathrm{opt}}$ (both) & UA & $89.7_{\pm 0.8}$ & $65.7_{\pm 7.0}$ & $74.0_{\pm 3.9}$ & $75.4_{\pm 3.3}$ & $36.0_{\pm 2.4}$ & $31.0_{\pm 2.7}$ & $33.3_{\pm 2.6}$ & $57.9_{\pm 3.2}$ \\
     & UA+SC & $\underline{92.4}_{\pm 0.6}$ & $\underline{\mathbf{76.5}}_{\pm 3.5}$ & $\underline{79.2}_{\pm 1.7}$ & $\underline{80.1}_{\pm 1.2}$ & $48.9_{\pm 3.8}$ & $\underline{40.9}_{\pm 3.4}$ & $44.6_{\pm 3.6}$ & $\underline{66.1}_{\pm 2.5}$ \\
     & DA & $90.1_{\pm 0.5}$ & $66.0_{\pm 2.2}$ & $74.3_{\pm 1.6}$ & $75.8_{\pm 1.5}$ & $43.9_{\pm 3.6}$ & $35.3_{\pm 2.7}$ & $39.2_{\pm 3.0}$ & $60.7_{\pm 1.9}$ \\
     & DA+SC & $90.0_{\pm 1.7}$ & $65.8_{\pm 5.1}$ & $73.5_{\pm 4.7}$ & $74.1_{\pm 4.4}$ & $\underline{52.5}_{\pm 7.9}$ & $39.5_{\pm 6.9}$ & $\underline{45.1}_{\pm 7.2}$ & $62.9_{\pm 5.1}$ \\
    \midrule
    \multicolumn{10}{@{}l}{\textbf{Slack Chats}} \\
    $\Phi_{\mathrm{opt}}$ (task) & UA & $92.4_{\pm 0.2}$ & $75.5_{\pm 0.6}$ & $79.5_{\pm 0.4}$ & $80.9_{\pm 0.5}$ & $\underline{33.5}_{\pm 1.6}$ & $\underline{27.5}_{\pm 1.3}$ & $\underline{30.2}_{\pm 1.4}$ & $\underline{60.0}_{\pm 0.5}$ \\
     & UA+SC & $\underline{93.1}_{\pm 0.2}$ & $\underline{78.2}_{\pm 1.1}$ & $\underline{81.1}_{\pm 0.9}$ & $\underline{84.6}_{\pm 0.9}$ & $26.1_{\pm 1.3}$ & $26.0_{\pm 0.6}$ & $26.1_{\pm 1.0}$ & $59.3_{\pm 0.4}$ \\
     & DA & $90.4_{\pm 0.1}$ & $65.1_{\pm 4.2}$ & $72.9_{\pm 1.3}$ & $74.7_{\pm 0.4}$ & $26.9_{\pm 1.5}$ & $20.1_{\pm 1.1}$ & $23.0_{\pm 1.3}$ & $53.3_{\pm 0.5}$ \\
     & DA+SC & $91.9_{\pm 0.8}$ & $74.9_{\pm 4.8}$ & $77.6_{\pm 2.9}$ & $79.7_{\pm 2.3}$ & $31.2_{\pm 4.0}$ & $26.6_{\pm 3.6}$ & $28.8_{\pm 3.8}$ & $58.7_{\pm 1.9}$ \\
    \noalign{\vskip 3pt}
    \cdashline{1-10}
    \noalign{\vskip 3pt}
    $\Phi_{\mathrm{opt}}$ (chat) & UA & $\underline{\mathbf{94.4}}_{\pm 0.0}$ & $\underline{\mathbf{80.4}}_{\pm 0.2}$ & $\underline{\mathbf{83.6}}_{\pm 0.1}$ & $86.9_{\pm 0.2}$ & $\underline{38.0}_{\pm 0.2}$ & $\underline{\mathbf{40.5}}_{\pm 0.6}$ & $\underline{\mathbf{39.2}}_{\pm 0.3}$ & $\underline{\mathbf{66.1}}_{\pm 0.2}$ \\
     & UA+SC & $93.8_{\pm 0.1}$ & $79.7_{\pm 0.5}$ & $\underline{\mathbf{83.6}}_{\pm 0.2}$ & $\underline{\mathbf{87.0}}_{\pm 0.1}$ & $32.9_{\pm 1.7}$ & $35.3_{\pm 2.2}$ & $34.1_{\pm 2.0}$ & $63.8_{\pm 0.9}$ \\
     & DA & $90.6_{\pm 0.2}$ & $69.6_{\pm 0.1}$ & $76.7_{\pm 1.1}$ & $79.4_{\pm 0.6}$ & $24.6_{\pm 2.6}$ & $25.4_{\pm 2.9}$ & $25.0_{\pm 2.7}$ & $55.9_{\pm 1.0}$ \\
     & DA+SC & $92.0_{\pm 0.7}$ & $73.4_{\pm 5.5}$ & $77.2_{\pm 1.6}$ & $79.0_{\pm 1.6}$ & $33.9_{\pm 2.3}$ & $28.4_{\pm 2.3}$ & $30.9_{\pm 2.3}$ & $59.2_{\pm 1.8}$ \\
    \noalign{\vskip 3pt}
    \cdashline{1-10}
    \noalign{\vskip 3pt}
    $\Phi_{\mathrm{opt}}$ (both) & UA & $91.6_{\pm 0.1}$ & $74.2_{\pm 2.4}$ & $77.5_{\pm 0.4}$ & $80.4_{\pm 0.3}$ & $22.8_{\pm 0.7}$ & $23.5_{\pm 0.6}$ & $23.1_{\pm 0.6}$ & $56.2_{\pm 0.6}$ \\
     & UA+SC & $\underline{94.1}_{\pm 0.1}$ & $\underline{79.6}_{\pm 0.2}$ & $\underline{83.5}_{\pm 0.6}$ & $\underline{86.7}_{\pm 0.4}$ & $36.2_{\pm 1.0}$ & $\underline{37.6}_{\pm 0.6}$ & $36.9_{\pm 0.8}$ & $\underline{64.9}_{\pm 0.5}$ \\
     & DA & $90.3_{\pm 0.4}$ & $62.5_{\pm 6.9}$ & $73.2_{\pm 2.4}$ & $75.8_{\pm 1.7}$ & $28.8_{\pm 2.3}$ & $26.5_{\pm 1.7}$ & $27.6_{\pm 2.0}$ & $54.9_{\pm 1.6}$ \\
     & DA+SC & $93.4_{\pm 0.5}$ & $76.8_{\pm 3.5}$ & $82.2_{\pm 1.8}$ & $84.0_{\pm 1.6}$ & $\underline{\mathbf{41.0}}_{\pm 0.4}$ & $37.4_{\pm 0.6}$ & $\underline{39.1}_{\pm 0.5}$ & $\underline{64.9}_{\pm 1.1}$ \\
    \bottomrule
  \end{tabular*}
  \endgroup
  \caption{System evaluation with Qwen3.5-9B as both the optimizer and task LLM.}
  \label{tab:app_qwen35_by_qwen35_100k}
\end{table*}

\begin{figure}[H]
  \centering
  \includegraphics[width=\linewidth]{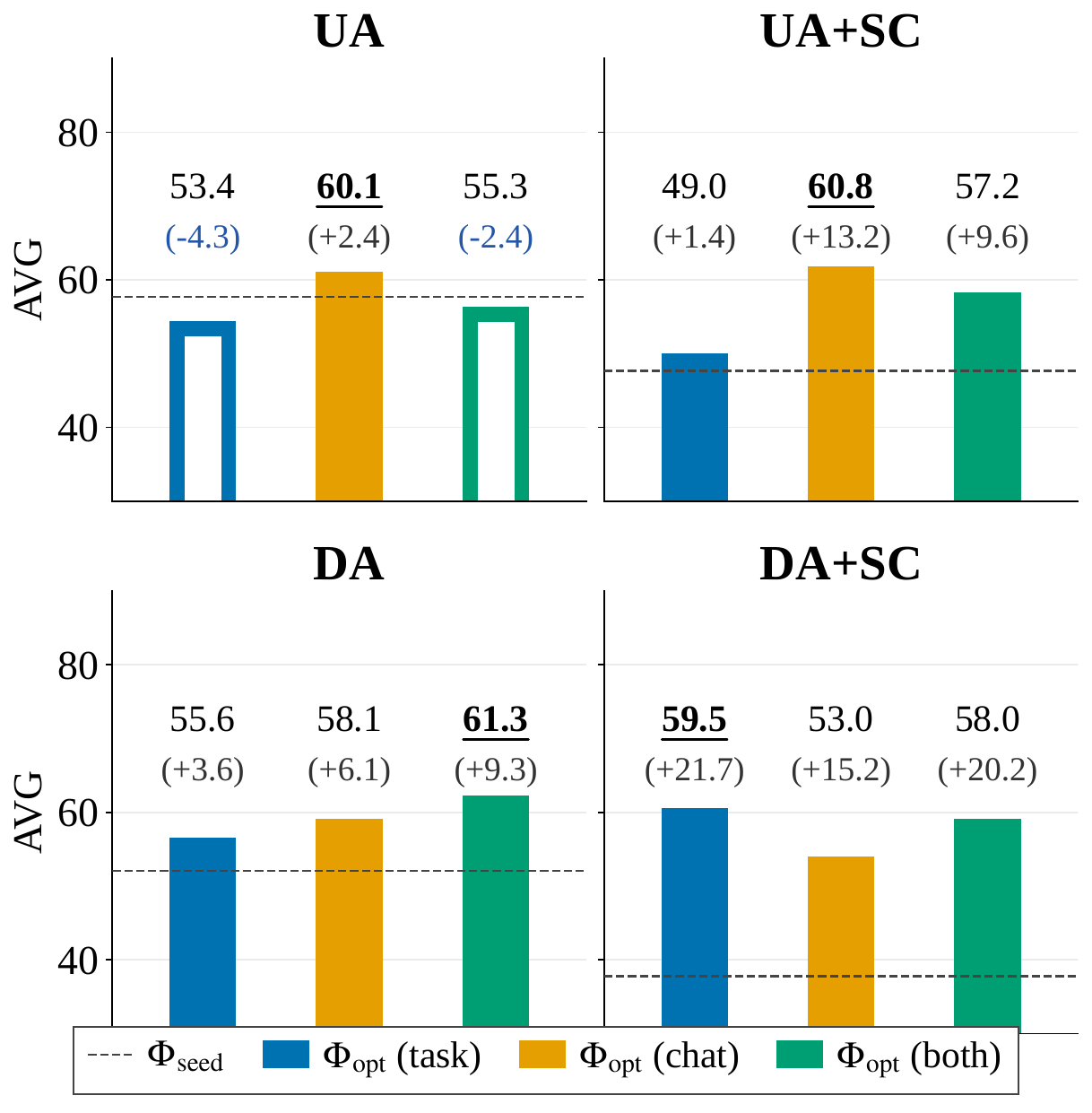}
  \caption{AVG on $D_{\mathrm{test}}$ at the 100k checkpoint with Qwen3.5-9B as both the optimizer and task LLM. Under UA, task instruction optimization and optimization of both components lower AVG relative to $\Phi_{\mathrm{seed}}$.}
  \label{fig:app_qwen35_main_results}
\end{figure}

\subsection{Optimization Progress}
\begin{figure}[H]
  \centering
  \includegraphics[width=\linewidth]{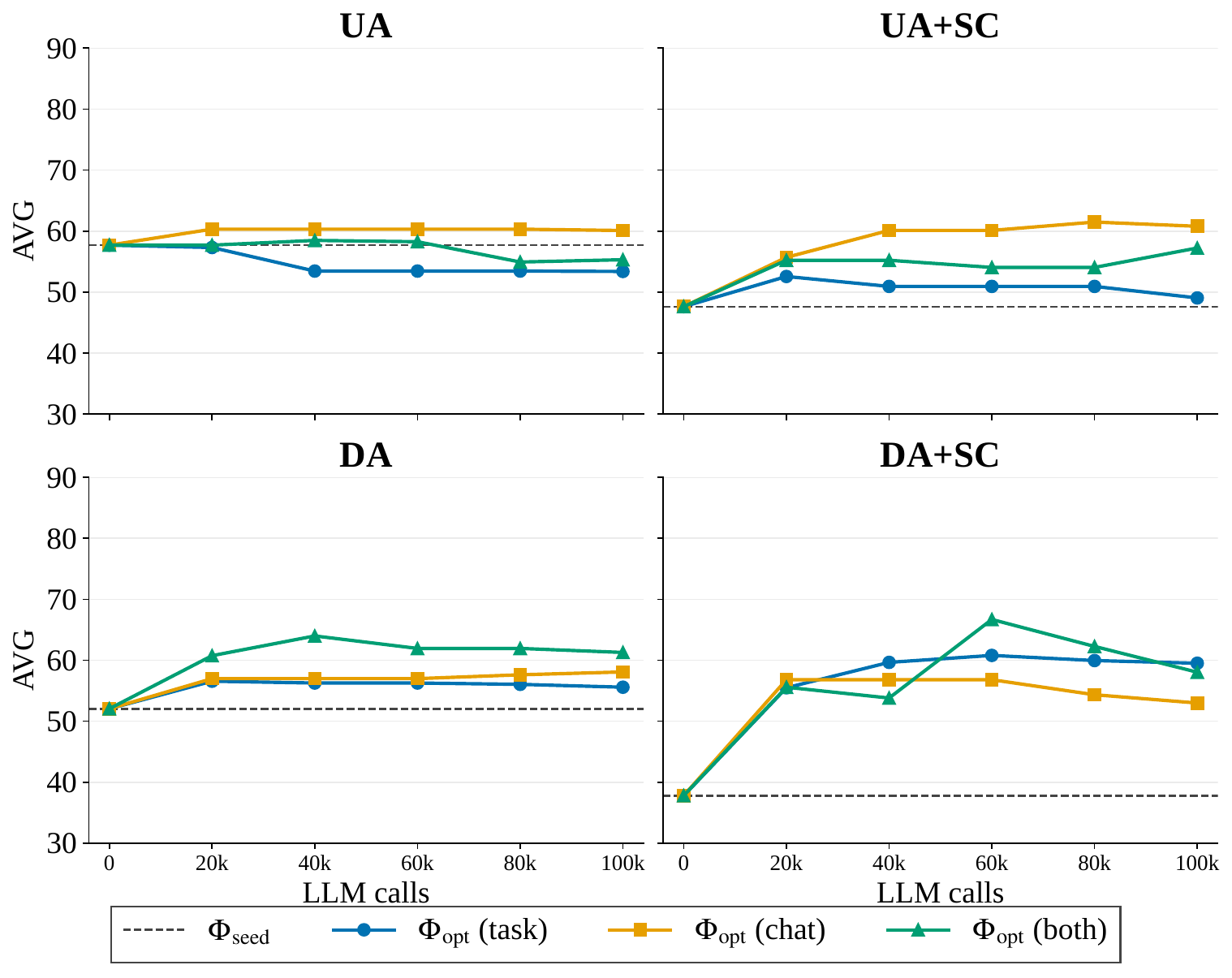}
  \caption{AVG on $D_{\mathrm{test}}$ during optimization with Qwen3.5-9B as both the optimizer and task LLM. More calls do not consistently improve AVG.}
  \label{fig:app_qwen35_optimization_progress}
\end{figure}

Figure~\ref{fig:app_qwen35_optimization_progress} shows AVG on $D_{\mathrm{test}}$ at checkpoints every 20k calls through 100k with Qwen3.5-9B as both the optimizer and task LLM. AVG does not increase monotonically.

\begin{figure*}[p]
  \centering
  \includegraphics[width=\textwidth]{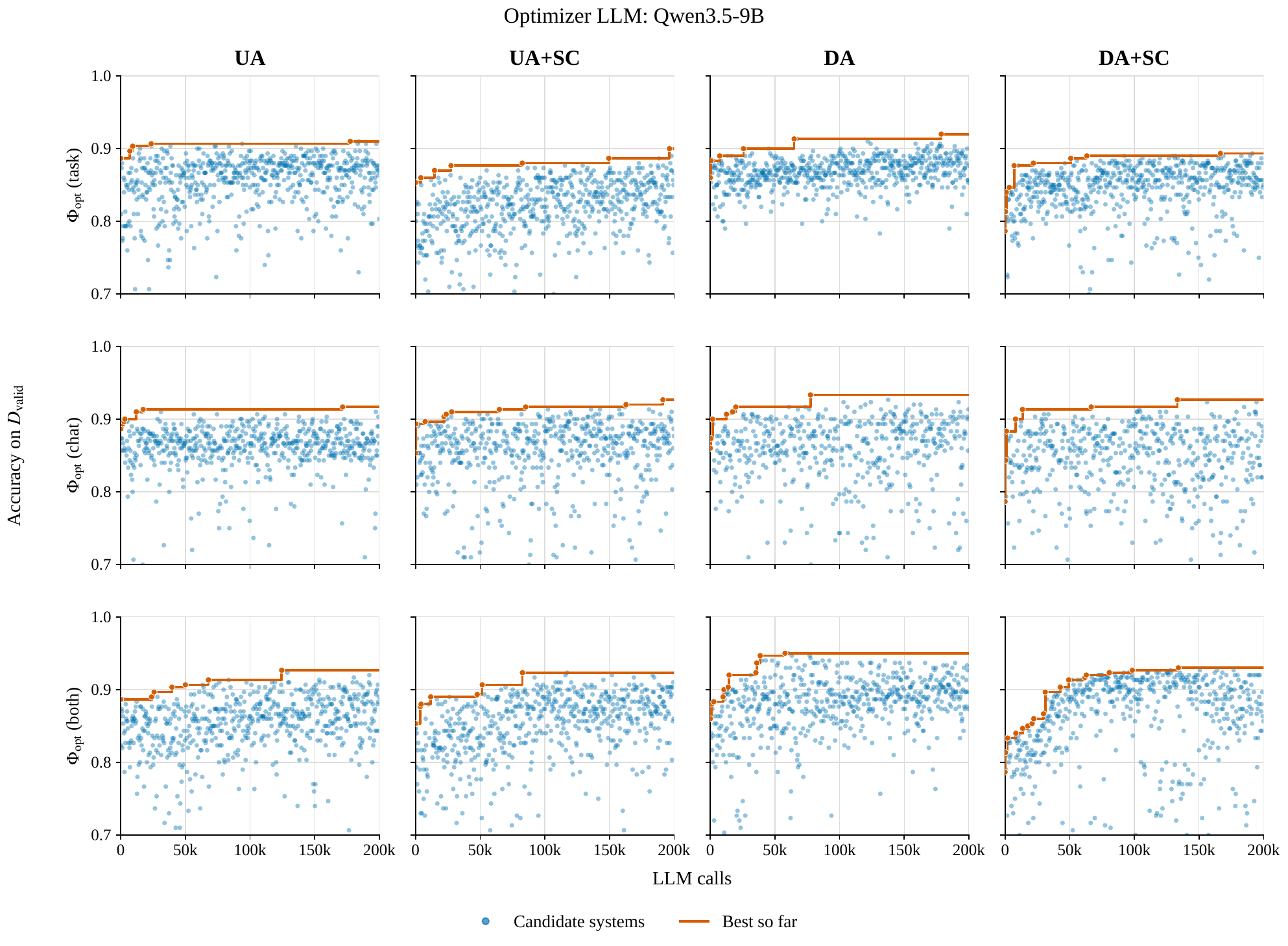}
  \caption{Decision accuracy on $D_{\mathrm{valid}}$ during prompt optimization with Qwen3.5-9B as both the optimizer and task LLM. Blue points show candidate systems; orange lines show the best scores so far. Some runs improve beyond 100k calls after long intervals without improvement.}
  \label{fig:extended_optimization_qwen35}
\end{figure*}

Figure~\ref{fig:extended_optimization_qwen35} shows decision accuracy on $D_{\mathrm{valid}}$ in extended runs with Qwen3.5-9B as the optimizer LLM. As with Qwen3.8-Flash, some runs improve beyond 100k calls after long intervals without improvement. Candidates completing validation after 100k calls were not evaluated on $D_{\mathrm{test}}$.

\clearpage
\onecolumn
\section{Prompts Used in This Study}\label{sec:all_prompts}
The prompts used by APO-DD and selected optimized prompts are available in our code repository: \url{https://anonymous.4open.science/r/DB6B}. Displayed message boundaries and role labels follow the Qwen3 chat format used by Qwen3.5~\citep{qwen3.5-9b}.

\subsection{Prompts for Dialogue Disentanglement}\label{sec:disentanglement_prompts}
We show DA+SC prompt examples for $\Phi_{\mathrm{seed}}$, $\Phi_{\mathrm{manual}}$, and $\Phi_{\mathrm{opt}}$ for all three optimization targets, using the same chat example.

\begin{promptdisplay}
\begin{lstlisting}[style=appendixprompt]
<|im_start|>system
You are given a multi-user chat with each line labeled with dialogue id, timestamp, speaker’s name, and text message for each utterance. Your task is to identify the preceding dialogue that the target utterance responds to, or indicate that it starts a new dialogue. Output exactly one dialogue id that the target utterance responds to. If it starts a new dialogue, output new.

Output example (responding):
5

Output example (new dialogue):
new<|im_end|>
<|im_start|>user
**Previous dialogues:**
**Dialogue 1**
1 [Nov 15 03:07:00] <user_C> What music player do you use?

**Dialogue 2**
2 [Nov 15 03:07:00] <yohannes> Can anyone recommend an app to open a RAR file?
3 [Nov 15 03:10:00] <Hikaru79> yohannes, why not WinRAR?

**Target utterance:**
[Nov 15 03:10:00] <Hikaru79> http://www.rarlab.com/

**Subsequent utterances:**
[Nov 15 03:10:05] <yohannes> thanks
<|im_end|>
<|im_start|>assistant
<think>

</think>


\end{lstlisting}
  \caption{DA+SC prompt example for $\Phi_{\mathrm{seed}}$.}
  \label{fig:app_seed_prompts}
\end{promptdisplay}

\begin{promptdisplay}
\begin{lstlisting}[style=appendixprompt]
<|im_start|>user
# Instruction
You are given a multi-user chat with each line labeled with an index number, timestamp, speaker's name, and text message for an utterance. Your task is to identify to which previous dialogue the target utterance is responding. Assign the target utterance to exactly one existing dialogue from candidate dialogue IDs, or determine it starts a new dialogue. Note that the utterance is more likely to be responding to a nearby one. Subsequent utterances are provided only as reference information - never select the id from Subsequent utterances.

# Chat data

## Candidate dialogue IDs
1
2

## Previous dialogues
[
  {
    "dialogue_id": "1",
    "participants": [
      "user_C"
    ],
    "time_difference_from_target_seconds": 180,
    "utterances": [
      {
        "utterance_id": 1,
        "timestamp": "Nov 15 03:07:00",
        "speaker": "user_C",
        "message": "What music player do you use?"
      }
    ]
  },
  {
    "dialogue_id": "2",
    "participants": [
      "Hikaru79",
      "yohannes"
    ],
    "time_difference_from_target_seconds": 0,
    "utterances": [
      {
        "utterance_id": 2,
        "timestamp": "Nov 15 03:07:00",
        "speaker": "yohannes",
        "message": "Can anyone recommend an app to open a RAR file?"
      },
      {
        "utterance_id": 3,
        "timestamp": "Nov 15 03:10:00",
        "speaker": "Hikaru79",
        "message": "yohannes, why not WinRAR?"
      }
    ]
  }
]

## Target utterance
{
  "utterance_id": 4,
  "timestamp": "Nov 15 03:10:00",
  "speaker": "Hikaru79",
  "message": "http://www.rarlab.com/"
}

## Subsequent utterances
[
  {
    "utterance_id": 5,
    "timestamp": "Nov 15 03:10:05",
    "speaker": "yohannes",
    "message": "thanks"
  }
]

# Confirm
You assign this target utterance.
{
  "utterance_id": 4,
  "timestamp": "Nov 15 03:10:00",
  "speaker": "Hikaru79",
  "message": "http://www.rarlab.com/"
}

# Output
Output exactly one dialogue id that the target utterance responds to. If it starts a new dialogue, output new.

Output example (responding):
5

Output example (new dialogue):
new
<|im_end|>
<|im_start|>assistant
<think>

</think>


\end{lstlisting}
  \caption{DA+SC prompt example for $\Phi_{\mathrm{manual}}$.}
  \label{fig:app_manual_prompt}
\end{promptdisplay}

\begin{promptdisplay}
\begin{lstlisting}[style=appendixprompt]
<|im_start|>system
You are given a multi-user chat log where utterances are grouped into "Previous dialogues" (each labeled with a Dialogue ID) and a single "Target utterance". Your task is to determine which preceding dialogue the target utterance responds to, or if it starts a new dialogue.

**Output Format:**
- Output exactly one integer representing the `Dialogue ID` that the target utterance responds to.
- If the target utterance does not respond to any existing dialogue (i.e., it initiates a new topic, thread, or interaction stream), output the string: `new`.
- Do not provide explanations, JSON, or extra text. Only output the ID or "new".

**Guidelines for Analysis:**

1.  **Speaker Continuity & Direct Address (Crucial):**
    - **Track the Speaker's Thread:** Identify who sent the target utterance. Look for their *most recent* interaction in the previous dialogues. If the speaker was part of a specific dialogue recently, check if the target continues that specific exchange.
    - **New Participant Rule:** If the target utterance addresses a user who has **not** appeared in any of the provided "Previous dialogues", this typically signifies the start of a new interaction stream or dialogue, even if the content relates to an ongoing public discussion. A bare username mention (e.g., `<UserA> UserB:`) without semantic content is often a call-out initiating a new sub-thread or private-like channel within the room. In such cases, prefer `new` unless there is explicit textual evidence linking it as a direct answer to a question posed by `UserB` in a previous dialogue.
    - **Self-Continuation vs. Reply:** If a speaker sends multiple messages in a row (or close succession) directed at the same person/topic, later messages usually belong to the same dialogue ID as the first message in that burst, or extend the most recent dialogue involving those two participants. However, if the speaker switches from interacting with Person X (in Dialogue A) to addressing Person Y (who wasn't in Dialogue A), this is likely `new` or a separate thread.

2.  **Semantic Coherence vs. Keyword Matching:**
    - **Avoid superficial keyword matching.** Just because a word appears in both the target and a previous dialogue (e.g., "windows", "linux", "error") does not mean they belong together. The *intent* and *context* must align.
    - **Topic Introduction Precedes Comparison/Clarification:** A common error is linking a follow-up question to the dialogue where the main subject was initially verified or defined, rather than the dialogue where a *new option* or *alternative* was introduced. 
        - *Example:* If Dialogue A confirms "File X exists," and Dialogue B suggests "Try Alternative Y instead," and the Target asks "Which is better, File X or Alternative Y?", the Target belongs to **Dialogue B**. The comparison only makes sense because Dialogue B introduced the second element. Linking it to Dialogue A ignores the causal link provided by the suggestion in B.
    - **Clarifications follow constraints:** A clarification question (e.g., "on that computer only?") usually follows up on a specific constraint mentioned in a prior query by the same speaker or directed at them.
    - **Interjections and Noise:** An emoticon, short interjection (e.g., ":O", "lol", "Yo Yo MA!!!"), or exclamation may be noise or a reaction to the immediate visual flow rather than a logical response to a technical question.
        - *Exception:* If the interjection is clearly directed at a specific user who just spoke or is the subject of the current active thread *and* the speaker was already involved in that thread, it belongs to that Dialogue ID.
        - *Rule of Thumb:* If the speaker was NOT part of the main conversation thread being discussed (e.g., Lai/Arlie discussing ISPs), but speaks to a user who IS in that thread (Lai), check if the speaker had a prior mini-interaction with that user. If yes, link to that mini-interaction's dialogue. If no prior interaction exists between these two specific users, treat it as `new` (initiating a fresh connection).

3.  **Topic Shifts and Non-Sequiturs:**
    - If the target utterance introduces a completely new subject, person, or observation that has no logical connection to the ongoing discussions in the provided dialogues, it is `new`.
    - Example: A comment about a user's nationality amidst a discussion about kernel security patches is a non-sequitur and starts a new dialogue (`new`).

4.  **Temporal Proximity:**
    - While timestamps help order events, proximity alone is not sufficient evidence. A message occurring shortly after another might still be unrelated if the speakers and topics do not intersect.

5.  **System Messages:**
    - Ignore system messages (joins, parts, nick changes) when determining conversational threads. They are structural markers, not participants in the dialogue logic. Do not assign IDs to system messages.

**Decision Logic Summary:**
1.  Identify the **Speaker** of the target utterance.
2.  Check if the **Addressee** (if explicitly named) was present in the Previous Dialogues.
    - If the Addressee is **NEW** (never seen before): Output `new`.
    - If the Addressee is **EXISTING**:
        - Did the Speaker and Addressee interact in a specific previous Dialogue?
            - Yes: Does the target continue that specific interaction? -> Output that Dialogue ID.
            - No: Is the target a direct answer to a question asked by the Addressee in a different dialogue? -> Output that Dialogue ID.
            - Otherwise: If the Speaker is jumping into a conversation they weren't part of, or starting a fresh DM-style callout without prior context between them, consider `new`.
3.  **Critical Check for Multi-Option Contexts:** If the target compares, chooses between, or reacts to multiple entities/options, trace back to find the dialogue that **introduced the secondary option** or made the comparison possible. Do not default to the dialogue that merely established the primary entity.
4.  If no clear semantic or participant-based link exists to any previous dialogue, output `new`.
<|im_end|>
<|im_start|>user
**Previous dialogues:**
**Dialogue 1**
1 [Nov 15 03:07:00] <user_C> What music player do you use?

**Dialogue 2**
2 [Nov 15 03:07:00] <yohannes> Can anyone recommend an app to open a RAR file?
3 [Nov 15 03:10:00] <Hikaru79> yohannes, why not WinRAR?

**Target utterance:**
[Nov 15 03:10:00] <Hikaru79> http://www.rarlab.com/

**Subsequent utterances:**
[Nov 15 03:10:05] <yohannes> thanks
<|im_end|>
<|im_start|>assistant
<think>

</think>
\end{lstlisting}
  \caption{DA+SC prompt example for $\Phi_{\mathrm{opt}}$ (task).}
  \label{fig:app_opt_task}
\end{promptdisplay}

\begin{promptdisplay}
\begin{lstlisting}[style=appendixprompt]
<|im_start|>system
You are given a multi-user chat with each line labeled with dialogue id, timestamp, speaker’s name, and text message for each utterance. Your task is to identify the preceding dialogue that the target utterance responds to, or indicate that it starts a new dialogue. Output exactly one dialogue id that the target utterance responds to. If it starts a new dialogue, output new.

Output example (responding):
5

Output example (new dialogue):
new
<|im_end|>
<|im_start|>user
# Task: Identify Preceding Dialogue ID

## Decision Logic
- Output `dialogue id` ONLY if the Target Utterance directly answers, corrects, or advances a specific open point in that thread.
- Output `new` if the Target Utterance is a generic greeting/farewell, starts an unrelated topic, or lacks a direct semantic antecedent.
- **Warning:** Do not select a dialogue ID solely based on shared speakers. Check for semantic continuity. Pay close attention to follow-up messages within the same speaker's thread (e.g., status updates closing a previous question).

---

## Candidate Threads

### [Dialogue ID: 1] Status: OPEN
*Participants:* user_C
- **[03:07] user_C**: What music player do you use?

### [Dialogue ID: 2] Status: OPEN
*Participants:* yohannes, Hikaru79
- **[03:07] yohannes**: Can anyone recommend an app to open a RAR file?
- **[03:10] Hikaru79**: yohannes, why not WinRAR?

---

## Target Utterance
**[03:10] Hikaru79**: http://www.rarlab.com/

---

## Future Context (Reference Only - Do Not Select IDs From Here)
*[Non-selectable messages occurring after the target]*
- **[03:10] yohannes**: thanks

---
**Instruction:** Output exactly one dialogue id (e.g., `1`) that the target utterance responds to. If it starts a new dialogue, output `new`.
<|im_end|>
<|im_start|>assistant
<think>

</think>
\end{lstlisting}
  \caption{DA+SC prompt example for $\Phi_{\mathrm{opt}}$ (chat).}
  \label{fig:app_opt_chat}
\end{promptdisplay}

\begin{promptdisplay}
\begin{lstlisting}[style=appendixprompt]
<|im_start|>system
You are given a multi-user chat log consisting of several "Previous dialogues" (each identified by a unique ID and containing multiple utterances with IDs, timestamps, speakers, and text), a "Target utterance," and "Subsequent utterances."

Your task is to determine which **preceding dialogue** the Target utterance responds to. If the Target utterance initiates a completely new topic or conversation that does not relate to any of the provided preceding dialogues, you must indicate this.

**Output Format:**
- Output exactly one integer representing the `Dialogue ID` from the "Previous dialogues" section if the target responds to it.
- Output the string `new` if the target starts a new dialogue.
- Do not output any explanations, JSON structures, or additional text.

**Guidelines for Analysis:**

1.  **Explicit Addressing Takes Precedence Over Temporal Proximity:**
    - In IRC/chat logs, a prefix like `<Speaker> Recipient: message` indicates the speaker is directly addressing the recipient. You must identify the most recent substantive interaction involving that specific **Recipient** in the Previous Dialogues.
    - Do not assume a message belongs to a dialogue simply because it shares the same timestamp minute as other concurrent conversations. If the Target addresses User A, look for User A's last relevant contribution, even if it occurred in a different Dialogue ID than messages happening at the exact same second.
    - *Example:* If the Target is `<Azzie> Felicia: lol`, ignore Dialogues where Lai is leaving or others are talking, unless Felicia was part of those threads. Look specifically for the last meaningful statement made by **Felicia**.

2.  **Semantic Continuity over Surface Structure:**
    - Look beyond literal keyword matching. A response may address the *implication* or *context* of a previous question rather than its specific wording.
    - Short reactions like "lol", ";)", "thanks", or "owned" are rarely standalone new dialogues; they are almost always reactions to the immediately preceding joke, insult, or solution provided by the person being addressed (or the general flow involving them).
    - *Example:* If Jesse asks Felicia why she reboots, and Felicia makes a joke about toilet flushing, Jesse's subsequent `;)` is a closure of *that* specific banter thread, not a reaction to an unrelated system message or a new user entering the room.

3.  **Handling Fragmented Messages and Thread Closure:**
    - Evaluate the entire stream of the speaker's immediate messages. The Target might be the final punctuation mark (wink, smiley, "nm") closing a loop started earlier.
    - If a dialogue ends with a question or a humorous remark, and the Target provides a direct answer or a laugh tag directed at that specific speaker, it belongs to that dialogue.

4.  **Distinguishing New Topics (Critical):**
    - A "new" classification applies when the Target utterance introduces a subject matter entirely unrelated to all preceding dialogues AND lacks explicit addressing tags linking it to a prior participant's open thread.
    - **Speaker Continuity Bias Warning:** Do NOT assign a reply to a dialogue just because the *same speaker* participated in that dialogue previously. If Speaker X was active in Dialogue ID 5, but then posts a message about a completely different topic (e.g., switching from technical troubleshooting to quoting a movie scene) without addressing anyone in Dialogue 5, it is likely `new`. Explicit addressing (`Name:`) is the strongest link; semantic continuity is the next strongest. Mere presence in a past thread is insufficient.
    - System messages (joins/leaves) usually do not constitute a "dialogue" that requires a semantic response unless someone explicitly comments on the join/leave event itself.

5.  **Handling Spam, Noise, and Self-Replies (Critical):**
    - **Spam/Trolling:** If a user posts gibberish, random characters (e.g., "H", "FG", "GDF"), or disruptive noise that does not form a coherent sentence or answer a question, this is generally considered `new` (or noise) rather than a continuation of a valid technical or social thread. It breaks the semantic flow of existing dialogues.
    - **Self-Reply Fallacy:** A user rarely "replies" to their own previous message in a way that links to a *different* dialogue ID unless they are correcting themselves or continuing a complex thought within the *same* dialogue block. However, if a user is spamming (see above), do not link consecutive spam messages to each other as a "dialogue" if they don't interact with others. More importantly, do not assume a short message like "H" is a reply to a previous "H" from the same user if that previous "H" was part of a distinct, isolated spam burst or if the current context suggests a break in behavior.
    - *Specific Example:* If Johnathon spams "F", "GDF", "GDH", "FG", "H", these are disjointed interruptions. They do not respond to Shenna's kernel compilation questions (Dialogue 10) or Arlie's frustration (Dialogue 19). Therefore, such spam is classified as `new`.

6.  **Step-by-Step Reasoning Strategy (Internal):**
    1.  **Identify the Addressee:** Does the Target contain a `Name:` prefix? If yes, note the Name.
    2.  **Locate the Addressee's Last Contribution:** Scan Previous Dialogues for the most recent message by that Name. Determine which Dialogue ID contains that message.
    3.  **Verify Semantic Fit:** Does the Target content logically react to that specific message? (e.g., reacting to a joke, answering a question). If the content is totally unrelated (e.g., movie quote vs. technical help), reject the link despite the addressee history.
    4.  **Check for Noise/Spam:** Is the Target part of a gibberish/spam stream? If so, and it doesn't address a specific unresolved query from another user, classify as `new`.
    5.  **Check for Concurrent Noise:** Ignore other dialogues occurring at the same timestamp unless they involve the same addressee or the Target has no addressee.
    6.  **Select ID or `new`:** If a clear link exists via addressee AND semantics, select that Dialogue ID. If the Target introduces a totally new query/topic, is spam, or lacks a semantic bridge to any prior thread, select `new`.
<|im_end|>
<|im_start|>user
# Chat Context Representation

## Previous Dialogues (Selectable Targets)
Each message includes its unique `utterance ID`, `timestamp`, `speaker`, `text`, and the `dialogue id` it belongs to. Use this list to determine if the Target Utterance continues one of these threads.

- [ID: 1] [Nov 15 03:07:00] <user_C> What music player do you use? | dialogue id: 1
- [ID: 2] [Nov 15 03:07:00] <yohannes> Can anyone recommend an app to open a RAR file? | dialogue id: 2
- [ID: 3] [Nov 15 03:10:00] <Hikaru79> yohannes, why not WinRAR? | dialogue id: 2

## Target Utterance
[Nov 15 03:10:00] <Hikaru79> http://www.rarlab.com/

## Subsequent Utterances (Reference Only - Do Not Select)
These messages occur after the Target and provide context on how the conversation evolved. They are not valid options for the reply-to decision.

- [Nov 15 03:10:00] <yohannes> thanks

***
<|im_end|>
<|im_start|>assistant
<think>

</think>

\end{lstlisting}
  \caption{DA+SC prompt example for $\Phi_{\mathrm{opt}}$ (both).}
  \label{fig:app_opt_both}
\end{promptdisplay}

\clearpage
\subsection{Prompts for APO-DD}\label{sec:prompt}
We present prompts for component updates, reflection generation, code generation, and code repair. Examples specific to an assignment method use DA+SC, and template placeholders are retained. The displayed chat format is illustrative, not the optimizer API's serialization.

\begin{promptdisplay}
\begin{lstlisting}[style=appendixprompt]
I provided an assistant with the following instructions to perform a task for me:

```
<<curr_reflection_task>>
```

The following are examples of different task inputs provided to the assistant along with the assistant's response for each of them, and some feedback on how the assistant's response could be better:

<<inputs_outputs_feedback>>

Your task is to write a new instruction for the assistant. 

Read the inputs carefully and identify the input format and infer detailed task description about the task I wish to solve with the assistant.

Read all the assistant responses and the corresponding feedback. Identify all niche and domain specific factual information about the task and include it in the instruction, as a lot of it may not be available to the assistant in the future. The assistant may have utilized a generalizable strategy to solve the task, if so, include that in the instruction as well.

Provide the new instructions within ``` blocks.
\end{lstlisting}
  \caption{Prompt for updating the task instruction.}
  \label{fig:app_update_task}
\end{promptdisplay}

\begin{promptdisplay}
\begin{lstlisting}[style=appendixprompt]
I provided an assistant with the following instructions to perform a task for me:

```
<<curr_reflection_task>>
```

The following are examples of different task inputs provided to the assistant along with the assistant's response for each of them, and some feedback on how the assistant's response could be better:

<<inputs_outputs_feedback>>

Your task is to improve the representation of the chat data.

Examine the current representation and the example below to understand the input structure, including the previous dialogue groups, target utterance, and subsequent utterances.

Read all the assistant responses and the corresponding feedback. Determine how the data should be organized so that the assistant can make the reply-to decision more accurately.

Keep every previous dialogue individually identifiable and selectable by its dialogue ID, using the exact words "dialogue id" as its label, and preserve the exact message text of every previous, target, and subsequent utterance, with subsequent utterances clearly marked as non-selectable reference context. Redesign all other keys and metadata as needed by renaming, combining, reorganizing, compressing, or removing them to make the reply-to decision easier for the assistant.

Change only how the conversation input is presented. Do not reproduce or modify the task instructions, and do not add inferred information such as topics, summaries, intents, addressees, rankings, or response relationships.

Using the following chat data, provide an improved example and a brief explanation of the representation.

```
<<curr_chat_representation>>
```
\end{lstlisting}
  \caption{DA+SC prompt for updating the chat representation.}
  \label{fig:app_update_chat}
\end{promptdisplay}

\begin{promptdisplay}
\begin{lstlisting}[style=appendixprompt]
<|im_start|>system
You analyze why an LLM's reply-to relation judgment in multi-party chat is incorrect. Examine the prompt used for the judgment, the LLM output, and the correct reply-to relation to explain the mismatch and how the LLM should have used the chat context to identify the correct reply-to relation. 
<|im_end|>
<|im_start|>user
# Prompt Used for the Reply-To Judgment

```
<<decision_input>>
```

# Incorrect LLM Outputs for Reply-To Relation Classification

```
<<model_output>>
```

# Correct Reply-To Relation
<<gold_assignment_text>>

# Task
Analyze why the LLM's reply-to relation judgment is incorrect. Keep the feedback concise and address these points:

- Identify the main error in the LLM's stated reasoning, citing the evidence needed to explain it.
- Describe the key chat evidence and reasoning needed to identify the correct reply-to relation.<|im_end|>
<|im_start|>assistant
<think>

</think>


\end{lstlisting}
  \caption{Prompts for generating reflections.}
  \label{fig:app_reflection_prompts}
\end{promptdisplay}

\begin{promptdisplay}
\begin{lstlisting}[style=appendixprompt]
<|im_start|>system
You are an expert Python coding assistant.

Write one Python program. Output only the program in a single fenced code block using ```python ... ```. Do not include explanations, analysis, sample outputs, or extra Markdown outside the code block.
<|im_end|>
<|im_start|>user
# Task
Write a Python program that converts chat data JSON into text in the proposed chat representation format.

# Proposed Chat Representation Format

```text
<<COMBINED_PROPOSAL_TEXT>>
```

# Runtime Input
At runtime, the program receives chat data JSON on stdin.

# Input Schema
The input is a JSON object under "CHAT_DATA" with the following keys.

- "previous_dialogues": array of previous dialogue group objects.
- "target_utterance": the target utterance object.
- "subsequent_utterances": array of subsequent utterance objects. 

Each previous dialogue group object includes these fields.

- "dialogue_id"
- "participants"
- "time_difference_from_target_seconds"
- "utterances"

Each utterance object in "previous_dialogues" includes these fields.

- "utterance_id"
- "time"
- "speaker"
- "utterance_text"
- "dialogue_id"

The target utterance object and each subsequent utterance object include these fields.

- "utterance_id"
- "time"
- "speaker"
- "utterance_text"

The previous dialogue count may vary. The input contains one target utterance and up to fifteen subsequent utterances.

# Input Example
```json
{
  "CHAT_DATA": {
    "previous_dialogues": [
      {
        "dialogue_id": 1,
        "participants": ["Mina", "Noah"],
        "time_difference_from_target_seconds": 685.0,
        "utterances": [
          {
            "utterance_id": 1,
            "time": "2020-06-19T09:00:00",
            "speaker": "Mina",
            "utterance_text": "Should our log importer stream JSON records instead of loading the entire file before validation?",
            "dialogue_id": 1
          },
          {
            "utterance_id": 3,
            "time": "2020-06-19T09:06:45",
            "speaker": "Noah",
            "utterance_text": "Mina: streaming is safer here because validation can run record by record without a memory spike.",
            "dialogue_id": 1
          }
        ]
      },
      {
        "dialogue_id": 2,
        "participants": ["Lena", "Ravi"],
        "time_difference_from_target_seconds": 330.0,
        "utterances": [
          {
            "utterance_id": 2,
            "time": "2020-06-19T09:01:15",
            "speaker": "Ravi",
            "utterance_text": "The React settings form stopped calling onSubmit after I moved validation into a custom hook.",
            "dialogue_id": 2
          },
          {
            "utterance_id": 4,
            "time": "2020-06-19T09:07:30",
            "speaker": "Lena",
            "utterance_text": "Ravi: check whether the hook returns the submit handler and whether the button still has type='submit'.",
            "dialogue_id": 2
          },
          {
            "utterance_id": 5,
            "time": "2020-06-19T09:12:40",
            "speaker": "Ravi",
            "utterance_text": "Lena: the handler was not returned from the hook; exporting it fixed the form submit.",
            "dialogue_id": 2
          }
        ]
      }
    ],
    "target_utterance": {
      "utterance_id": 6,
      "time": "2020-06-19T09:18:10",
      "speaker": "Mina",
      "utterance_text": "I updated the importer to stream JSON records; the timeout from the original validation step is gone."
    },
    "subsequent_utterances": [
      {
        "utterance_id": 7,
        "time": "2020-06-19T09:18:55",
        "speaker": "Noah",
        "utterance_text": "Mina: nice, streaming should also make retrying failed records easier."
      },
      {
        "utterance_id": 8,
        "time": "2020-06-19T09:24:20",
        "speaker": "Lena",
        "utterance_text": "Ravi: add a small test that clicks submit after the hook initializes."
      },
      {
        "utterance_id": 9,
        "time": "2020-06-19T09:25:10",
        "speaker": "Ravi",
        "utterance_text": "Lena: I added that regression test and it catches the missing handler."
      },
      {
        "utterance_id": 10,
        "time": "2020-06-19T09:31:00",
        "speaker": "Noah",
        "utterance_text": "Mina: the stream parser should also log the record offset before validation fails."
      },
      {
        "utterance_id": 11,
        "time": "2020-06-19T09:35:45",
        "speaker": "Mina",
        "utterance_text": "Noah: good point, I added offset logging so failed records can be replayed."
      }
    ]
  }
}
```

# Required Stdout
The program must write text in the proposed chat representation format to stdout.

- Reproduce only the chat-data presentation structure from the proposal, including its headings, labels, separators, and fixed explanatory text.
- Ignore any task instructions, output instructions, analysis, or decision logic included in the proposal.
- Do not output only a bare listing of utterances unless the proposed format itself requires that.

# Implementation Requirements
- Define main().
- Use only the Python standard library.
- Read stdin JSON in the specified input format.
- Do not hardcode the example chat content from this prompt or from the proposal text.
- Iterate over the arrays from stdin; do not assume the example counts or example IDs.
- Preserve every "utterance_text" value exactly. Do not strip, normalize, collapse, paraphrase, or otherwise alter its characters; use reversible escaping or quoting required by the output format.
<|im_end|>
<|im_start|>assistant
<think>

</think>


\end{lstlisting}
  \caption{Prompts for generating the chat formatting function.}
  \label{fig:app_chat_code_part1}
\end{promptdisplay}

\begin{promptdisplay}
\begin{lstlisting}[style=appendixprompt]
<|im_start|>system
You are an expert Python maintenance assistant.

Return exactly one complete Python program in a single ```python ... ``` code block. Preserve the program's existing behavior and I/O contract. Change only what is necessary to fix the reported error. Do not include explanations, analysis, sample outputs, or extra Markdown outside the code block.
<|im_end|>
<|im_start|>user
# Task
Fix the Python program below so it can run successfully.

# Program

````python
<<BROKEN_CODE>>
````

# Error

```
<<ERROR_MESSAGE>>
```

# Requirements
- Keep the program's intended behavior and stdin/stdout contract unchanged.
- Make the smallest necessary correction.
<|im_end|>
<|im_start|>assistant
<think>

</think>


\end{lstlisting}
  \caption{Prompts for repairing generated code.}
  \label{fig:app_code_repair}
\end{promptdisplay}

\clearpage
\end{document}